# GWTC-4.0: An Introduction to Version 4.0 of the Gravitational-Wave Transient Catalog

The LIGO Scientific Collaboration, the Virgo Collaboration, and the KAGRA Collaboration
(See the end matter for the full list of authors)

## ABSTRACT

The Gravitational-Wave Transient Catalog (GWTC) is a collection of short-duration (transient) gravitational-wave signals identified by the LIGO–Virgo–KAGRA Collaboration in gravitational-wave data produced by the eponymous detectors. The catalog provides information about the identified candidates, such as the arrival time and amplitude of the signal and properties of the signal's source as inferred from the observational data. GWTC is the data release of this dataset and version 4.0 extends the catalog to include observations made during the first part of the fourth LIGO–Virgo–KAGRA observing run up until 2024 January 31. This paper marks an introduction to a collection of articles related to this version of the catalog, GWTC-4.0. The collection of articles accompanying the catalog provides documentation of the methods used to analyze the data, summaries of the catalog of events, observational measurements drawn from the population, and detailed discussions of selected candidates.

## 1. OVERVIEW

The Laser Interferometer Gravitational-Wave Observatory (LIGO; Aasi et al. 2015a) and the Virgo (Acernese et al. 2015) and KAGRA (Akutsu et al. 2021) observatories form an international network of ground-based gravitational-wave (GW) detectors. This paper is an introduction to the collection of articles describing the contents of the LIGO–Virgo–KAGRA Collaboration (LVK) Gravitational-Wave Transient Catalog (GWTC) version 4.0, hereafter GWTC-4.0, along with reviews of the methods used in various aspects in the construction of this catalog, astrophysical and cosmological implications of the observations, and tests of general relativity (GR) that are performed on the observed transients. This paper provides details on the network of GW detectors, the observing runs, observatory evolution, and a review of the transient signals that have been identified. In addition, we describe conventions and notations that are used throughout the collection of papers accompanying the catalog.

### 1.1. *The GWTC Sources and Science*

Transient GW signals may be produced by a variety of astrophysical sources, including compact binary coalescences (CBCs) of compact objects such as black holes (BHs) and neutron stars (NSs), core-collapse supernovae, and other explosive phenomena (Abbott et al. 2020a). The first observed GW transient, GW150914, was a binary black hole (BBH) coalescence (Abbott et al. 2016a), and we have since observed a binary neutron star (BNS) coalescence (Abbott et al. 2017a) that had associated electromagnetic counterparts (Abbott et al. 2017b), and neutron star–black hole binary (NSBH) coalescences (Abbott et al. 2020d).

This GWTC-4.0 collection of papers describes the GW transient candidates observed by the LVK from the first observing run (O1) through the end of the first part of the fourth observing run (O4a) and the astrophysical implications of these observations. The paper collection includes:

- "GWTC-4.0: Methods for Identifying and Characterizing Gravitational-wave Transients" (Abac et al. 2025a) reviews the procedures used to go from the calibrated output of the detectors to a list of transient candidates that includes measurements of the statistical significance and inferences on each of the corresponding astrophysical sources.

- "GWTC-4.0: GWTC-4.0: Updating the Gravitational-Wave Transient Catalog with Observations from the First Part of the Fourth LIGO–Virgo–KAGRA Observing Run" (Abac et al. 2025b) describes the primary observational results contained in GWTC-4.0: the significant GW transient candidates observed through the end of the O4a observing run and the inferred source parameters under the hypothesis that these transients arise from GWs emitted by CBCs (Section 5.2).

Corresponding author: LSC P&P Committee, via LVK Publications as proxy
lvc.publications@ligo.org

- "GWTC-4.0: Population Properties of Merging Compact Binaries" (Abac et al. 2025c) describes the underlying population of CBCs inferred using GWTC-4.0 data, and related astrophysical implications.
- "GWTC-4.0: Tests of General Relativity I — Overview and General Tests" (Abac et al. 2025d) presents an overview of the methods and tests of general relativity performed on the subset of signals suitable for such tests, and focuses on the general and consistency tests.
- "GWTC-4.0: Tests of General Relativity II — Parameterized Tests" (Abac et al. 2025e) describes the parameterized tests of GR performed on the signals.
- "GWTC-4.0: Tests of General Relativity III — Tests of the Remnant" (Abac et al. 2025f) describes the tests of the coalescence remnants.
- "GWTC-4.0: Constraints on the Cosmic Expansion Rate and Modified Gravitational-wave Propagation" (Abac et al. 2025g) describes the methods used to determine the Hubble constant and related parameters, including parameterized deviations from GR on cosmological scales, using GWTC-4.0 candidates.
- "GWTC-4.0: Searches for Gravitational Wave Lensing Signatures" (Abac et al. 2025h) describes the searches for lensed GW signals in the geometric and wave optics regime in the GWTC-4.0 dataset. It also sets constraints on the merger rate at high redshift and the relative rate of strongly lensed signals compared to unlensed ones.
- "Open Data from LIGO, Virgo, and KAGRA through the First Part of the Fourth Observing Run" (Abac et al. 2025i) describes the publicly accessible data and other science products that can be freely accessed through the Gravitational Wave Open Science Center (GWOSC). These datasets include the raw GW strain time series, details of the calibration and cleaning process, efforts to remove instrumental noise artifacts, and details of the online GWTC-4.0.
- "GW230814: Investigation of a Loud Gravitational-wave Signal Observed with a Single Detector" (Abac et al. 2025j) describes the analysis of the loudest event in the GWTC-4.0 catalog, GW230814_230901, which was detected on 2023 August 14. This event is notable for its high signal-to-noise ratio (SNR) and its potential implications for our understanding of GW signals and GR.
- "GW231123: a Binary Black Hole Merger with Total Mass 190–265 $M_\odot$" (Abac et al. 2025k) describes the analysis of the candidate GW231123_135430, detected on 2023 November 23. The candidate's source is exceptional, having the highest inferred total mass of any high-confidence BBH observations to date.

To reference the whole GWTC-4.0 collection, we encourage citing this introductory paper.

### 1.2. *The Electronic Catalog: GWTC*

Abac et al. (2025i) documents the released open data, including the GWTC dataset. The catalog contains *candidates* (sometimes called *events*) identified in observational data that are deemed likely to be caused by GW signals, as well as *triggers* corresponding to times selected by searches of the data for GW transient signals that potentially contain an identifiable signal but with lower confidence of being caused by a GW.

#### 1.2.1. *The Catalog Naming Convention*

The LVK GWTC is a cumulative dataset containing data on all transient candidates reported by the LVK. Released versions of the catalog have major and minor numbers in the format:

`GWTC-`*<major>*`.`*<minor>*

The major number is determined by the span of time containing all candidates in the catalog as described below.

Prior to GWTC-4.0, the minor number was routinely omitted when describing a catalog version when that minor number was 0, so GWTC-1.0, GWTC-2.0, and GWTC-3.0 were referred to as GWTC-1, GWTC-2, and GWTC-3 in the papers that described those catalog versions. In this paper, and in the future, we will include the .0 when referring to those catalog versions. We also say that GWTC-*<major>* can refer to GWTC-*<major>*.*<minor>* for any minor version having that major version number.

Each catalog version is a superset of the previous (apart from retracted candidates), so that, for example, GWTC-3.0 (Abbott et al. 2023) contains all the candidates in GWTC-2.1 (Abbott et al. 2024). Since GWTC-2.1 provided a deeper list of candidates observed over the same period as GWTC-2.0 (Abbott et al. 2021b), the minor version numbers of these two releases differ while their major version numbers remain the same. In general:

- The major number is incremented when the span of time over which observational data were searched for transients is increased.
- The minor version resets to 0 when the major version number is increased.
- The minor version is incremented when there is a change in the data describing the transients (additional data, modified data, or removed data) contained in the catalog within the current timespan covered.

The time span covering the transient candidates in the catalog indicated by the major number is as follows:

GWTC-1: Contains candidates occurring in data taken before 2018 October 01 00:00:00. The GWTC-1.0 dataset is described in Abbott et al. (2019a).

GWTC-2: Contains candidates occurring in data taken before 2019 October 01 15:00:00. The GWTC-2.0 dataset is described in Abbott et al. (2021b) and the GWTC-2.1 dataset in Abbott et al. (2024).

GWTC-3: Contains candidates occurring in data taken before 2020 May 01 00:00:00. The GWTC-3.0 dataset is described in Abbott et al. (2023).

GWTC-4: Contains candidates occurring in data taken before 2024 January 31 00:00:00. The GWTC-4.0 dataset is described in Abac et al. (2025b).

In addition to GWTC, other catalogs of GW transients include the Open Gravitational-wave Catalog (OGC), the most recent version 4-OGC includes observations from 2015 to 2020 (Nitz et al. 2023), as well as catalogs of candidate signals identified by the IAS pipeline (Venumadhav et al. 2019; Olsen et al. 2022; Wadekar et al. 2024; Cheung et al. 2025). Companion paper Abac et al. (2025i) provides details on the GWOSC event portal,[1] a database of published GW transient events, including Community Catalogs (Kanner et al. 2025) containing catalog results from communities outside of the LVK.

#### 1.2.2. *Candidate Naming Conventions*

The naming of our GW candidates follows the format

`GW`*<YY><MM><DD>*`_`*<hh><mm><ss>*

encoding the date and Coordinated Universal Time (UTC) of the signal. For example, GW200105_162426 was the transient observed on 2020 January 5 at 16:24:26 UTC. For transients signals spanning multiple second intervals, the time assigned to a signal is an estimate of the time of peak GW amplitude.

GW candidates reported prior to the release of GWTC-2.0 were designated by the abbreviated form

`GW`*<YY><MM><DD>*

including candidates first appearing in GWTC-1.0 (Abbott et al. 2019a) as well as GW190412 (Abbott et al. 2020e), GW190425 (Abbott et al. 2020b), GW190521 (Abbott et al. 2020f), and GW190814 (Abbott et al. 2020d). These candidates retain their legacy names.

### 1.3. *Outline*

An outline of the remainder of this article is: We briefly describe the network of ground-based GW detectors in Section 2 and their observing runs that have contributed to the GWTC-4.0 in Section 3. These sections are followed by short reviews of the evolution of the various observatories in Section 4 and of the nature of the transient sources observed in Section 5. A list of common acronyms is provided in Appendix A. Mathematical conventions used throughout the articles in this compendium are described in Appendix B.

[1] GWOSC event portal https://gwosc.org/eventapi

## 2. THE INTERNATIONAL GW OBSERVATORY NETWORK

The international ground-based GW observatory network currently comprises four primary observatories employing laser interferometric GW detectors. The four observatories are the two US-based LIGO detectors, LIGO Hanford Observatory (LHO) in Washington and LIGO Livingston Observatory (LLO) in Louisiana (Aasi et al. 2015a), the European Virgo detector (Acernese et al. 2015), and the Japanese KAGRA detector (Akutsu et al. 2021; Aso et al. 2013; Somiya 2012). All these detectors are enhanced Michelson interferometers that sense relative changes in the lengths $L_1$ and $L_2$ of their two 3 km to 4 km long arms caused by passing GWs in the high-frequency band ∼10 Hz to ∼1000 Hz (Thorne 1987). Other GW frequency bands include the very-low-frequency band ∼1 nHz to ∼100 nHz observed by pulsar timing arrays such as the European Pulsar Timing Array (EPTA; Desvignes et al. 2016), the North American Nanohertz Observatory for Gravitational Waves (NANOGrav; Brazier et al. 2019), the Parkes Pulsar Timing Array (PPTA; Kerr et al. 2020), the Indian Pulsar Timing Array (InPTA; Joshi et al. 2018), and their combined consortium the International Pulsar Timing Array (IPTA; Verbiest et al. 2016); and the low-frequency band ∼0.1 mHz to ∼10 mHz that will be observed by the Laser Interferometer Space Antenna (*LISA*; Colpi et al. 2024).

The fractional change in the relative lengths of the two optical paths of interferometric detectors, $\Delta(L_1 - L_2)$, induced by a GW is known as the detector strain, $h = \Delta(L_1 - L_2)/L$, where $L$ is the average arm length (Section 5.1). The sensitivity of ground-based detectors is fundamentally limited below ∼1 Hz by ground motion noise (Saulson 1984) and at high frequencies by shot noise (Forward 1978; Krolak et al. 1991). Significant noise sources at intermediate frequencies include thermal noise in the optics and their suspensions and quantum readout noise (Weiss 2022; Saulson 2017; Buonanno & Chen 2001). In the frequency domain, the overall detector sensitivity is characterized by the (one-sided) noise power spectral density in strain-equivalent units, $S_n(f)$, with dimensions of time (Appendix B).

The GEO 600 GW detector (GEO) is a British–German instrument with 600 m arms located near Hannover, Germany (Luck et al. 2010; Affeldt et al. 2014; Dooley et al. 2016). This instrument is a laboratory for prototyping advanced interferometry techniques, but also is operated in data-taking *astrowatch* mode when not being used for instrument-science research (Grote 2010; Dooley et al. 2016). Astrowatch provides GW observing coverage for times when the larger detectors are not observing between observing runs and when the detectors are not taking scientific data; e.g., GEO data was used to constrain post-merger signals following the first BNS detection (Abbott et al. 2017c).

## 3. OBSERVING RUNS

The GW observing schedule is divided into observing runs, down time for construction and commissioning, and transitional engineering runs between commissioning and observ-ing runs (Abbott et al. 2020a). Figure 1 shows a timeline of GW observations up to the end date of the time period covered by GWTC-4.0. Indicated are the observing periods of each observing run, and the times when each detector was in operation. Also shown are the times when GW transient signals were detected.

In order to quickly compare sensitivities of detectors, the GW community uses a fiducial range, to which a typical BNS can generally be detected. This fiducial distance assumes that a SNR of at least 8 is needed for a detection, and it approximates the BNS inspiral waveform at Newtonian order (Section 5.2). The BNS inspiral range is a volume-averaged measure of sensitivity to a signal from two $1.4\,M_\odot$ bodies in a quasi-circular inspiral at a single-detector SNR threshold of 8 (Finn & Chernoff 1993; Chen et al. 2021). When a homogeneous BNS population of is assumed and cosmological effects are ignored, the BNS inspiral range for a detector is determined by its noise power spectrum as

$$R = 1.016 \times 10^{-20}\,\mathrm{Mpc\,s^{-1/6}} \sqrt{\int_0^\infty \frac{f^{-7/3}}{S_n(f)}\,df}\,, \qquad (1)$$

and the sensitive volume of the detector (also when neglecting cosmological effects) is given by $V = (4\pi/3)R^3$ (Appendix B). (This measure is taken as a simple figure of merit of sensitivity to CBCs; it does not attempt to account for the true underlying astrophysical distribution describing such systems.) If the number of BNS mergers per unit time per unit volume of space, the *merger rate density* of BNSs, is $\mathcal{R}$, then the expected number of BNS signals seen with SNR greater than 8 in time $T$ would be $\mathcal{R}VT$. Figure 1 also gives the typical BNS inspiral range, as given in Equation (1), for each detector during each observing run.

The *amplitude* strain noise spectra is the square-root of the (one-sided) noise power spectral density in strain equivalent units $S_n^{1/2}(f)$ having dimensions of time$^{1/2}$. The amplitude strain noise spectra of LHO, LLO, and Virgo during the various observing runs are shown in Figure 2. There is an overall reduction in the detector noise levels with successive observing runs resulting in increased sensitivity. Figure 2 also shows the fraction of the run duration during which different combinations of detectors were observing.

Figure 3 shows the cumulative number of candidates detected vs. the estimated effective time–volume hypervolume $VT$ for the detector network. For the first two observing runs (described below), only data when two detectors were operating were searched for GWs. In this case the rate at which $VT$ is accumulated at any observing time is given by the sensitive volume $V$ for the *second* most sensitive instrument observing at that time. Beginning with the third observing run, periods during which only a single detector was observing were included in the search. During such time, the rate at which $VT$ is accumulated is again given by the sensitive volume $V = (4\pi/3)R^3$ but where $R$ is computed from Equation (1) divided by 1.5, representing an effective SNR threshold for detection of 12 rather than 8 for single-detector observation (Abbott et al. 2021b). This simple estimate of $VT$, derived from the BNS inspiral range, is an approximate one done for a quick and convenient overview. In particular, it makes a crude approximation of whether a signal is detectable and its numerical value is only representative of sensitivity to sources in a small region of mass space. Actual *measured* sensitive hypervolume $\langle VT \rangle$ values for various CBC mass regions and search methods are reported in Abac et al. (2025b).

### 3.1. *O1: The First Observing Run*

O1 consists of the time-period from 2015 September 12 to 2016 January 19. O1 includes short time-periods which were originally planned to be engineering time (2015 September 12 to 2015 September 18 and 2016 January 12 to 2016 January 19), but which were of sufficient quality to be included in O1. This was the first observing run with the Advanced LIGO (aLIGO) interferometers, in progress toward full aLIGO design sensitivity (Abbott et al. 2016b,c), LHO achieving a BNS range of 80 Mpc and LLO a range of 70 Mpc.

Of the 129.7 d duration of O1, there were only 49.0 d (38%) when both LHO and LLO were observing jointly, and there were 36.2 d (28%) when neither detector was observing. The largest non-observing periods were due to locking, the time spent bringing the interferometers from an uncontrolled state to their low-noise configuration (Staley et al. 2014), and environmental issues such as earthquakes, wind, and microseismic noise arising from ocean storms (Effler et al. 2015; Abbott et al. 2016d). Wind and microseismic noise have seasonal variation as storms are more prevalent in winter months; LLO was more susceptible to these than LHO, mainly due to its local geophysical environment (Daw et al. 2004).

Overall, a total effective hypervolume $VT = 1.59 \times 10^{-4}\,\mathrm{Gpc^3\,yr}$ was accumulated during joint LHO–LLO observing during O1.

### 3.2. *O2: The Second Observing Run*

The O2 run from 2016 November 30 to 2017 August 25. It was preceded by an engineering run that began on 2016 October 31 at LLO and on 2016 November 14 at LHO. The LHO and LLO detectors achieved a typical BNS range sensitivity of 80 Mpc and 100 Mpc respectively (Abbott et al. 2017d, 2019a). However, on 2017 July 06 LHO was severely affected by a 5.8 magnitude earthquake in Montana resulting in a post-earthquake sensitivity drop of approximately 10 Mpc in BNS range for the remainder of the run (Abbott et al. 2019a).

The Advanced Virgo (AdV) interferometer (Acernese et al. 2015) joined O2 on 2017 August 01, forming a three-detector network for the last month of the run. A vacuum contamination issue required AdV to use steel wires rather than fused silica fibers to suspend the test masses, limiting the sensitivity of AdV (Abbott et al. 2019a). In O2, a 30 Mpc BNS range was achieved.

The LIGO detectors saw some improvement in duty factors during non-winter months with an almost 50% reduction in downtime due to environmental effects at both sites, though LLO lost over twice as much observing time as LHO to earthquakes, microseismic noise, and wind. O2 had a planned mid-run engineering break to effect needed repairs and to attempt improvements to the sensitivity. The Virgo instrument operated with a duty factor of approximately 85% after joining O2. There were 15 d of all three detectors observing simultaneously.

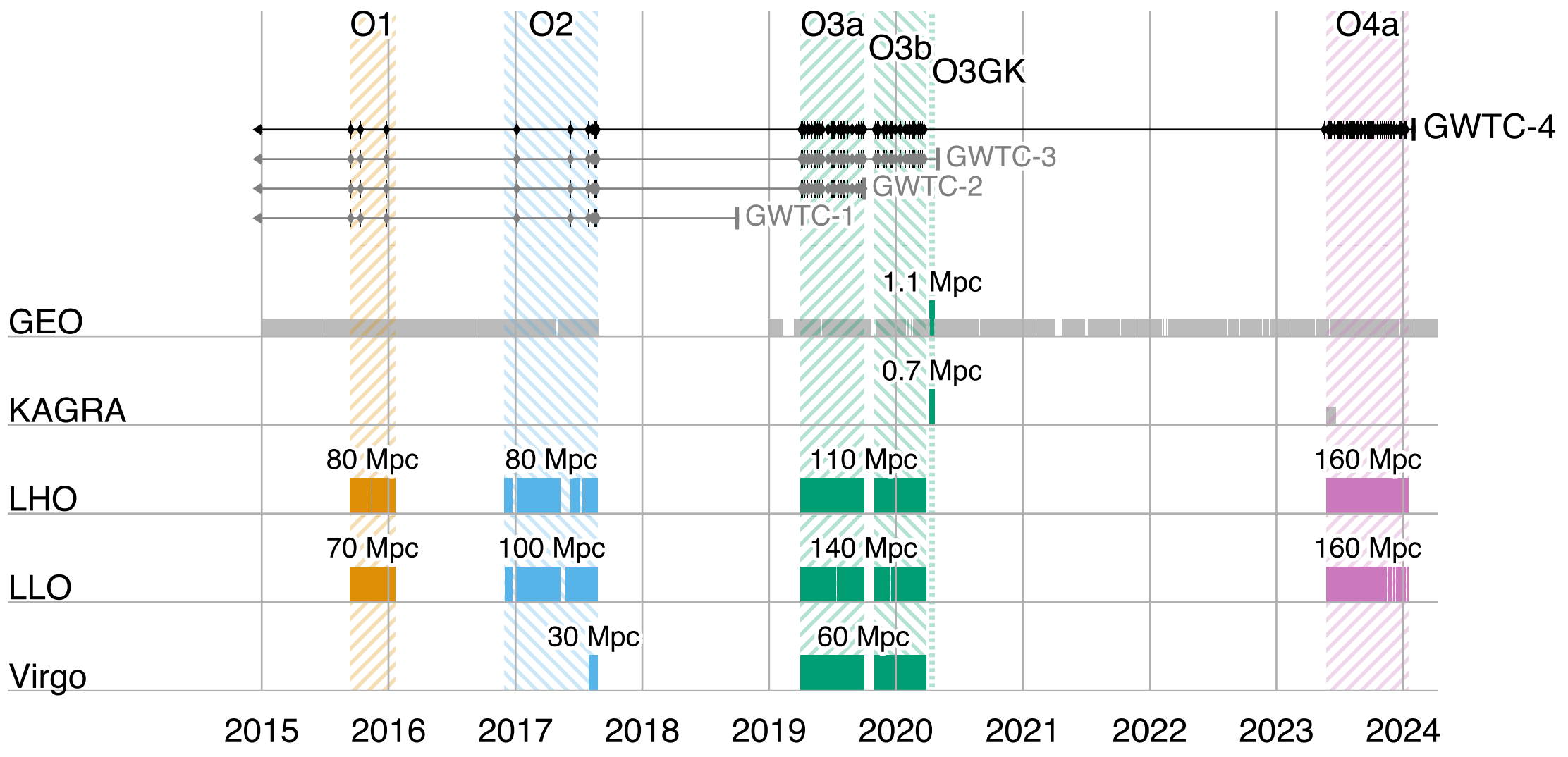

**Figure 1.** The timeline of observing runs covering a timespan starting from 2015 and lasting up to the beginning of O4b on 2024 April 10. The periods in which the various detectors in the network were observing are shown in this timeline, along with the typical BNS inspiral ranges for those detectors during the observing run. GEO astrowatch observing periods are shown in light gray. KAGRA observing periods during O4a, also shown in light gray, were not used for GW observational analyses. In O1 and O4a, only LHO and LLO were participating. Virgo joined these two detectors for the last month of O2 and was observing alongside them throughout O3a and O3b. At the end of O3 there was a short joint observing run, O3GK, which included GEO and KAGRA. Also shown is a timeline of the observed candidates contained in GWTC-1.0, GWTC-2.1, GWTC-3.0, and GWTC-4.0 with a probability of astrophysical origin greater than or equal to 50%. The time intervals covered by the various versions of the GWTC are bounded from above but not from below, as indicated by the arrows pointing left (see Section 1.2.1).

Overall, a total effective hypervolume $VT = 3.52 \times 10^{-4}\,\mathrm{Gpc^3\,yr}$ was accumulated during O2; of this, $3.27 \times 10^{-4}\,\mathrm{Gpc^3\,yr}$ was accumulated during joint LHO–LLO observing, $2.41 \times 10^{-5}\,\mathrm{Gpc^3\,yr}$ was accumulated while all three detectors were observing, and only $3.62 \times 10^{-7}\,\mathrm{Gpc^3\,yr}$ and $4.80 \times 10^{-7}\,\mathrm{Gpc^3\,yr}$ were accumulated during joint LHO–Virgo and LLO–Virgo observing respectively.

### 3.3. *O3: The Third Observing Run*

O3 started on 2019 April 01, with a commissioning break from 2019 October 01 to 2019 November 01. This observing run was planned to continue to 2020 April 30 but the COVID-19 pandemic resulted in a suspension of observing on 2020 March 27 (Abbott et al. 2023). The period of O3 prior to the commissioning break is referred to as O3a while the period after the break is referred to as O3b. KAGRA had intended to join LIGO and Virgo at the end of O3 but the early end made this impossible. Instead, KAGRA and GEO jointly observed for a two week period from 2020 April 07 to 2020 April 21 after LIGO and Virgo had suspended their observing. This joint GEO–KAGRA run (distinct from the O3 run described previously) is referred to as O3GK (Abbott et al. 2022).

In O3, the LHO and LLO detectors achieved a BNS range of 110 Mpc and 140 Mpc respectively (Buikema et al. 2020). This increase in sensitivity arose from a variety of improvements, chief among them an increase in the input laser power, the addition of a squeezed vacuum source at the interferometer output (Tse et al. 2019), and mitigation of noise arising from scattered light (Soni et al. 2020). In addition, end test-mass optics with lower-loss coatings, along with new reaction masses, were installed in each LIGO interferometer (Granata et al. 2020; Aston et al. 2012).

The steel wires in AdV were replaced with fused silica fibers in preparation for O3. Along with other improvements such as reduction of technical noises, an increase in laser power, and the installation of a squeezed vacuum source, Virgo achieved a BNS range of 60 Mpc (Acernese et al. 2019).

Over all of O3a and O3b, 361.1 d combined, there were 154.3 d (43%) of three-detector observation and only 42.1 d (12%) during which no detector was observing. The total effective hypervolume $VT$ accumulated was $3.21 \times 10^{-3}\,\mathrm{Gpc^3\,yr}$. Of this, $2.27 \times 10^{-3}\,\mathrm{Gpc^3\,yr}$ was accumulated during three-detector observations, $7.20 \times 10^{-4}\,\mathrm{Gpc^3\,yr}$ when LHO and LLO were observing, $4.09 \times 10^{-5}\,\mathrm{Gpc^3\,yr}$ when LHO and Virgo were observing, $5.03 \times 10^{-5}\,\mathrm{Gpc^3\,yr}$ when LLO and Virgo were observing. The amount accumulated with only a single detector observing was $4.47\times10^{-5}$ Gpc$^3$ yr, $7.47\times10^{-5}$ Gpc$^3$ yr, and $9.72\times10^{-6}$ Gpc$^3$ yr for LHO, LLO, and Virgo, respectively.

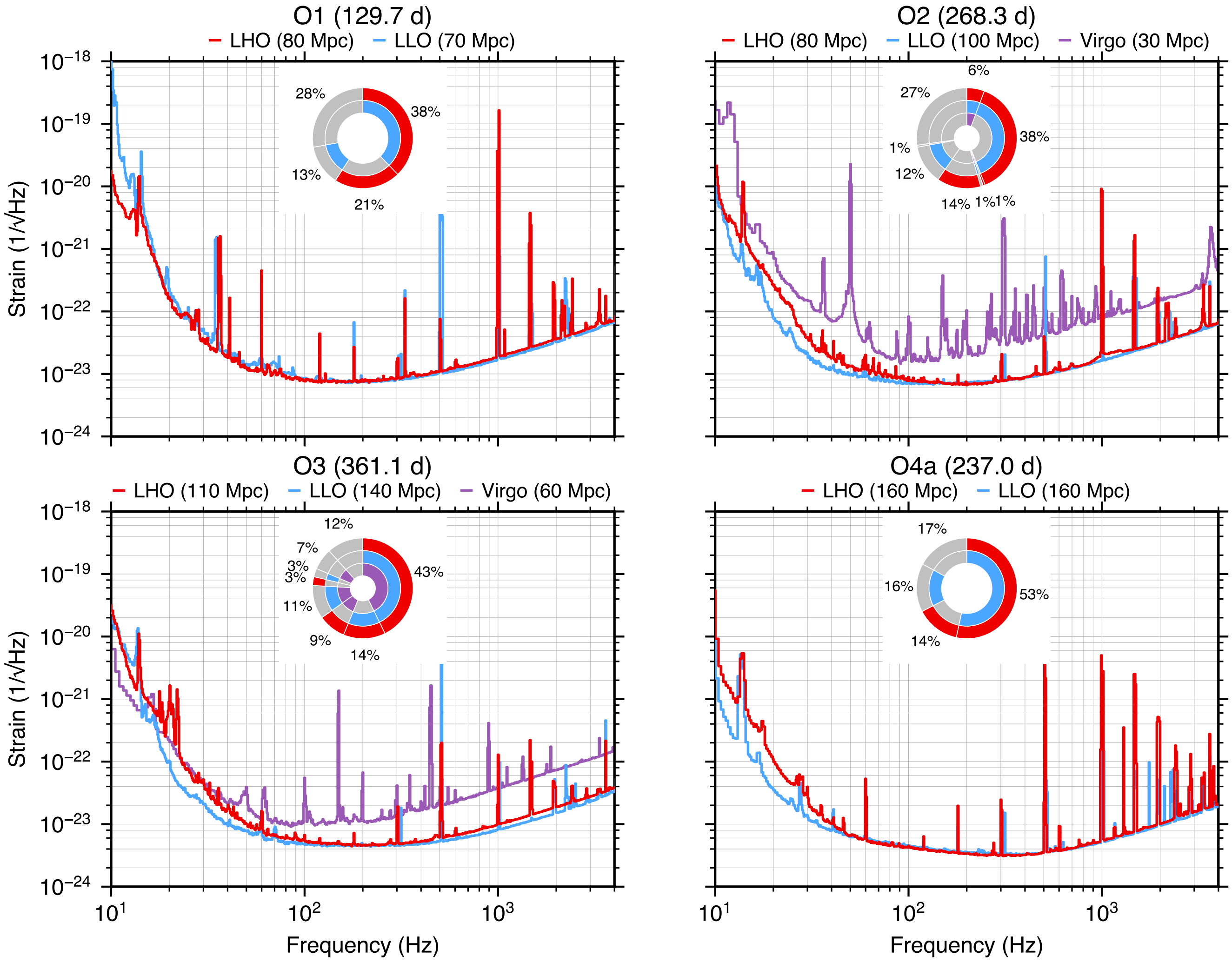

**Figure 2.** Representative noise amplitude spectral densities for LHO, LLO, and Virgo during O1 (LHO, LLO: 2015 October 24), O2 (LHO: 2017 June 10, LLO: 2017 August 06, Virgo: from Acernese et al. (2023a)), O3 (LHO: 2020 January 04, LLO: 2019 April 29, Virgo: 2020 February 09), and O4a (LHO: 2024 January 11, LLO: 2023 November 19). The BNS inspiral ranges, defined by Equation (1), for these noise curves are given in the legend. Inset sunburst charts show the fraction of the run duration during which different combinations of detectors were observing. Gray regions in each ring indicates portions when a detector is not operating. The segments of the sunburst chart, clockwise from 12 o'clock, are: LHO–LLO, LHO alone, LLO alone, and neither for observing runs involving only LHO and LLO; and LHO–LLO–Virgo, LHO–LLO, LHO– Virgo, LLO–Virgo, LHO alone, LLO alone, Virgo alone, and none for observing runs involving LHO, LLO, and Virgo.

The first operation of the KAGRA detector in an initial configuration with a simple Michelson interferometer occurred in March 2016 (Akutsu et al. 2018). In August 2019, the first lock of the Fabry–Perot Michelson interferometer was achieved, with power recycling accomplished in January 2020. By the end of March 2020, KAGRA obtained a BNS range of approximately 1 Mpc (Abe et al. 2023) and, although the LIGO and Virgo instruments had ended their O3 run, KAGRA was operated jointly with GEO, which had a comparable BNS range, in O3GK yielding 6.4 d of joint observing time.

### 3.4. *O4: The Fourth Observing Run*

O4 began on 2023 May 24 at 15:00:00 UTC. This run is again divided into parts: the first part of the fourth observing run (O4a) ended on 2024 January 16 at 16:00:00 UTC and was followed by a commissioning break; the second part of the fourth observing run (O4b) started on 2024 April 10 at 15:00:00 UTC. The O4b period continued until 2025 January 28 17:00:00 UTC, the original intended end of O4; however it was decided to continue observing into a third part of the fourth observing run (O4c). The period covered by GWTC-4.0 contains events that occurred in O4a and earlier observing runs only (see Section 1.2.1). O4b and O4c analyses are underway and will be included in future versions of the GWTC.

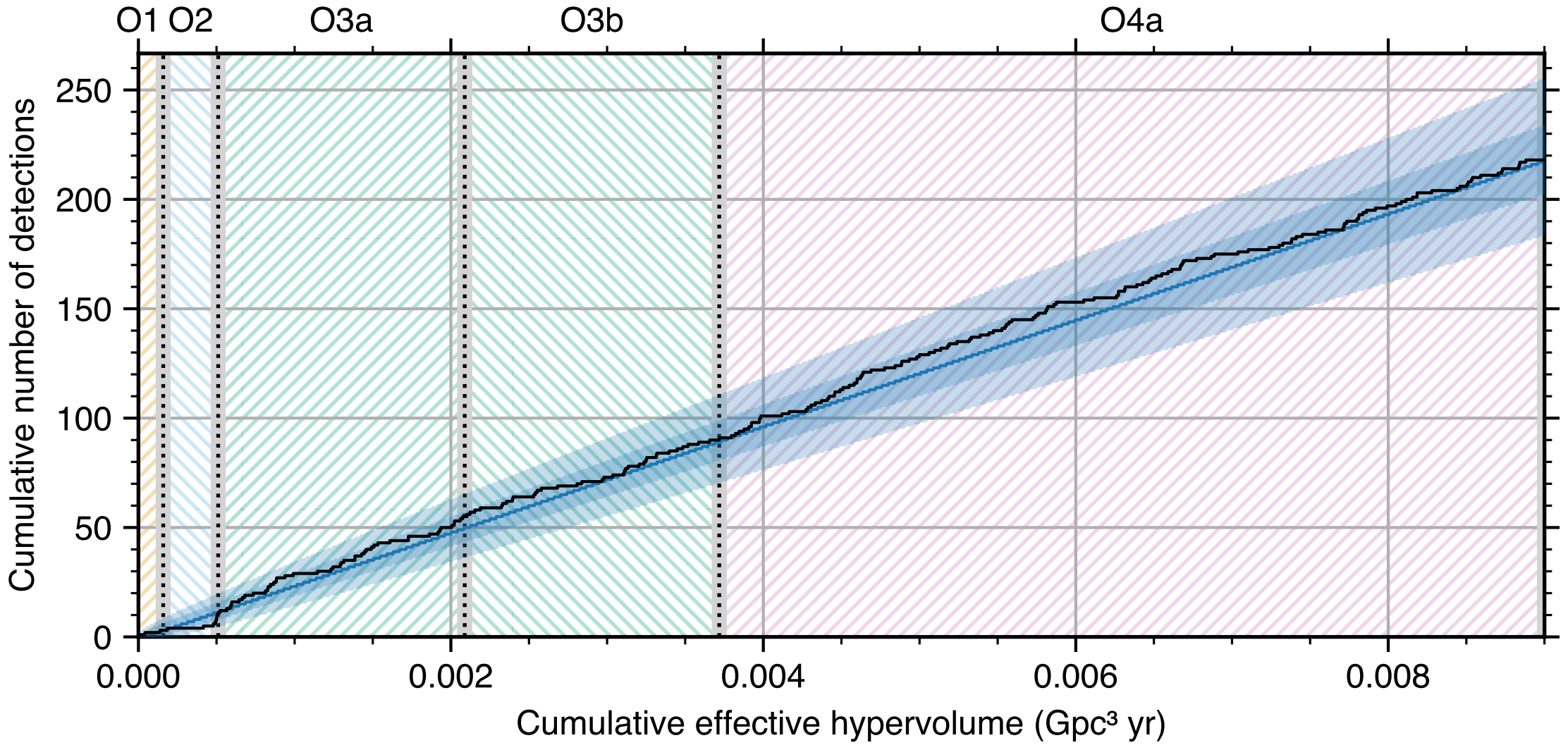

**Figure 3.** The number of CBC detection candidates with a probability of astrophysical origin greater than or equal to 50% versus the detector network's effective surveyed hypervolume for BNS coalescences (Abbott et al. 2021b). The BNS effective surveyed hypervolume is a valid proxy for overall sensitivity to CBCs, though its scale is set to the case of canonical BNS signals. The colored bands indicate the different observing runs. The final data sets for O1, O2, O3a, O3b, and O4a consist of 49.0 d, 122.2 d, 149.6 d (177.1 d), 124.6 d (141.9 d), and 126.5 d (196.8 d) with at least two detectors (one detector) observing, respectively. The cumulative number of probable candidates is indicated by the solid black line, while the blue line, dark blue band and light blue band are the median, 50% confidence interval and 90% confidence interval for a Poisson distribution fit to the number of candidates at the end of O4a.

The two LIGO detectors were observing during O4a, both having a BNS range of approximately 160 Mpc. During the 237.0 d there were 126.5 d (53%) of two-detector joint observation and 40.2 d (17%) when neither of the LIGO detectors were observing. Virgo did not join joint observation until O4b in order to continue commissioning to address a damaged mirror that limited performance and to improve sensitivity. KAGRA also continued commissioning to improve sensitivity with the goal of joining O4 toward the end of the run.

During O4a, the total effective hypervolume $VT$ accumulated was $5.28 \times 10^{-3}\,\mathrm{Gpc}^3\,\mathrm{yr}$. This is divided into $3.85 \times 10^{-4}\,\mathrm{Gpc}^3\,\mathrm{yr}$ during which LHO alone was observing, $4.57 \times 10^{-4}\,\mathrm{Gpc}^3\,\mathrm{yr}$ during which LLO alone was observing, and $4.44 \times 10^{-3}\,\mathrm{Gpc}^3\,\mathrm{yr}$ during which both detectors were observing.

At the time of writing, O4 is expected to continue until 2025 November 18 at 16:00 UTC with no further planned breaks in observing. The timeline for a fifth observing run (O5) is being assessed in order to maximize the scientific output of the global network. Updates to the planned observing schedule will be provided as soon as such decisions are made.[2]

## 4. OBSERVATORY EVOLUTION

The advanced detector era is characterized by a series of technological improvements from the initial detectors that deliver higher sensitivity and greater BNS range that made possible the era of GW observation. Some of the key instrument science elements of the advanced era detectors are: (i) increases in the input laser power entering the interferometer, and to the circulating power in the interferometer cavities (a higher power in the arms produced a lower quantum shot noise limited sensitivity above ~200 Hz); (ii) increases in test mass mirror size to accommodate larger beams which mitigates coating thermal noise and heavier masses to reduce inertial and quantum back-action effects; (iii) implementation of signal recycling (Meers 1988) in addition to power recycling (Drever 1983), which alters the frequency band of the detectors' sensitivity (typically to give broader-band sensitivity); (iv) implementation of monolithic test-mass suspensions, which reduces the suspension thermal noise in the detectors' sensitivity band by using the same low mechanical loss material (fused silica for LIGO and Virgo) for the suspension fibers as for the mirror substrate, and low loss jointing techniques and thermo-elastic nulling (Aston et al. 2012; Travasso 2018); (v) improved passive and active seismic iso-lation systems, and sensors to reduce ground motion coupling to the detector and to damp suspension modes (Braccini et al. 2005; Matichard et al. 2015; Cooper et al. 2023); (vi) improved low-thermal-noise, low-absorption, high-reflectivity mirror coatings (Harry et al. 2007; Granata et al. 2020).

[2] LVK observing run plans https://observing.docs.ligo.org/plan

Throughout the advanced-detector era of GW observation, the LIGO and Virgo detectors have undergone a series of performance-improving detector upgrades and commissioning activities of which detail is given in this section. Detector upgrades include the installation of new hardware or upgrades to existing hardware in a detector. Examples of detector upgrades include the installation of new laser systems to provide higher power into the interferometer, installation of baffles to mitigate scattered light and the injection of squeezed light to manipulate the quantum-noise limited sensitivity of the detectors (Tse et al. 2019; Acernese et al. 2019). Commissioning activities cover a range of improvements to sensitivity and observing uptime of the instruments from targeted noise-hunting activities to remove glitches, lines and broadband noise, and improved control schemes to mitigate instabilities and improve detector robustness.

Alongside this has been the effort to build and commission the KAGRA detector utilizing advanced technologies such as cryogenic cooling of the test-masses and an underground location. This schedule of planned upgrades and commissioning activities between observing runs ensures that the maximal science output is achieved from the network. In terms of valuable scientific output, a successful upgraded detector that has been offline for a period of time rapidly overtakes a non-upgraded detector in continuous observational mode in terms of number of significant detections, and the resolution and sky-localization of high interest signals.

The aLIGO and AdV detectors are designed to be dual-recycled Fabry–Perot Michelson interferometers with orthogonal kilometer-scale arms (Aasi et al. 2015a; Acernese et al. 2015). Each arm contains a Fabry–Perot optical cavity, and a beam splitter at the corner between the arms forms a Michelson interferometer that measures the change in the relative phase of the light induced by changes in the lengths of these cavities (Thorne 1987; Vinet et al. 1988). Additional power-recycling and signal-recycling cavities are created by adding mirrors in the symmetric and antisymmetric ports of the interferometer. These improve sensitivity by building up the light power on the beam splitter and beneficially modifying the response of the interferometer respectively (Meers 1988). The input and end mirrors on each of the Fabry–Perot cavities are the test masses whose separations are affected by GWs. The mirrors are isolated by multistage pendulums that suppress the ground motion by more than 10 orders of magnitude at frequencies around 10 Hz. Monolithic fused-silica fibers are used on the bottom stage of the suspension system to suppress thermal noise and the mirrors themselves are fused-silica substrates with low-loss, highly reflective coatings (Aston et al. 2012).

Ground-based interferometers generally have the same fundamental limiting noise sources (Weiss 2022; Saulson 2017), with the response of each detector and the exact extent to which each noise limits sensitivity being specific to the detailed design of each detector. At low observational frequency below ~10 Hz the detectors are limited by a combination of seismic noise, gravity-gradient noise, suspension thermal noise and quantum radiation-pressure noise. Thermal noise in the mirror optical coatings is a significant noise source at mid frequencies ~50 Hz to ~200 Hz (Harry et al. 2007), and at high frequencies above ~200 Hz, sensitivity is limited by the quantum shot noise.

In addition to these fundamental noise sources the detectors are also limited by technical noise. This includes scattered-light noise, which occurs when some fraction of light is deflected from the interferometer beam path and is incident on another moving surface varying the phase of the light; this couples noise into the interferometer readout if part of this light is reflected back into the main beam (Accadia et al. 2010; Ottaway et al. 2012). Interferometer controls-system noise is when signals couple between the multiple feedback loops that control the degrees of freedom of the interferometer and requires complicated optimization of control-loop parameters to mitigate (Buikema et al. 2020). Laser noise due to fluctuations in the frequency, intensity and pointing of the laser beam entering the interferometer is reduced with dedicated multi-stage stabilization systems to a level such that is does not impact the sensitivity of the detectors, however suboptimal tuning of these stabilization systems can lead to laser noise affecting sensitivity (Cahillane et al. 2021). Environmental noise is caused when environmental effects in the vicinity of the interferometer (e.g., seismic activity) couple into the measurement of the interferometer strain signal (Acernese et al. 2006; Effler et al. 2015; Fiori et al. 2020; Nguyen et al. 2021; Helmling-Cornell et al. 2024). Detector commissioning seeks to mitigate such non-fundamental noise sources.

The key parameters of the LIGO, Virgo, KAGRA and GEO detectors across the advanced era observing runs are given in Table 1. The specific evolution of each detector in terms of detector upgrades and improvements is detailed in the remainder of this section.

### 4.1. *LIGO Hanford & Livingston Observatories*

LIGO is a US national facility comprising two US-based interferometric detectors in Hanford, Washington (LHO), and Livingston, Louisiana (LLO), each with 4 km arms. LIGO construction began in 1994. From 2002 to 2010, initial power-recycled Fabry–Perot Michelson interferometers were operated at these sites in a series of science runs S1 through S6 (Abbott et al. 2009; Aasi et al. 2015b). During this period, LIGO also operated a second interferometer with 2 km arms at the Hanford site. Subsequently the aLIGO project resulted in a major overhaul of the interferometers to improve the capabilities of the detectors (Aasi et al. 2015a) leading up to O1 and the first observation of GWs.

Across the observing runs certain areas have been the main focus of much of the detector improvement effort: (i) increas-ing the arm cavity power by increasing the injected laser power and the power-recycling gain while achieving stable operation; (ii) mitigation of scattered-light sources and coupling mechanisms; (iii) reduction of quantum noise with addition of a squeezed-light system for O3 and the following improvements to the quantum-enhancement factor.

**Table 1.** Selected optical and physical parameters of the LIGO Hanford (LHO), LIGO Livingston (LLO), Virgo, KAGRA, and GEO 600 (GEO) interferometers throughout the advanced-detector era. The input laser power is the power that would be measured at the power recycling mirror (after the input mode cleaner) and is an estimate of the maximum level typically achieved during an observing period. Suspension types are monolithic fused silica fibers, sapphire fibers, or steel wires.

| Observing period | Interferometer | Input laser power | Power recycling gain | Signal recycling | Squeezing | Suspension type |
|---|---|---|---|---|---|---|
| O1 | LHO | 21 W | 38 | ✓ | × | Silica |
| | LLO | 22 W | 38 | ✓ | × | Silica |
| O2 | LHO | 26 W | 40 | ✓ | × | Silica |
| | LLO | 25 W | 36 | ✓ | × | Silica |
| | Virgo | 10 W | 38 | × | × | Steel |
| O3a | LHO | 34 W | 44 | ✓ | ✓ | Silica |
| | LLO | 44 W | 47 | ✓ | ✓ | Silica |
| | Virgo | 18 W | 36 | × | ✓ | Silica |
| O3b | LHO | 34 W | 44 | ✓ | ✓ | Silica |
| | LLO | 40 W | 42 | ✓ | ✓ | Silica |
| | Virgo | 26 W | 34 | × | ✓ | Silica |
| O3GK | GEO | 3 W | 1000 | ✓ | ✓ | Silica |
| | KAGRA | 5 W | 12 | × | × | Sapphire |
| O4a | LHO | 57 W | 50 | ✓ | ✓ | Silica |
| | LLO | 64 W | 35 | ✓ | ✓ | Silica |

Both aLIGO detectors are operated with a lower injected laser power and lower power-recycling gain than the design goal (Aasi et al. 2015a). The full amount of available laser power cannot be fully utilized due to issues with maintaining long-duration stable locking of the interferometer due to angular instabilities and point absorbers in the test-mass mirrors (Brooks et al. 2021). This issue was the focus of commissioning efforts to continually improve the operating power in the cavity by optimizing the interferometer control loops (Buikema et al. 2020) and reducing the presence of point absorbers in the mirrors. Stray-light control can be achieved by the addition of baffles to block unwanted beam paths and with active control of known scattered-light paths. The addition of a squeezed vacuum source at the interferometer's output alters the quantum noise in the interferometer and with the inclusion of a filter cavity can produce frequency dependent squeezing which can be used to surpass the standard quantum limit on sensitivity of a laser interferometer (Tse et al. 2019; Ganapathy et al. 2023).

#### 4.1.1. *O1*

The sensitivity and limiting noise sources of the LIGO detectors during O1 is described in Abbott et al. (2016c). Figure 2 shows a representative amplitude spectral density of the strain noise and the BNS range. In O1, the typical input power entering the power-recycling cavity was 21 W in LHO and 22 W in LLO, circulation of laser light in the power recycling cavity increases the power on the beam splitter to be a factor of 38 times greater (the power recycling gain), and a further increase in circulating power by a factor of 144 is achieved in the arms by the Fabry–Perot cavities. The laser input power and power-recycling gain during O1 and the later observing runs is given in Table 1 alongside other detector parameters. An example of commissioning improvement is the investigation at LLO during O1 of recurring changes in the BNS range from 65 Mpc to 60 Mpc. By searching for correlation between the detector range and the hundreds of data channels recorded by aLIGO it was found that the issue was caused by a malfunctioning temperature sensor. This sensor was replaced resulting in a more stable increased range (Walker et al. 2018).

#### 4.1.2. *O2*

After O1, several improvements were made to both LIGO instruments (Abbott et al. 2017d). Detector upgrades included installation of new mass dampers on the end test-mass suspensions to dampen mechanical modes, improving the stabilization of laser intensity, and installing a new output Faraday isolator and higher quantum-efficiency photodiodes at the output port to improve signal-detection efficiency in the readout system. Mitigation of scattered light sources and other improvements to the detector sensitivity through-out O2 resulted in a BNS range improvement to 100 Mpc by the end of the run (Davis et al. 2021). Commissioning tests during O2 on the LHO detector to increase in the laser power to 50 W did not result in an overall improvement in performance of 80 Mpc BNS range at the end of O1, owing to point absorbers on one of the input test-mass optics, so the detector operated with 30 W input power. After O2, it was demonstrated that the use of witness channels to perform noise subtraction on the strain data was able to increase the BNS range by 20% (Davis et al. 2019; Driggers et al. 2019).

#### 4.1.3. *O3*

Leading up to O3, several upgrades were made to the LIGO instruments (Buikema et al. 2020). The most significant was the installation of an in-vacuum squeezed-light injection system at each site to inject squeezed vacuum into the interferometers to reduce shot noise at frequencies above 50 Hz (Tse et al. 2019). The squeezer works by optically pumping a non-linear crystal to modify the distribution of the quantum vacuum state that enters the interferometer (Caves 1981; Barsotti et al. 2019).

Between O3a and O3b, adjustments to the squeezing subsystem produced large sensitivity improvements. Among these were the installation of higher power laser amplifiers with stable operation and output power over 70 W (Bode et al. 2020). A program of installation of optical baffles was completed to improve stray light control. The correlation of microseismic activity with scattered-light noise was determined to be primarily caused by a scattered-light path arising from large relative motion between the end test mass and the reaction mass that is immediately behind it (Soni et al. 2020). A control loop that makes the reaction mass follow the end mass, implemented on 2024 January 07 at LLO and 2024 January 14 2024 at LHO, reduced the relative motion and mitigated the scattered-light noise (Davis et al. 2021). At LHO, wind fences were installed to mitigate ground tilt induced by wind on the buildings (Nguyen et al. 2021).

#### 4.1.4. *O4a*

Several upgrades were implemented at LHO and LLO to improve the quantum-limited sensitivity of the detectors via improved quantum squeezing and higher intracavity power (Abac et al. 2024a). Further upgrades to the laser amplification system were implemented with stable operation and output power over 140 W (Bode et al. 2020). A new vacuum system to house a 300 m filter cavity was built at both detectors along with an upgraded squeezing injection system to allow the injection of frequency-dependent squeezed vacuum to achieve quantum noise reduction across the detection frequency band (Ganapathy et al. 2023; Jia et al. 2024). Squeezing levels in O4a reached 5.8 dB at LLO and 4.6 dB at LHO, compared to the 2 dB to 3 dB achieved in O3 (Capote et al. 2025). Test-mass mirrors were replaced at both observatories to remove point defects on the mirrors that contributed to controls challenges and excess noise (Buikema et al. 2020). This involved a replacement of both end test masses at LLO and the input $y$-arm test mass at LHO. Replacing these test masses allowed both observatories to approximately double the input power compared to O3, further improving the quantum-limited sensitivity of the detectors due to higher circulating power in the Fabry–Perot arm cavities (Capote et al. 2025; Buikema et al. 2020).

Other upgrades to the LIGO detectors include improvements to the electronics in the GW signal readout chain, damping of baffles to mitigate scattered light, and improvements to electronics grounding (Capote et al. 2025; Soni et al. 2025). The photodetector transimpedance amplifiers were improved ahead of O4a using a design tested at GEO, resulting in a factor of ten reduction in dark noise compared to O3 (Grote et al. 2016). At both LIGO detectors, a septum window separating two vacuum volumes housing the output optics was removed, significantly reducing the coupling of acoustic noise. Baffles along the arm cavity and around vacuum pumps were previously identified to couple excess scattered light in O3, and were damped to reduce their motion and therefore shift the frequency of up-converted scattered light out of the sensitive band. Finally, injections into the building electronics ground demonstrated that many spectral features in the strain at LHO were the result of a fluctuating ground potential (Capote et al. 2025; Soni et al. 2025). The resistance to ground was reduced for several electronics chassis around the detector. Additionally, the voltage biases of the test-mass electrostatic drives were adjusted to minimize the electronics noise coupling further (Capote et al. 2025).

Detector commissioning ahead of O4a also focused on optimization of the auxiliary controls to reduce technical noise that limited the detectors at low frequency in O3 (Buikema et al. 2020). Alignment controls noise was reduced by a factor of ten and length controls noise by a factor of two at both detectors near 20 Hz (Capote et al. 2025; Buikema et al. 2020). Significant improvements to the controls included the upgrade to a camera servo system that requires no line injection to sense the alignment of the main detector optics (Capote et al. 2025). Suspension local control loops were re-optimized to focus on noise suppression above 5 Hz, reducing both noise directly coupled to the strain, and noise that couples indirectly through the length and alignment controls (Capote et al. 2025). Both detectors were also limited by unmitigated beam-jitter noise that was well-witnessed by auxiliary sensors (Capote et al. 2025). As such, front-end infrastructure using the non-stationary estimation and noise subtraction (NonSENS) code (Vajente 2018) was implemented to perform noise cleaning in low latency, increasing detector sensitivity by up to 5 Mpc in BNS range (Vajente et al. 2020; Vajente 2022; Capote et al. 2025).

#### 4.1.5. *Beyond O4*

Looking to the future there is ongoing construction of LIGO-India (Souradeep et al. 2017), a third LIGO interferometer to be built in the Hingoli district of Maharashtra, India. This facility will be based on aLIGO hardware and design, and its location will provide a significant improvement in the sky localization of GW sources (Pankow et al. 2020; Saleem et al. 2022; Pandey et al. 2025).

In parallel, there are plans underway to upgrade the existing LIGO detectors (and eventually LIGO-India) to Advanced+ LIGO (A+) sensitivity (Abbott et al. 2020a; Cooper et al. 2023). The A+ upgrade to the LIGO detectors is a series of detector upgrades utilizing improved technology that has been developed in parallel to the observing runs. The inclusion of frequency dependent squeezing was originally planned as an A+ upgrade but was implemented ahead of O4a at both sites (Capote et al. 2025). Other A+ upgrades, which will be implemented for future observing runs, include new optics with lower noise and loss, improved sensors for controlling the mirrors, a new pre-mode cleaner to reduce beam-jitter noise, improved output mode cleaners with lower loss, and a balanced homodyne readout system that allows for better readout control of the interferometer signal.

A post-O5 upgrade, referred to as LIGO $A^\sharp$ ($A^\sharp$), explores more transformative changes in detector design with the goal of increasing the sensitivity to the limits of what is possible with the existing infrastructure of the LIGO detectors (Fritschel et al. 2024). Detector improvements that facilitate the achievement of the $A^\sharp$ sensitivity include the upgrade of the laser injection system to deliver more power into the interferometer, and an improved system for the thermal compensation of the test-mass mirrors. The test-mass mirrors will be replaced with heavier masses with improved optical coatings, and $A^\sharp$ targets an improved exploitation of the quantum noise reduction from the squeezed-light system. The $A^\sharp$ configurations are natural outgrowths of A+ configurations, and will serve as pathfinders for the next-generation Cosmic Explorer concept (Evans et al. 2021). Additionally, it has much technological overlap with Advanced Virgo+ (AdV+) and Virgo_nEXT (Section 4.2), which presents the possibility of collaborating on developing these technologies.

### 4.2. *Virgo Observatory*

The Virgo interferometer, located in Cascina (Italy), is the largest European GW detector, designed in its AdV phase I as a 3 km dual-recycled Fabry–Perot Michelson interferometer (Acernese et al. 2015). Construction of Virgo started in 1997 and was completed in 2003 (Acernese et al. 2005). Four science runs of the initial Virgo interferometer, VSR1 through VSR4, took place between 2007 and 2011. These were followed by upgrades leading to the AdV design operated during O2 and O3. Subsequently, further upgrades leading to AdV+ were planned to take place in two phases, the first for operation during O4 and the second for operation during O5. A proposed next-generation upgrade planned post-O5, Virgo_nEXT, would provide further sensitivity by pushing current facilities to their limit and would serve as a pathfinder for future ground-based GW detectors.

The first-generation Virgo detector (Accadia et al. 2012a) observed jointly with the initial LIGO detector's fourth and fifth science runs. After several years of commissioning, from 2007 May to 2007 October the first scientific data run VSR1 (along with LIGO) took place, for which a BNS range of 4 Mpc was achieved (Acernese et al. 2008). At this stage, Virgo was a power-recycled Fabry–Perot Michelson interferometer with a 20 W laser source. The second Virgo science run, VSR2 (also along with LIGO), from 2009 July to 2010 January (Accadia & Swinkels 2010), was preceded by set of major improvements to mitigate scattered light and to improve the light-injection system.

The replacement of the four payloads in the Fabry–Perot cavities was the major improvement in preparation for the third Virgo science run VSR3 from 2010 July to 2010 October (Accadia et al. 2012b). Issues arising from thermal noise due to improperly-aligned suspension wires and degraded contrast resulting from differing radii of mirror curvature were addressed leading up to VSR4, from 2011 June to 2011 October, during which Virgo achieved a BNS range of 12 Mpc. While the three previous VSR were aligned with initial LIGO science runs, Virgo took data during this run together with GEO. The main upgrade consisted on the installation of the central heating radius of curvature correction (CHRoCC) on both end mirrors, which allowed to control the radius of curvature of the mirrors in real-time (Accadia et al. 2013). This system was designed to correct the thermal lensing effect in the mirrors, which had been a significant source of noise in the interferometer. Virgo stopped observing in 2011 for the AdV upgrade.

#### 4.2.1. *O2*

After these four science runs, major modifications were made to the optical layout to increase the broadband sensitivity by up to an order of magnitude (Abbott et al. 2017e). These upgrades marked the transition from Virgo, a first-generation interferometer, to AdV, a second-generation GW detector (Acernese et al. 2015). The installation of AdV started in 2011 and was completed in 2016. AdV was planned as a dual-recycled interferometer with 125 W entering the interferometer, though signal recycling was not implemented until O4. The main improvements included a $\sim$10-fold increase in the arm-cavity finesse (a measure of how long light stays within the cavity), 42 kg fused silica test masses with ultra low absorption and high homogeneity, new stray light control using diaphragm baffles and a vibration isolation system, an improved thermal compensation system with double axicon $CO_2$ laser projectors and ring heaters, an improved output mode cleaner with two cascaded monolithic bow-tie resonators, and a new design of payloads triggered by the need to suspend heavier mirrors, baffles and compensation plates.

The several months of commissioning that started at the end of 2016 October achieved the target early-stage BNS range of 8 Mpc in 2017 April with 13 W input laser power. After an intense campaign of noise investigations, AdV sensitivity was considered sufficient to join aLIGO during the O2 observing run in 2017 August (Acernese et al. 2018). During O2, the AdV BNS range reached 30 Mpc. As noted in Section 3.2, the low-frequency Virgo sensitivity during O2 was limited by thermal noise from metallic suspension wires, which were implemented as a fallback option due to the fre-quent failure of monolithic suspensions after the installation of the main AdV upgrades.

#### 4.2.2. *O3*

The most important Virgo upgrades for O3 were the mitigation of suspension thermal noise by installation of monolithic suspensions, and the mitigation of quantum noise by increase of input laser power and by injection of frequency-independent squeezing. An in-air optical parametric amplifier was implemented in the Virgo interferometer before the start of O3a, and squeezing injections were maintained during the whole of O3, with a 3 dB gain in sensitivity at high frequency (Acernese et al. 2019, 2020).

Throughout O3, work was continuously carried out to improve the Virgo sensitivity in parallel with the ongoing data taking. Dedicated tests were made during planned breaks in operation (commissioning, calibration and maintenance), in-depth data analysis of these tests was performed between breaks to ensure continual improvement. In particular, the one-month commissioning break between the O3a and O3b observing periods was used to get a better understanding of the Virgo sensitivity and of some of its main limiting noises (Abbott et al. 2023). This effort culminated during the last three months of O3b.

The most significant change to the Virgo configuration between O3a and O3b was the increase of the input power from 18 W to 26 W. As with the LIGO detectors, it was found that the optical losses of the arms increased following the increase of the input power.

New high quantum-efficiency photodiodes that had been installed at the output (detection) port of the interferometer prior to the start of O3a were found to increase the electronics noise at low frequency. These were improved at the end of 2020 January during a maintenance period, by replacing pre-amplifiers. The electronic noise disappeared completely, leading to a BNS inspiral range gain of $\sim$2 Mpc.

Finally, in the period between the end of 2020 January to the beginning of 2020 February the alignment was improved for the injection of the squeezed light into the interferometer (Acernese et al. 2019, 2020), a critical parameter of the low-frequency sensitivity. By mitigating scattered-light noise, the BNS range increased by 1 Mpc to 2 Mpc.

#### 4.2.3. *O4*

The AdV+ interferometer layout (Acernese et al. 2023b) was designed as a two-step project, for O4 (Phase I) and O5 (Phase II), with the aim to reduce quantum and thermal noise, respectively. The main upgrades for O4 included: a new high-power fiber laser amplifier replacing the former solid state amplifier, to reach 125 W; the implementation of an additional recycling cavity at the output of the interferometer, the signal recycling cavity, to broaden the sensitivity band; an output mode cleaner with increased finesse; a frequency-dependent squeezing system to reduce quantum noise at all frequencies; a network of seismic and acoustic sensors for Newtonian noise monitoring; and a Newtonian calibrator for improved calibration accuracy. These upgrades, while meant to improve the detector sensitivity, also increased the difficulties in controlling the interferometer in presence of optical defects (both from thermal aberration and cold defects), due to the marginal stability of the Virgo recycling cavities (Acernese et al. 2023b). Efforts were put forward to control the dual-recycled interferometer's sensitivity to small defects. For instance, a CHRoCC (Accadia et al. 2013) was installed in 2022 to create a thermal lens on the pick-off plate so as to match the power recycling cavity to the arm cavities. These turned out not to be enough to have a stable interferometer working at the targeted laser power. High-order modes were resonant in the cavities, and these strongly complicated stable operations. The laser input power was decreased to 18 W to improve interferometer control and stability. The various changes on the configuration and attempts to reach stable operations prolonged the anticipated commissioning period between runs. Thus, AdV+ could not join for the O4a observing run. Instead, continued commissioning allowed AdV+ to reach a BNS range of 54 Mpc, with which it joined O4b.

### 4.3. *KAGRA Observatory*

The KAGRA interferometer, situated in Japan's Kamioka mine, is the only large-scale GW detector in East Asia. It is designed as a cryogenic, 3 km, dual-recycled Fabry–Perot Michelson interferometer. The KAGRA project was funded in 2010, construction begin in 2012 and tunnel excavation was completed in 2014 (Akutsu et al. 2021). Following installation and assembly in the tunnel, two operations using temporary detector configurations served as key project milestones: the initial-phase KAGRA (iKAGRA) operation in 2016 April (Akutsu et al. 2018) and the baseline-design KAGRA (bKAGRA) phase-1 operation in 2018 April. During the bKAGRA phase-1 operation, both cryogenic technology and the large-scale vibration isolation systems of KAGRA were successfully demonstrated (Akutsu et al. 2019). By the summer of 2019, the primary installation of instruments was completed, allowing for the commissioning of the detector to begin immediately. In 2019 October, a memorandum of agreement forming the LVK was signed and the LVK international observation network was launched (Brady et al. 2019). After that the commissioning phase continued until 2020 March, marking the commencement of the detector's scientific operation.

#### 4.3.1. *O3GK*

O3GK was a joint observation conducted with the GEO detector in 2020 April (Abe et al. 2023) just after the early termination of O3b. The O3GK operation marked the first joint observation between KAGRA and GEO. This collaboration aimed to improve the detection capabilities by combining data from both detectors. The optical configuration used during O3GK was a power-recycled Fabry–Perot Michelson interferometer, with one room-temperature sapphire test mass and the others set around 250 K.

During the O3GK operation, KAGRA observed for approximately 7.3 d, with a strain sensitivity of $3.0 \times 10^{-22}$ Hz at 250 Hz. The BNS range was about 0.7 Mpc (Abbott et al. 2022). The sensitivity of KAGRA during O3GK was influenced by various noise sources, including sensor noise from local controls of the vibration isolation systems, acoustic noise, shot noise, and laser frequency noise (Abe et al. 2023). Understanding these noise contributions was crucial for planning future improvements to the detector's sensitivity. To enhance its performance, KAGRA plans to implement hardware upgrades and refine its noise mitigation strategies. These improvements aim to extend the detection range and increase the precision of GW observations.

#### 4.3.2. *O4*

On 2024 January 01 a 7.5 magnitude earthquake struck near the KAGRA site, marking the most significant seismic event in the area in the past century. As a result, 10 seismic noise isolators sustained damage but have since been restored. While further investigation and improvements were still needed for some vacuum and facility-related components, partial commissioning began in 2024 July. By 2024 October, all earthquake-related repairs were completed, followed by noise-reduction efforts across multiple domains. During the October commissioning, KAGRA achieved a significant improvement on the BNS range using a power-recycled Fabry–Perot Michelson interferometer configuration with DC readout. Further commissioning tasks have been performed, including: reduction of suspension local control noise through updates to the control filters; reduction of photodiode dark noise below the shot noise level by mitigating electrical coupling from other electronic devices; reduction of quantum shot noise by increasing the laser power to above 10 W; reduction of thermal noise by cooling the mirrors and their suspensions to below 100 K; and reduction of frequency noise and acoustic noise through hardware improvements and control system updates. Following these improvements, KAGRA began operating in O4c on 2025 June 11.

### 4.4. *GEO Observatory*

The GEO detector is a Michelson interferometer with two nearly-orthogonal 600 m arms (Willke et al. 2002). Rather than Fabry–Perot cavities, GEO uses folding in the arms, in which the light traverses each arm twice, to give an optical length of 1200 m for each arm. GEO is sensitive to GWs in the 50 Hz to 1.5 kHz frequency range. GEO began operation in 2001. From 2009 to 2014, it underwent a series of upgrades, the GEO-HF program, that resulted in a factor of 4 improvement in sensitivity at high frequencies (Grote 2010; Dooley et al. 2016). In 2010, squeezed vacuum injection was first applied in GEO (Abadie et al. 2011), and the first long-term application of squeezing was demonstrated in GEO in 2011 (Grote et al. 2013). Subsequently, 6 dB of squeezing (equivalent to a factor of 4 increase in light power) has been achieved (Lough et al. 2021).

GEO has served as an advanced development center and testbed for technologies that were subsequently incorporated in larger detectors (Affeldt et al. 2014) such as dual-recycling (Heinzel et al. 2002), monolithic suspension (Goßler 2004), thermal compensation (Luck et al. 2004), homodyne detection (DC readout) (Hild et al. 2009), and squeezed-light injection (Abadie et al. 2011).

#### 4.4.1. *Astrowatch*

Following the first-generation LIGO and Virgo science runs, GEO embarked on an astrowatch program of near continual data collection (when the detector is not being used for instrument science research) as the sole observing detector (Dooley 2015). This mode of operation has continued since 2007 and allows for searches for GWs associated with external events such as gamma-ray bursts, neutrino detections, or nearby supernovae, occurring outside of other detectors' observing periods (e.g., Abac et al. 2024b).

#### 4.4.2. *O3GK*

As described in Section 3.3, a two-week-long joint observing run with the GEO and KAGRA detectors took place in 2020 April, during which GEO operated with a 80% duty cycle (10.9 d of operation) and a BNS range of 1.1 Mpc (Abbott et al. 2022). The laser power injected was about 3 W, which led to about 3 kW of circulating power in the power recycling cavity, or 1.5 kW circulating power per arm (Affeldt et al. 2014; Dooley et al. 2016). Bilinear noise subtraction resulted in modest improvement in sensitivity and data quality (Mukund et al. 2020). Since GEO and KAGRA had similar sensitivity during O3GK, this joint run enabled searches for GW transient signals occurring simultaneously in both detectors, though no significant events were observed (Abbott et al. 2022).

## 5. REVIEW OF OBSERVED TRANSIENT SOURCES

The GWTC includes all observed transient GW candidates reported by the LVK. It is most likely that the significant candidates in GWTC-4.0 have an astrophysical origin and were produced by CBC sources (the remaining less-significant candidates are largely non-astrophysical). This section provides a foundational overview of transient GW signals, especially those from CBCs, for use in interpreting the catalog's contents and for reference in companion papers. We first provide a short overview of the basic physics of GWs and then provide an introduction to the CBC sources to be used as a reference for other papers in the collection of articles. Additional detail can also be found in Maggiore (2007, 2018) and Creighton & Anderson (2011).

### 5.1. *Gravitational Waves*

In metric theories of gravity, such as GR, the local gravitational field can be described in terms of 6 independent degrees of freedom that represent the relative accelerations of a collection of nearby freely-falling observers (Pirani 1956; Misner et al. 1973). Plane wave solutions to the linearized gravitational field equations (Einstein 1916) represent the weak GWs in the far-field region (where the observer is far from the source, and the gravitational field is treated as a perturbation to Minkowski spacetime) that are observed by GW detectors. The vacuum Einstein field equations of GR then further restrict the degrees of freedom of the plane wave solutions to 2 transverse polarizations that propagate at the speed of light (Eddington 1922). These are called the *plus* (+) polarization and the *cross* (×) polarization. In a suitably-chosen set of coordinates, known as the *transverse-traceless gauge* (Misner et al. 1973; Thorne 1987) which is akin to the radiation gauge in classical electromagnetism, the perturbation to the Minkowski metric for these two polarizations is given by the two functions of spacetime $h_+$ and $h_\times$, respectively. These polarizations represent two spin-2 purely-transverse tensor modes (Weinberg 1972). The transverse-traceless gauge is a useful choice because worldlines that are the histories of fixed points in these spatial coordinates are geodesics of the perturbed spacetime (Hartle 2021). Thus, changes in time in the metrical distance between fixed spatial coordinate locations, which is described by the time derivatives of $h_+$ and $h_\times$, represent the deviation of the geodesics at these locations. Therefore, $h_+$ and $h_\times$ are the physical (observable) degrees of freedom of a GW.

From an observational point of view, GW signals are broadly classified as *persistent* or *transient*. The main classes of persistent GWs include quasi-monochromatic signals, e.g., as produced by rotating NSs having a non-axisymmetric mass distribution (Zimmermann & Szedenits 1979), and continuous stochastic superpositions of GWs from numerous unresolved independent sources (Romano & Cornish 2017). Here we focus on the transient signals that are cataloged in GWTC.

### 5.1.1. *Transient GW Signals*

A *transient* GW is one that registers a signal of short duration (much less than the duration of the observing run) within the sensitivity band of the GW detectors. Such GWs can be characterized by their *geocentric arrival time* $t_{\rm geo}$, the time at which some fiducial point in the GW's waveform (such as its peak amplitude, for example) passes through the Earth's center. We expect that transient GWs will be observed as plane waves originating from a particular point in the sky, usually given in terms of the equatorial celestial coordinate system of right ascension $\alpha$ and declination $\delta$, with a normal vector $-\boldsymbol{N}$ along this line of sight. A key task for multimessenger astronomy with GWs is the reconstruction of the source location, which facilitates follow-up with other astronomical facilities (Abbott et al. 2020a).

A network of detectors spaced at different locations on the Earth can observe the difference in the time of arrival of the fiducial point in the waveform arising from the propagation of the plane wave across the Earth, and thereby reconstruct the direction of propagation $\boldsymbol{N}$ (Fairhurst 2009, 2011; Creighton & Anderson 2011; Singer & Price 2016). Such triangulation is the main way in which the source of transient GWs are localized. Hence, uncertainty in the sky location of the source, $\Delta\Omega$, partially results from the measurement uncertainty of the arrival time in each detector (Fairhurst 2011). A single detector provides no ability to determine the sky location of a source for transient signals lasting much less than a day and having wavelengths much longer than the size of the detector (as is the case for all candidates reported in GWTC-4.0); however, with two detectors, the difference in times of arrival identifies a circle on the celestial sphere, centered on the axis separating the detectors, on which the wave's origin may lie. The presence of a third detector whose location is not colinear with the other two then identifies the source position to one of two points on the sky mirrored across the plane containing these three detectors. A fourth detector, not coplanar with the other three, finally resolves the location of the source to a single point on the sky. Additional localization information can be provided by coherently combining the observed GW signals from an array of detectors as described below. For some types of transient sources having known GW emission, such as CBCs, it is also possible to estimate the distance to the source from measurements of the wave amplitude (Cutler & Flanagan 1994). In such cases, there is a volume-localization uncertainty $\Delta V$ as well (Singer et al. 2016; Del Pozzo et al. 2018).

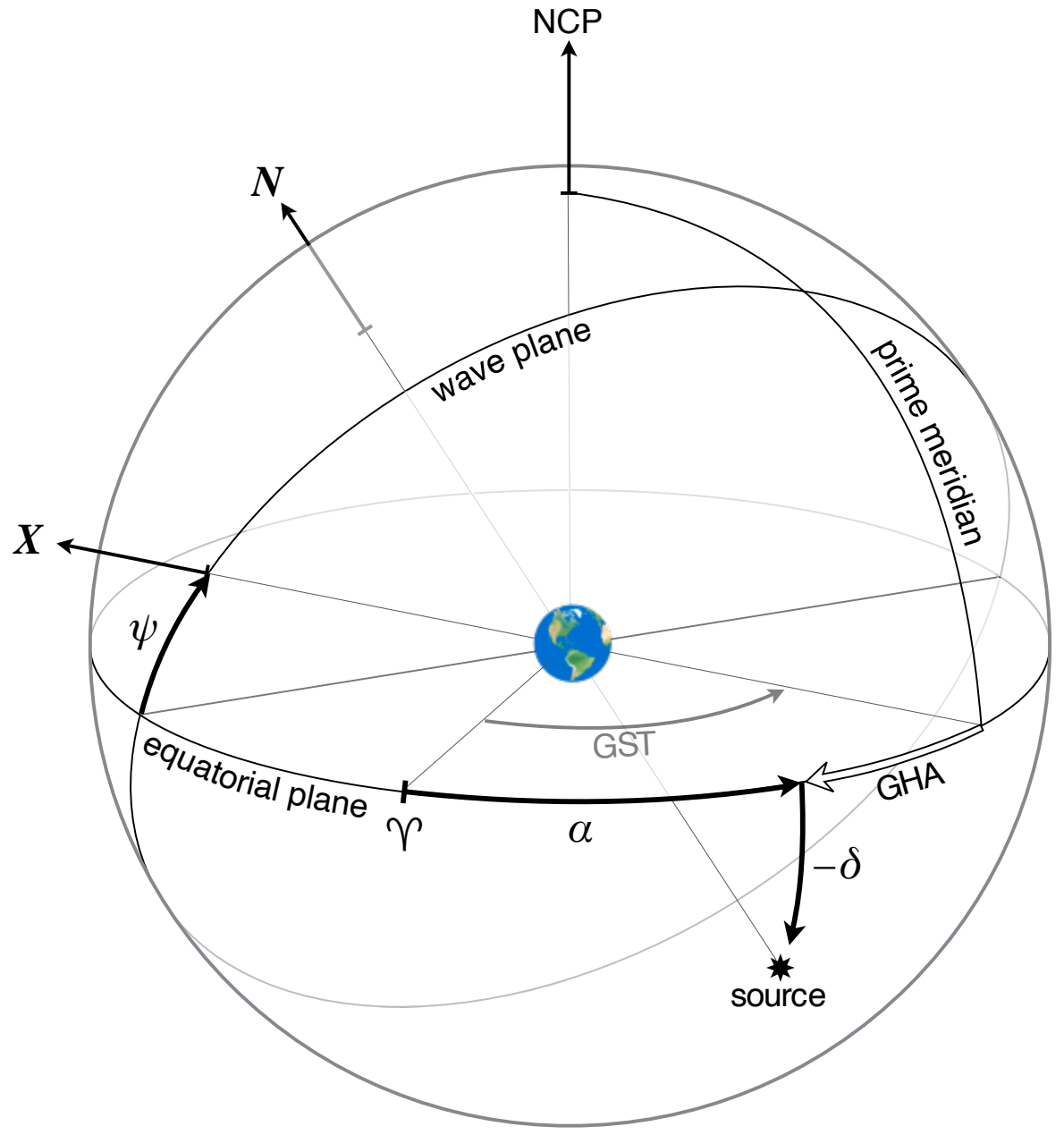

**Figure 4.** Relationship between the sky location in equatorial coordinates, the polarization angle, and the GW coordinate frame. The direction from the source to the Earth is $\boldsymbol{N}$ and the vector $\boldsymbol{X}$ defines a reference direction on the transverse plane called the wave plane. The location on the source on the sky in the equatorial coordinate system is given by its right ascension $\alpha$ and declination $\delta$. The polarization angle $\psi$ is the angle counterclockwise about $\boldsymbol{N}$ between the equatorial plane and $\boldsymbol{X}$. Also shown is the Greenwich sidereal time (GST), the angle between the first point of Aries ♈ and the prime meridian, and the Greenwich hour angle (GHA) of the source, GHA = GST − $\alpha$. NCP is the North celestial pole.

GW detectors such as the LIGO, Virgo, and KAGRA detectors are designed to sense changes in the difference of the lengths of their orthogonal arms, $\Delta L = \Delta(L_1 - L_2)$, caused by GWs, via laser interferometry. These L-shaped Michelson interferometers measure the difference in phase of coherent light, split at a beam-splitter located at the vertex of the L, after traversing the arms and recombining at the beam-splitter $\Delta\phi = 2\pi\Delta L/\lambda_*$, where $\lambda_*$ is the wavelength of the laser light (Moss et al. 1971; Weiss 2022; Forward 1978). For GW transients having durations much less than a day and wavelengths much greater than the length $L$ of the detector arms, the strain induced on the arms is a linear combination of plus- and cross-polarizations of the metric perturbation (Forward 1978; Rudenko & Sazhin 1980; Schutz & Tinto 1987; Thorne 1987)

$$h = \frac{\Delta L}{L} = F_+ h_+ + F_\times h_\times. \tag{2}$$

Here, $F_+$ and $F_\times$ are the detector's beam pattern functions, which depend on the position on the sky from which the GW source is located, a polarization angle that defines the axes of the plus- and cross-polarization in the wave frame, the Earth rotation angle at the time of the signal's arrival, and the location, orientation, and geometry of the detector on the Earth's surface (Anderson et al. 2001). Figure 4 shows the sky coordinate conventions used. For long-duration signals, effects of the Earth's rotation need to be included; for short-wavelength signals, the beam pattern functions also depend on the wavelength of the GWs (Rakhmanov 2009; Rakhmanov et al. 2008). Neither of these effects are significant in any of the transient signals detected to date. The amplitudes of the strains measured in a network of detectors provide additional information about the location of the source of the GW if the polarization of the GW is known owing to the dependence of the beam pattern functions on the position of the source on the sky (e.g., Singer & Price 2016). This information helps to break degeneracies in sky localization; for instance, with only two detectors, the source is typically localized to an extended arc or ring on the sky, but the amplitude response can help reduce this uncertainty to specific regions along that ring.

Table 2 summarizes the parameters associated with a general transient plane GW, a detector's response to such a GW, and the accuracy of localization of the wave's source.

*LVK Catalog of Observed Transient GW Signals*—In the companion paper Abac et al. (2025b), we describe the significant transient GWs candidates in GWTC-4.0, highlighting those observed in O4a. GWTC-4.0 also provides the inferred properties of the GWs as well as their sources, e.g., the masses and spins of the binary components under the assumption than the GWs were produced by CBCs. The GWTC dataset, along with other open data products, is detailed in the companion paper Abac et al. (2025i).

### 5.1.2. *Gravitational Lensing of GWs*

Like electromagnetic waves, GWs can be gravitationally lensed by massive objects, e.g., galaxies, interposed between the GW source and the observer. The GW polarization tensor is parallel-propagated along geodesics (Misner et al. 1973) and is little affected by the gravitational potential of the lensing mass so it is sufficient to consider scalar diffraction theory (Takahashi & Nakamura 2003). In a thin-lens approximation, the bending of the trajectory of the GW propagation occurs on a lens plane orthogonal to the line of sight and at the distance of the lensing body. With $\xi_1$ and $\xi_2$ as the coordinates of the lens plane, at each point on this plane there is an observed time delay $T(\xi_1, \xi_2)$ relative to straight-line motion with no lens, corresponding to the path from the source to that point on the lens plane to the observer. This delay accounts for the gravitational field of the lens. GWs are deflected by a gravitational lens with the time delay field on the lens plane determining the complex phases of the interfering partial waves used to compute a frequency-dependent complex-valued magnification factor. This factor is given by the Fresnel–Kirchhoff diffraction formula

$$F(f) = i\frac{D_{\rm OS}}{D_{\rm OL}D_{\rm LS}}\frac{(1+z_{\rm L})f}{c}\iint \exp\left[-2\pi i f T(\xi_1, \xi_2)\right]\, d\xi_1\, d\xi_2\,, \tag{3}$$

where the integral is over the lens plane, $f$ is the observed GW frequency, $(1 + z_{\rm L})f$ is the blue-shifted frequency of the GW on the lens plane ($z_{\rm L}$ is the redshift of the lens), and the distances $D_{\rm OS}$, $D_{\rm OL}$, and $D_{\rm LS}$ are the distances between the observer (us) and the GW source, between the observer and the gravitational lensing object, and between the lensing object and the source respectively (Schneider et al. 1992). In a cosmological setting, these are *angular diameter* distances (Hogg 1999). The geometric optics limit corresponds to Fermat's principle in which the geodesic paths taken by GWs are those passing through the lens plane at extrema of this two-dimensional time-delay field $T(\xi_1, \xi_2)$, which may be local minima, which produce *Type I images*, local maxima, which produce *Type III images*, or saddle points, which produce *Type II images* (Schneider et al. 1992). Equation (3) is evaluated in this high-frequency limit by use of the stationary phase approximation to obtain

$$F_j(\pm|f|) = \sqrt{\mu_j}\exp\left(\mp 2\pi i|f|t_j \pm i\pi n_j\right)\,, \tag{4}$$

where $\sqrt{\mu_j}$ and $t_j$ are the magnification amplitude and observed time delay of image $j$, and $n_j$ is 0, 1/2, or 1 for Type I, Type II, and Type III images respectively. Therefore, such images are magnified or demagnified by a factor that is positive for Type I images and negative for Type III images, while the gravitational waveform of Type II images is additionally distorted, appearing as the Hilbert transform of the original waveform (Dai & Venumadhav 2017; Ezquiaga et al. 2021). For GW transients, the images are a set of repeated signals from the same event observed at different times, the delays determined by the differences in the time-delay field on the lens plane of the different images. These delays are typically minutes to months for galaxy lenses (Li et al. 2018; Ng et al. 2018; Oguri 2018) and up to years for galaxy cluster lenses (Smith et al. 2017, 2018; Robertson et al. 2020; Ryczanowski et al. 2020). The images also appear at different points on the sky, with arcminute-scale separation, but GW detectors have insufficient sky-localization capabilities to distinguish them in this way. When gravitational lensing can be described in this geometric optics limit it is referred to as *strong lensing*.

**Table 2.** Parameters describing a transient plane GW, a detector's instantaneous antenna response in the long-wavelength limit, and measures of inferred localization of the signal on the sky.

| Parameter name | Symbol | Notes [Dimensions] |
|---|---|---|
| Plus (+) and cross (×) polarizations | $h_+, h_\times$ | Functions describing the plus polarization ($h_+$) and cross polarization ($h_\times$) of the metric perturbation [dimensionless] |
| Geocentric arrival time | $t_{\mathrm{geo}}$ | Time of arrival at the center of the Earth of some fiducial point in the GW's waveform, normally close to the peak amplitude of the waveform [time] |
| Propagation direction | $\boldsymbol{N}$ | Direction of propagation of the GW, the unit vector normal to the planar wavefronts; the direction to the source of the wave is $-\boldsymbol{N}$ [dimensionless] |
| Right ascension | $\alpha$ | Azimuth of the sky location of the source of the GW in the equatorial coordinate system (see Figure 4) [angle] |
| Declination | $\delta$ | Latitude of the sky location of the source of the GW in the equatorial coordinate system (see Figure 4) [angle] |
| Polarization angle | $\psi$ | Orientation of the axes defining the plus- and cross-polarization on the transverse plane of the GW relative to the line-of-nodes of this plane and the Earth's equatorial plane (see Figure 4) [angle] |
| Plus and cross beam patterns | $F_+, F_\times$ | Antenna response of a detector to the plus-polarization ($F_+$) and the cross-polarization ($F_\times$), functions of the sky location of the source, the polarization angle, the geocentric arrival time of the signal, and the location, orientation, and geometry of the detector on the Earth (Anderson et al. 2001) [dimensionless] |
| Detector strain | $h$ | Gravitational-wave induced strain on a detector, Equation (2); the GW readout of the detector is proportional to this quantity [length/length] |
| Sky area | $\Delta\Omega$ | Localization area, typically taken as the 90% credible area; if results at different CLs are quoted, these are indicated with a subscript, e.g., $\Delta\Omega_{50}$ is the 50% credible area [solid angle] |
| Volume localization | $\Delta V$ | Localization volume (for signals where the distance to the source can be estimated), typically taken as the 90% credible volume; if results at different CLs are quoted, these are indicated with a subscript, e.g., $\Delta V_{50}$ is the 50% credible volume [volume] |

However, when the wavelength of the GW is comparable to the Schwarzschild radius of the gravitational lens, the geometric optics limit of Fermat's principle is no longer valid, and the Fresnel–Kirchhoff diffraction formula of Equation (3) must be used to determine the complex-valued and frequency dependent magnification factor. Such lensing effects can result from objects having masses up to $10^5\,M_\odot$ and searches can be done in a modeled (e.g., Wright & Hendry 2021) or phenomenological (Liu et al. 2023) way.

*Searches for Gravitational-lensing Signatures in GW Signals*— In the companion paper Abac et al. (2025h), we present searches for gravitational-lensing signatures in GWTC-4.0. Such signatures sought include multiple images from strong lensing, individual Type-II strongly lensed images, and waveform distortions induced by point-mass lensing.

### 5.1.3. *GW Polarization and Propagation in Alternative Theories of Gravity*

In GR, plane GW perturbations to flat spacetime propagate at the speed of light and contain two transverse polarizations. However, in modified theories of gravity extending beyond GR, additional polarizations may be present, including two transverse-longitudinal spin-1 vector modes, a transverse spin-0 scalar mode, and a longitudinal spin-0 scalar mode (Eardley et al. 1973a,b; Will 2018). With multiple detectors, it is possible to test for such additional polarizations (Schutz 1986). A linear combination of strain data from three detectors can be formed in which any GW signal from a known sky location and containing only plus- and cross-polarizations is canceled (Guersel & Tinto 1989; Klimenko et al. 2008; Sutton et al. 2010; Creighton & Anderson 2011; Wong et al. 2021). Any residual GW signal found in such a null-space would provide evidence for the presence of vector or scalar non-GR polarizations.

In addition, in alternative Lorentz invariance violating theories of gravity or in which the graviton is massive, GWs are dispersive. Certain theories of dark energy also result in dispersive GW propagation (de Rham & Melville 2018; Baker et al. 2022; Harry & Noller 2022). The GW dispersion relation between the frequency $f$ and the wavelength $\lambda$ (one where they are not inversely proportional) leads to phase speeds and/or group speeds that differ from the speed of light. Such propagation effects can be measured for a known waveform by the anomalous arrival times of different frequency components. A common parameterized dispersion relationship is motivated by a modified energy–momentum relationship for the graviton of the form (Mirshekari et al. 2012)

$$E^2 = (pc)^2 + A_\alpha (pc)^\alpha \,, \tag{5}$$

where $A_\alpha$ is a GR-violating parameter having dimensions of $(\text{energy})^{2-\alpha}$. For de Broglie waves, $E = 2\pi\hbar f$ and $p = 2\pi\hbar/\lambda$, where $2\pi\hbar$ is the Planck constant. Such a modified energy–momentum relation leads to a dispersion relation in which the phase velocity $v_\mathrm{p}$ is given by

$$\left(\frac{v_\mathrm{p}}{c}\right)^2 = 1 + A_\alpha \left(\frac{2\pi\hbar c}{\lambda}\right)^{\alpha-2} \,, \tag{6}$$

where the phase velocity is related to the frequency and the wavelength of the GW, $v_\mathrm{p} = \lambda f$. The group velocity, $v_\mathrm{g} = v_\mathrm{p} - dv_\mathrm{p}/d\ln\lambda$, determines the difference in arrival times of different frequency components of the GW after propagation from its source to the observer. For small deviations from GR ($v_\mathrm{p} \approx c$), the group velocity is frequency dependent with

$$\frac{v_\mathrm{g} - c}{c} \approx \frac{1}{2}(\alpha - 1) A_\alpha (2\pi\hbar f)^{\alpha-2} \,. \tag{7}$$

Special cases include (i) a graviton of mass $m_g \neq 0$ for which $\alpha = 0$, $A_0 = m_g^2 c^4$, and

$$\frac{v_\mathrm{g} - c}{c} \approx -\frac{1}{2}\left(\frac{\lambda_g f}{c}\right)^{-2} \,, \tag{8}$$

where $\lambda_g = 2\pi\hbar/(m_g c)$ is the Compton wavelength of the graviton, and (ii) the case in which GWs are non-dispersive but propagate at a speed different than the speed of light for which $\alpha = 2$ and

$$v_\mathrm{g} = c\sqrt{1 + A_2} \,. \tag{9}$$

Stringent bounds on the latter are provided by the close temporal association of the BNS signal GW170817 and the gamma-ray burst GRB 170817A (the gamma rays arriving less than 2 s after the BNS GW merger signal), resulting in $|A_2| \lesssim 10^{-14}$ (Abbott et al. 2017f).

The Einstein–Hilbert action of GR contains second derivatives of the spacetime metric (Weinberg 1972; Misner et al. 1973; Wald 1984; Carroll 2019). Standard model extensions having modified actions containing third derivatives of the metric can produce CPT-violating terms in the gravitational field equations, which can produce birefringence effects in which different GW helicities propagate with different phase velocities (Kostelecky 2004; Kostelecký & Mewes 2016; Mewes 2019; Haegel et al. 2023). Other theories of gravitation also have GWs with birefringent propagation (Zhu et al. 2024). Such birefringence leads to a frequency-dependent rotation of the GW polarization angle.

Both GW birefringence and the modified GW dispersion relation can potentially be anisotropic, where the magnitude of the observed effect depends on the direction to the source.

*Tests of GR: GW Polarization and Propagation*—In the companion paper Abac et al. (2025d), we test the GR prediction of the polarizations of GWs by searching for evidence of vector- or scalar-polarization modes in observed GW signals. Meanwhile, Abac et al. (2025e) presents tests of a modified dispersion relation and of anisotropic birefringence using GW signals from CBCs, for which it is assumed that the GW near the source is described by GR to a good approximation, but the waveform is affected during propagation.

## 5.2. *Compact Binary Coalescence*

Binaries consisting of two BHs (BBH systems), two NSs (BNS systems), or in which one component is a NS and the other a BH (NSBH systems), have all been observed by the LVK (Abbott et al. 2016a, 2017a, 2021c). The detectable signal produced by such systems arises from the late stage of orbital decay, driven by GW emission, and by the ensuing merger of the binary components and the settling of the resulting object (a NS or BH) to a final, stationary configuration (Chatziioannou et al. 2024).

Table 3 provides a list of parameters used to describe CBCs.

### 5.2.1. *Newtonian Inspiral*

At early stages of the inspiral, when the magnitude of the difference in velocity vectors of the two components of the binary, $v$, is much smaller than the speed of light, the orbit is determined approximately by Newtonian mechanics while the gravitational radiation is described by the quadrupole formula (Einstein 1916, corrected by Eddington 1922, page 279). For a quasi-circular orbit that is inclined an angle $\iota$ relative to the direction to an observer, $h_+$ and $h_\times$ are sinusoidal and are 90° out of phase,

$$h_+ = -2(1 + \cos^2\iota)\frac{GM\eta}{c^2 r}\left(\frac{v}{c}\right)^2 \cos 2\phi \tag{10a}$$

and

$$h_\times = -4\cos\iota \frac{GM\eta}{c^2 r}\left(\frac{v}{c}\right)^2 \sin 2\phi \,, \tag{10b}$$

where $r$ is the distance between the source and the observer, $M = m_1 + m_2$ is the total mass of the system, $\eta = m_1 m_2/M^2$ is the symmetric mass ratio, $M\eta$ is the reduced mass of the system, and $\phi$ is the orbital phase relative to the ascending node (Peters & Mathews 1963; Thorne 1987; Finn & Chernoff 1993; Will & Wiseman 1996). See Figure 5. When $\iota = 0$ or $\iota = \pi$ (face on and face off respectively), the amplitudes of the sinusoidal functions $h_+$ and $h_\times$ are equal and the GW is circularly polarized; when $\iota = \pi/2$ (edge on), $h_\times = 0$, and the GW is linearly polarized.

The GW luminosity of such a system, i.e., the power in gravitational radiation, is

$$\dot{E}_{\rm GW} = \frac{32}{5}\frac{c^5}{G}\eta^2\left(\frac{v}{c}\right)^{10} . \tag{11}$$

This radiation gives rise to a secular orbital decay. Since the (Newtonian) energy of the bound system is $E_{\rm orb} = -(1/2)\eta M v^2$, and equating $\dot{E}_{\rm GW} = -\dot{E}_{\rm orb}$, we deduce that the period of the orbit, $P = 2\pi GM/v^3$ by Kepler's third law, evolves according to

$$\dot{P} = -\frac{192\pi}{5}\eta\left(\frac{v}{c}\right)^5 . \tag{12}$$

At fixed orbital period (or orbital frequency), the orbital velocity is proportional to the cube root of the total mass, $v \propto M^{1/3}$. It can be seen then that $h_+, h_\times \propto \eta M^{5/3}$, $\dot{E}_{\rm GW} \propto (\eta M^{5/3})^2$, $E_{\rm orb} \propto \eta M^{5/3}$ and $\dot{P} \propto \eta M^{5/3}$. At the Newtonian level of approximation, a single combination of the component masses,

$$\mathcal{M} = \eta^{3/5} M = \frac{(m_1 m_2)^{3/5}}{(m_1 + m_2)^{1/5}} , \tag{13}$$

known as the *chirp mass*, solely determines both the amplitude of a GW at fixed orbital frequency and its frequency evolution (Kafka 1988; Cutler et al. 1993; Finn & Chernoff 1993). This chirp mass is normally the most accurately-measured mass parameter for low-mass systems in which most of the signal observed arises from the pre-merger phase.

#### 5.2.2. *Post-Newtonian Inspiral and Other Effects*

Additional terms in the GW amplitude and frequency evolution appear at higher orders in $v/c$ in a post-Newtonian (PN) expansion in the equations of motion and in the gravitational emission (Blanchet 2014). At the Newtonian order, the frequency of the GW is twice the frequency of the orbital motion, $f = 2f_{\rm orb} = 2/P$, and

$$v^3 = \pi G M f . \tag{14}$$

At $O(v/c)$ beyond this, additional components to the GW at frequencies at one and three times the orbital frequency arise from current quadrupole and mass octupole radiation, and other components occur at $O(v^2/c^2)$ beyond Newtonian order from current octupole and mass hexadecapole radiation (Thorne 1980); the amplitudes of these *higher-order multipole moments* of radiation are proportional to a different combination of component masses (Kidder 2008). The frequency evolution also gains additional terms at $O(v^2/c^2)$ beyond the leading-order Newtonian term, again having a different dependence on the component masses from the leading Newtonian order (Wagoner & Will 1976). Spin effects from rotating binary components also appear in post-Newtonian corrections to the quadrupole waveform due to $O(v^3/c^3)$ spin–orbit and $O(v^4/c^4)$ spin–spin effects (Kidder et al. 1993), and to precession of the orbital plane if the spin angular momentum vectors of the bodies are not aligned (or anti-aligned) with the orbital angular momentum vector (Apostolatos et al. 1994). The dimensionless parameter $\chi_{\rm eff}$,

$$\chi_{\rm eff} = \frac{c}{GM}\frac{\boldsymbol{L}\cdot(\boldsymbol{S}_1/m_1 + \boldsymbol{S}_2/m_2)}{|\boldsymbol{L}|} \tag{15}$$

where $\boldsymbol{S}_1$ and $\boldsymbol{S}_2$ are the spins of the two binary components and $\boldsymbol{L}$ is the orbital angular momentum about the center of mass, is an effective inspiral spin parameter that is conserved under the orbit-averaged precession equations of motion at $O(v^4/c^4)$ (Damour 2001; Racine 2008; Ajith et al. 2011; Santamaria et al. 2010; Kesden et al. 2010). Whereas $\chi_{\rm eff}$ de-pends on the spin components aligned with the orbital angular momentum, a dimensionless effective precession spin parameter that depends on in-orbital-plane components of the spins,

$$\chi_{\rm p} = \frac{c}{Gm_1} \max\left\{ \frac{|\boldsymbol{L}\times\boldsymbol{S}_1/m_1|}{|\boldsymbol{L}|}, \frac{3m_1+4m_2}{3m_2+4m_1} \frac{|\boldsymbol{L}\times\boldsymbol{S}_2/m_2|}{|\boldsymbol{L}|} \right\}, \tag{16}$$

captures the dominant precession effects (Schmidt et al. 2015).

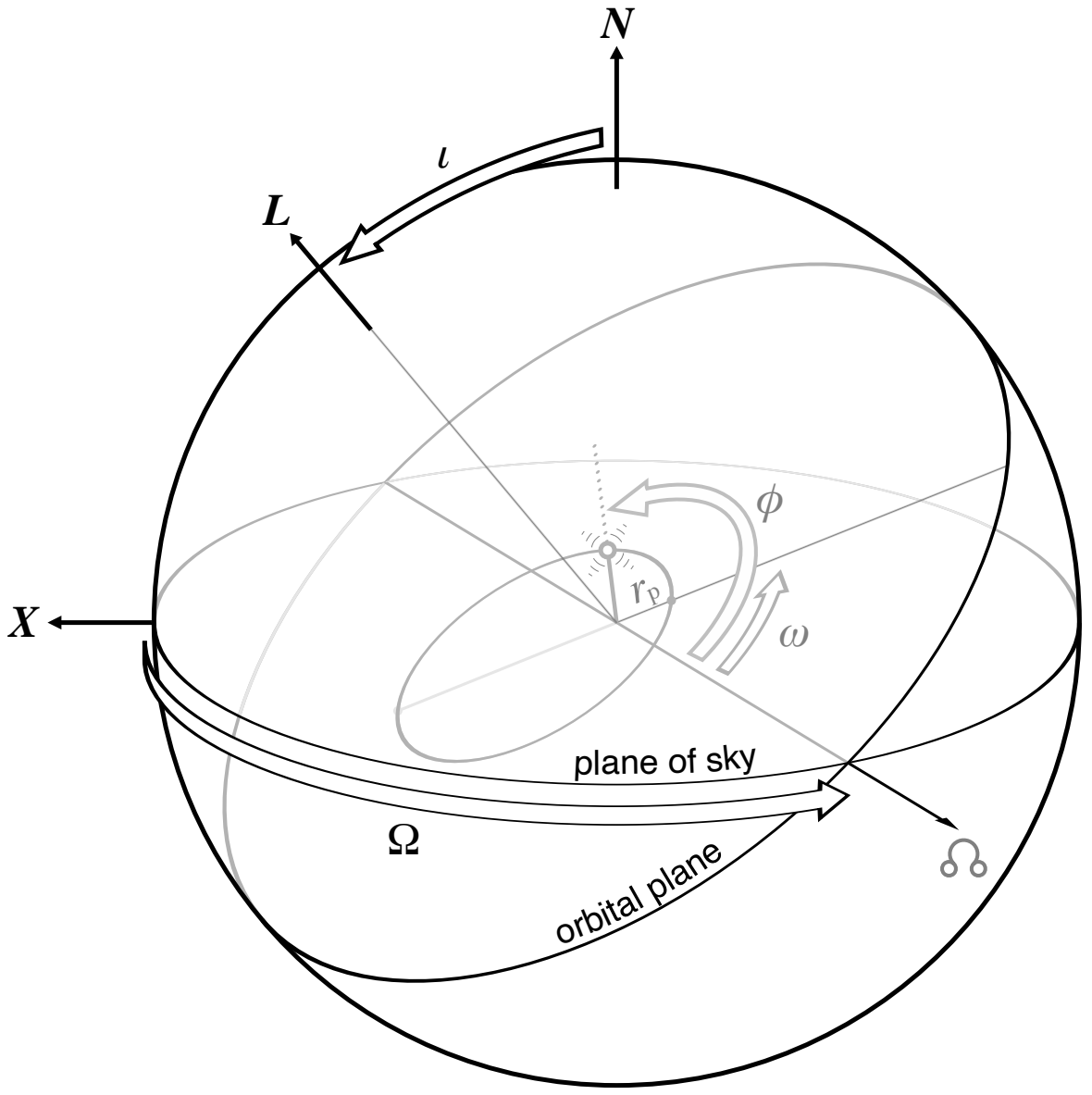

**Figure 5.** Relationship between the orbital elements and the GW coordinate frame. The direction from the source to the Earth is $\boldsymbol{N}$ and the vector $\boldsymbol{X}$ defines a reference direction on the transverse plane (the plane of the sky). The inclination $\iota$ is the angle between $\boldsymbol{N}$ and the orbital angular momentum vector $\boldsymbol{L}$. The longitude of the ascending node of the orbit $\Omega$ is the angle on the plane of the sky between $\boldsymbol{X}$ the ascending node ☊, $\boldsymbol{N}\times\boldsymbol{L}$. The angle $\Omega$ is degenerate with the polarization angle $\psi$. The orbit of the primary about the center of mass of the system is shown. The orbital phase $\phi$ is the angle on the orbital plane between the ascending node and position vector of the primary relative to the center of mass. For an eccentric orbit, the distance of the primary from the center of mass at periapsis is $r_{\rm p}$ and the argument of the periapsis $\omega$ for the primary is the angle on the orbital plane between the ascending node and the position vector of the primary at periapsis.

Deformable binary components (NSs but not BHs) suffer an induced quadrupole deformation $Q_{ij}$ under an external tidal field $\mathcal{E}_{ij}$, where these quadrupole tensors are those appearing in a multipole expansion of the Newtonian potential centered on the body of mass $m$ as (Thorne 1998)

$$\Phi(\boldsymbol{x}) = -\frac{Gm}{|\boldsymbol{x}|} - \frac{1}{2} G Q_{ij} \frac{3x^i x^j - |\boldsymbol{x}|^2 \delta^{ij}}{|\boldsymbol{x}|^5} + \frac{1}{2}\mathcal{E}_{ij} x^i x^j + \cdots. \tag{17}$$

The *dimensionless tidal deformability*, $\Lambda$, of a body of mass $m$ is defined in terms of the ratio of the induced deformation to the external tidal field as

$$\frac{GQ_{ij}}{(Gm/c^2)^5} = -\Lambda \mathcal{E}_{ij}, \tag{18}$$

where BHs have $\Lambda = 0$. Newtonian tidal interactions of deformable components appear as effective $O(v^{10}/c^{10})$ corrections to the binding energy and GW luminosity (Flanagan & Hinderer 2008). At this order, the dimensionless combination of tidal parameters given by (Favata 2014)

$$\tilde{\Lambda} = \frac{16}{13} \frac{(m_1 + 12m_2) m_1^4 \Lambda_1 + (m_2 + 12m_1) m_2^4 \Lambda_2}{(m_1+m_2)^5} \tag{19}$$

appears, where $\Lambda_1$ and $\Lambda_2$ are the dimensionless tidal deformabilities of the two bodies, and it is this parameter that is most measurable in the waveforms produced by binaries with deformable companions (Poisson 2021). Spinning bodies experience quadrupole deformation of the form

$$Q_{ij} = \mathrm{diag}\left( -\frac{1}{3}Q, -\frac{1}{3}Q, \frac{2}{3}Q \right), \tag{20}$$

where $Q$ is the spin-induced mass quadrupole moment scalar. The quadrupole deformations induced by an object's spin results in Newtonian quadrupole–monopole effects as an effective $O(v^4/c^4)$ correction, the same order as the spin–spin coupling relativistic effects (Poisson 1998). The size of the spin-induced deformation depends on the nature of the body, where the ratio of the quadrupole scalar to the square of the body's spin magnitude is given by the dimensionless parameter $\kappa$ as

$$Q = -\kappa \frac{|\boldsymbol{S}|^2}{mc^2}, \tag{21}$$

where $m$ is the mass of the body. For a BH, $\kappa = 1$ (Thorne 1980).

Binaries detected by ground-based observatories are commonly assumed to have negligibly small orbital eccentricity remaining by the time the orbital period has decayed to the point that the GW frequencies have entered the high-frequency sensitivity band of the detectors. This decay in eccentricity happens because orbital eccentricity is efficiently reduced by GW emission during the orbital decay (Peters 1964). However, there are channels of compact binary formation that could result in non-negligible orbital eccentricity being present even at the last stages of inspiral observed by ground-based detectors (e.g., Mapelli 2020). The leading order effects of orbital eccentricity would appear at the Newtonian level (Peters & Mathews 1963). Two additional parameters are needed to describe an eccentric binary system, the eccentricity $e$, and the argument of the periapsis $\omega$. Although these are well defined for Newtonian two-body systems, there are different ways to generalize their definitions for relativistic systems, and there is not yet a settled convention for these parameters (Shaikh et al. 2023).

*Tests of GR from CBC Inspiral*—PN theory in GR predicts the relative amplitudes of subdominant modes of GW radiation (Blanchet 2014), which depend on the binary's masses and spins (e.g., Arun et al. 2009). Thus, allowing for freedom in these amplitudes and checking if they are consistent with those predicted by GR provides a consistency test of the agreement of the signal with the waveform model used to analyze it (Puecher et al. 2022). In the companion paper Abac et al. (2025d), this test is carried out for BBH signals, considering deviations, $\delta A_{\ell m}$, in the amplitude of the $(\ell = 2, m = \pm 1)$ or $(\ell = 3, m = \pm 3)$ subdominant multipole moments relative to the dominant $(\ell = 2, m = \pm 2)$ and other multipole moments.

The PN expansion of the orbital energy and GW energy loss makes a prediction of how the GW phase evolves with time as the orbit decays (Blanchet 2014). The PN formalism expresses this phase evolution with a set of coefficients in a series expansion of the GW phase in terms of powers $(v/c)^{n-5}$ and $(v/c)^{n-5}\log(v/c)$ for integer $n$ (with $n = 0$ for the leading-order Newtonian inspiral) that depend on the binary components' masses and spins for point particles. Violations of GR can lead to differences in the values of the PN coefficients from those predicted by GR (e.g., Yunes & Pretorius 2009; Tahura & Yagi 2018) which could be observed in a GW signal (Blanchet & Sathyaprakash 1994, 1995; Arun et al. 2006; Mishra et al. 2010; Li et al. 2012). In the companion paper Abac et al. (2025e), we present parameterized tests for such violations.

Effects arising from the finite size of the component masses of a binary include spin-induced multipole moments, most importantly their spin-induced quadrupole moments $Q$, which also affect the orbital evolution. For a BH, there is a fixed relation between its spin-induced quadrupole moment and its mass and spin (Poisson 1998). Deviations from this predicted value, as observed in the phase evolution of a GW signal, can be used to distinguish a BBH from a compact binary containing exotic, non-BH components. Some examples of exotic alternatives to BHs, being compact objects capable of having masses greater than the maximum mass of a NS, include boson stars (Kaup 1968; Ruffini & Bonaz-zola 1969), gravastars (Mazur & Mottola 2004), fuzzballs (Mathur 2005), and firewalls (Almheiri et al. 2013). The companion paper Abac et al. (2025e) presents such parameterized tests of the nature of the components of CBCs.

#### 5.2.3. *Compact Binary Merger and Ringdown*

The final stages of GW emission from CBCs that result in a BH remnant can be modeled as a linear gravitational perturbation to a Kerr BH spacetime (Press & Teukolsky 1973). Remarkably, the partial differential equations for the outgoing GW content of such a perturbation *decouple* from the other gravitational modes, and those decoupled equations are *separable* into a radial equation, an angular equation, and an exponential function of time with a complex frequency (Teukolsky 1972, 1973). The separation results in a spectrum of complex eigenfrequencies of the GW perturbations to the BH spacetime indexed by integer degree $\ell$ and order $m$ numbers, $\ell \geq 2$ and $|m| \leq \ell$, and integer overtone $n$ with $n \geq 1$ (Leaver 1985; Berti et al. 2006). The angular eigenfunctions, which depend on $\ell$ and $m$, also depend on the dimensionless complex frequency and the dimensionless spin parameter of the remnant BH. The complex eigenfrequencies describe the spectrum of exponentially-decaying sinusoidal GW *quasinormal modes* that make up what is called the *BH ringdown*. GR therefore provides a prediction for the relationship between the frequency and the decay constant for the spectrum of such quasinormal modes which depend solely on the mass and spin of the final BH, and thus the BH ringdown radiation can be used to test these predictions of GR.

Spanning the region between the portion of the waveform that can be computed by PN calculations at early time and by a superposition of quasinormal modes at late time is what is known as the *merger phase* of the compact binary. Due to the non-perturbative nature of this phase, numerical relativity (NR) solutions to Einstein's field equations are sought (Lehner & Pretorius 2014; Duez & Zlochower 2019). Such solutions both interpolate these early and late phases and also provide the information about the quasinormal mode amplitudes and phases excited as well as the mass and spin of the remnant BH (Hofmann et al. 2016; Healy & Lousto 2017; Jiménez-Forteza et al. 2017).

When at least one component of the binary in the compact binary coalescence is not a BH (i.e., a NS), the merger and ringdown phases might be considerably more complex due to the presence of matter in the system. NR is typically required to compute the entire post-inspiral phase of the GWs emitted from such systems (Faber & Rasio 2012; Kyutoku et al. 2021). One important piece of such simulations is to determine if disruption of a NS component occurs, particularly in the case of NSBH systems in which the NS might be swallowed whole by the BH (typical for large mass and low spin BHs) or might be tidally disrupted by the BH (typical for small mass or high spin BHs). Such NS disruption would be expected to produce electromagnetic emission that could be observed by electromagnetic astronomical observatories. Guided by numerical simulations, one can estimate whether a system having particular parameters inferred from the inspiral phase will be electromagnetically bright and so a candidate for electromagnetic follow-up observations (Foucart et al. 2018; Chatterjee et al. 2020; Berbel et al. 2024).

Depending on the masses of the initial components, the product of the merger of two NSs might be another NS, a supramassive NS (a uniformly-spinning NS that is more massive than the highest allowed mass for non-spinning NS, which remains a NS until its angular momentum is dissipated, resulting in its collapse to a BH), a hypermassive NS (a NS more massive that would be allowed for any stationary, spinning, configuration, but which is temporarily supported by differential rotation, and which will ultimately collapse to a BH), or there might be a direct collapse on a dynamical timescale to form a BH after the merger (Baumgarte et al. 2000; Piro et al. 2017). Both the electromagnetic and GW emission from these different scenarios are expected to vary considerably (Abbott et al. 2017c).

*Tests of GR from CBC Merger*—NR simulations of BBHs in GR provide predictions for the GW waveform spanning the inspiral, merger, and final ringdown phases of evolution. Tests for violations of GR can be performed by subtracting the best-fit GR waveform from the observed data and testing whether the remaining residual is consistent with detector noise or if there is remaining signal present. Alternatively, since NR predicts how a final BH mass and spin are related to the initial BH masses and spins for a BBH CBC (Hofmann et al. 2016; Healy & Lousto 2017; Jiménez-Forteza et al. 2017), a test of consistency between the initial orbital parameters and the final BH mass and spin can be performed. Here, the initial component BH masses and spins can be determined from the early inspiral phase of the GW signal, while the mass and the spin of the final BH are found from the late-time ringdown radiation. In practice, such a consistency test divides the GW signal into low- and high-frequency portions (below and above a given cutoff frequency) that are independently modeled with full inspiral–merger–ringdown waveforms (Ghosh et al. 2016, 2018). Companion paper Abac et al. (2025d) presents results from such residual and inspiral–merger–ringdown consistency tests.

In GR, a BH remnant produced by a CBC will rapidly settle to a stationary Kerr BH (Kerr 1963), uniquely characterized by its mass and spin (Israel 1967; Carter 1971), through emission of ringdown radiation in a spectrum of quasinormal modes, as described earlier. These quasinormal modes have a discrete spectrum of complex-valued eigenfrequencies (the imaginary part of which determines the decay timescale), so a possible non-BH remnant (e.g., Macedo et al. 2013), or modifications of the spectrum in alternative theories of GR (e.g., Cano et al. 2024), can be tested by looking for deviations in the observed ringdown radiation from the anticipated spectrum of quasinormal modes (Berti et al. 2025).

Furthermore, if the remnant object does not possess an event horizon, ingoing GW radiation can be reflected off of a surface or scattered off of an inner potential and reemerge as an *echo* signal observed within ~seconds after the merger (Cardoso et al. 2016; Cardoso & Pani 2019; Siemonsen 2024). The companion paper Abac et al. (2025f) presents tests of the nature of the remnant resulting from CBC through observed quasinormal mode spectra and searches for GW echoes.

The post-inspiral portion of a BBH signal can be phenomenologically modeled with various parameters that are fitted to NR simulations (Pratten et al. 2020). The companion paper Abac et al. (2025e) explores possible deviations of these parameters from their nominal values (Meidam et al. 2018; Roy et al. 2025).

#### 5.2.4. *Redshift and Cosmological Effects*

GWs can be redshifted, just as electromagnetic waves are. These changes are caused by the Doppler effect due to relative motion of the emitter and the observer (often described in terms of *peculiar velocities* relative to the rest frame of the cosmological microwave background radiation), the expansion of space between the emitter and the observer, or due to gravitational redshift if the emitter and observer have different gravitational potentials. For sources beyond the nearby universe (having redshifts $\gtrsim 0.1$), cosmological expansion is the dominant source of redshift (Peterson et al. 2022).

The redshift is the fractional difference between the frequency of a wave at emission at its source $f_{\rm src}$, and its observed frequency at a detector $f_{\rm det}$, $z = (f_{\rm src} - f_{\rm det})/f_{\rm det}$ (Hogg 1999). Thus, the observed frequency of a wave is related to its emitted frequency by $f_{\rm det} = f_{\rm src}/(1+z)$. Similarly, an interval in time in the source-frame $dt_{\rm src}$ is related to an observed interval in time by a detector $dt_{\rm det}$ by $dt_{\rm det} = (1 + z)dt_{\rm src}$. Equation (10) and (12) are both parameterized in terms of the orbital velocity $v$, which is related to the GW frequency $f$ in the dominant mode by Equation (14). At a fixed moment in a GW waveform, where the binary has some instantaneous value of $v$, we have

$$v^3 = \pi G M f_{\rm src} = \pi G M (1 + z) f_{\rm det}. \qquad (22)$$

That is: a redshifted signal, observed at frequency $f_{\rm det}$, produced by a system with intrinsic mass $M$ has identical morphology to an un-redshifted signal produced by a system with intrinsic mass $(1 + z)M$ (Krolak & Schutz 1987). If the redshift is unknown, then the observable mass parameters are the various combinations of $(1 + z)m_1$ and $(1 + z)m_2$, e.g., $(1 + z)\mathcal{M}$ and $(1 + z)M$. These mass parameters with the $1 + z$ scale factor are referred to as *detector-frame masses*, $m_1^{\rm det} = (1 + z)m_1$, $m_2^{\rm det} = (1 + z)m_2$, $M^{\rm det} = (1 + z)M$, and $\mathcal{M}^{\rm det} = (1 + z)\mathcal{M}$. PN corrections to the waveform preserve this degeneracy for point particles (and BHs). However, for NSs, a functional relationship between the mass of a NS and its tidal deformability means that a measurement of $\tilde{\Lambda}$ can break the degeneracy between mass and redshift, allowing the two to be independently measured (Messenger & Read 2012).

The amplitudes of $h_+$ and $h_\times$ given in Equation (10) also depend on the total mass through the factor $M/r$. If the factor $(1 + z)M$ is determinable from the rate of decay of the orbital period, Equation (12), then the amplitude factor can be written $[(1 + z)M]/[(1 + z)r]$ suggesting that the measurable amplitude distance parameter is $(1 + z)r$.

The parameter $r$ that appears in inverse proportion to the GW amplitude in Equation (10) represents the areal radius, i.e., spheres centered on the GW source have area $4\pi r^2$. Within a cosmological setting, this parameter is the *transverse comoving distance* $D_{\rm M}$ (Hogg 1999). Then, if the redshift is entirely due to the cosmological expansion of spacetime, the combination $(1 + z)D_{\rm M}$ is equal to the *luminosity distance* of the source, and this becomes the observable distance parameter from the GW amplitude. In this sense, then, given a known cosmology (i.e., the values of the Hubble constant, matter density, and the spatial curvature) the functional relationship between luminosity distance and redshift allows the determination of the latter from the former, and the intrinsic masses, e.g., $M$, can then be deduced from the observed mass–redshift combined parameters, e.g., $M^{\rm det} = (1 + z)M$. However, if other redshift effects are present, e.g., due to peculiar motion of the source or the observer relative to the Hubble flow, the combination $(1 + z)D_{\rm M}$ is no longer equal to the luminosity distance.

Nevertheless, when reporting the parameters of a CBC, we normally assume that cosmological expansion is the only significant source of redshift, and so the observed amplitude parameter $(1 + z)D_{\rm M}$ is referred to as luminosity distance $D_{\rm L}$, while a dimensionful intrinsic mass parameter such as the primary mass $m_1$ is derived from observed detector frame mass parameters as $m_1 = m_1^{\rm det}/[1 + z(D_{\rm L})]$, where the relationship between the redshift and the luminosity distance, $z(D_{\rm L})$, is obtained by some standard cosmological model. The only case where this was not done was for GW170817 where the measured geocentric redshift to its host galaxy NGC 4993 was used (Abbott et al. 2019b). The main uncertainty in the chirp mass of the system comes from the unknown peculiar velocity of the system relative to its host galaxy. Unless otherwise specified, the reference cosmology used to relate luminosity distance to redshift throughout the works is a ΛCDM model (Peebles & Ratra 2003) corresponding to a spatially-flat Friedman–Lemaître–Robertson–Walker spacetime (Friedmann 1999a,b; Lemaitre 1931; Robertson 1935a,b, 1936; Walker 1937; Weinberg 1972; Misner et al. 1973) with Hubble constant $H_0 = 67.9\,{\rm km\,s^{-1}\,Mpc^{-1}}$, matter density parameter $\Omega_{\rm m} = 0.3065$, and cosmological constant density parameter $\Omega_\Lambda = 1 - \Omega_{\rm m} = 0.6935$ (Ade et al. 2016, column TT+lowP+lensing+ext of Table 4).

*Constraints on Cosmic Expansion from GW Observations* —If the redshift of a GW source can be determined independently of its distance then a distance-redshift relationship can be obtained and used to infer cosmological parameters (Schutz 1986; Krolak & Schutz 1987). Here, the CBC is called a *standard siren* (akin to the *standard candles* such as Cepheid variables and Type Ia supernovae used to measure distances to their galaxy hosts), where the luminosity distance of the CBC is inferred from the amplitude of the GWs (Holz & Hughes 2005). The most direct method of determining the redshift of a GW source is if there is an electromagnetic counterpart in which spectroscopic measurements of the redshift of its host galaxy can be made (Krolak & Schutz 1987; Holz & Hughes 2005; Dalal et al. 2006; Chen et al. 2018). This method is known as the *bright sirens* method. For example, the BNS coalescence GW170817 (Abbott et al. 2017a) was associated with the optical kilonova AT 2017gfo in the galaxy NGC 4993 (Abbott et al. 2017b), which allowed for a measurement of the maximum a posteriori value of the Hubble constant with 68.3% CL highest density interval $69^{+17}_{-8}$ km s$^{-1}$ Mpc$^{-1}$ (Abbott et al. 2021a).

If no electromagnetic counterpart to a GW is observed, various methods are available to deduce the associated redshift (BBHs are not normally expected to produce any electromagnetic radiation, unless there is matter present in their environment). One such method, the *galaxy catalog* method, also called the *dark siren* method, is to obtain statistical association of GW sources with potential host galaxies observed in surveys (Schutz 1986; MacLeod & Hogan 2008). This is usually done simultaneously with information obtained from another method, the *spectral siren* method. In the spectral siren method, a known feature in the mass distribution of the population of CBCs is used to statistically infer the redshift of a number of sources at a given distance by how the observed detector-frame mass distribution is shifted with respect to the local (zero redshift) distribution (Chernoff & Finn 1993; Taylor et al. 2012; Farr et al. 2019; Mastrogiovanni et al. 2021). Finally, future observations of BNS mergers may be capable of directly inferring the redshift of the source from the GW signal alone through measurements of the NS tidal deformations (Messenger & Read 2012; Chatterjee et al. 2021).

The companion paper Abac et al. (2025g) reports on constraints on the cosmic expansion history based on combined CBC bright- and dark-sirens, including both the galaxy catalog method and the spectral siren approach. If GWs propagate through cosmological backgrounds differently from electromagnetic waves in a manner that produces a different distance–amplitude relation, the modified propagation effects can be observed using standard siren methods (Belgacem et al. 2018). The companion paper Abac et al. (2025g) also reports on constraints on such effects of modified GW propagation.

#### 5.2.5. *Populations of compact binaries*

With a multitude of observed CBCs, one can infer the underlying population of these sources. In doing so, one needs to account for the detector selection effects, e.g. the fact that farther away events are less probable to be detected as compared to events that are nearby. One key element of the population that can be measured is the local merger rate density, $\mathcal{R}$, representing the number of compact binary coalescences occurring per unit time per unit volume in the local universe (Phinney 1991; Kim et al. 2003; Brady et al. 2004; Biswas et al. 2009; Farr et al. 2015; Abbott et al. 2016e), or its evolution with cosmic redshift $\mathcal{R}(z)$, which is the number of coalescences per unit source-frame time per unit comoving volume at a cosmological redshift of $z$ (Fishbach et al. 2018). Another is the population distribution of masses and spins of merging compact binaries, $p(m_1, m_2, \boldsymbol{S}_1, \boldsymbol{S}_2)$, which might also evolve over cosmic history, $p(m_1, m_2, \boldsymbol{S}_1, \boldsymbol{S}_2|z)$. The measurement uncertainty in single-event parameters and the total number of detected events dictate the measurability of features in the population. These inferences are important in understanding the underlying astrophysical formation channels of compact binaries (e.g., Stevenson et al. 2017; Farr et al. 2017; Zevin et al. 2021; Mandel & Broekgaarden 2022).

*Inference of the Population of CBCs*—In the companion paper Abac et al. (2025c), we present measurements of the local rate of BNS, NSBH, and BBH mergers, inference of the evolution of the CBC rate over cosmological time, and inference of the distribution of masses and spins of CBCs.

**Table 3**. Parameters describing a CBC system with quasi-circular orbits.

| Parameter name | Symbol | Notes [Dimensions] |
|---|---|---|
| Primary and secondary masses | $m_1, m_2$ | Mass of the more massive ($m_1$) and less massive ($m_2$) body in system, $m_1 \geq m_2$ [mass] |
| Chirp mass | $\mathcal{M}$ | See Equation (13) [mass] |
| Total mass | $M$ | $M = m_1 + m_2$ [mass] |
| Final mass | $M_\mathrm{f}$ | Mass of the remnant [mass] |
| Mass ratio | $q$ | $q = m_2/m_1 \leq 1$ [dimensionless] |
| Symmetric mass ratio | $\eta$ | $\eta = m_1 m_2/(m_1 + m_2)^2 \leq 1/4$ [dimensionless] |
| Energy radiated | $E_\mathrm{rad}$ | $E_\mathrm{rad} = (M - M_\mathrm{f})c^2$ [energy] |
| Peak luminosity | $\ell_\mathrm{peak}$ | Peak GW luminosity, typically 0.1% of the Planck luminosity ($\ell_\mathrm{Planck} = c^5/G$) for BBH coalescences [power] |

**Table 3** *continued*

**Table 3** *(continued)*

| Parameter name | Symbol | Notes [Dimensions] |
|---|---|---|
| Primary and secondary spin vectors | $\boldsymbol{S}_1, \boldsymbol{S}_2$ | Spin angular momentum of the primary ($\boldsymbol{S}_1$) and secondary ($\boldsymbol{S}_2$) [angular momentum] |
| Primary and secondary dimensionless spin magnitudes | $\chi_1, \chi_2$ | $\chi_{1,2} = c\lvert\boldsymbol{S}_{1,2}\rvert/(Gm_{1,2}^2)$; $\chi_{1,2} \leq 1$ for Kerr BHs primary/secondary [dimensionless] |
| Remnant dimensionless spin magnitude | $\chi_\mathrm{f}$ | $\chi_\mathrm{f} = cS_\mathrm{f}/(GM_\mathrm{f}^2)$ where $S_\mathrm{f}$ is the magnitude of the remnant's spin angular momentum; $\chi_\mathrm{f} \leq 1$ for a Kerr BH remnant [dimensionless] |
| Newtonian orbital angular momentum | $\boldsymbol{L}$ | Instantaneous orbital angular momentum about the center of mass; defines $z$-direction for spin coordinates [angular momentum] |
| Total angular momentum | $\boldsymbol{J}$ | $\boldsymbol{J} = \boldsymbol{L} + \boldsymbol{S}_1 + \boldsymbol{S}_2$ [angular momentum] |
| Primary and secondary tilt angle | $\theta_1, \theta_2$ | Angle between $\boldsymbol{S}_{1,2}$ and $\boldsymbol{L}$ [angle] |
| Spin azimuthal angle difference | $\phi_{12}$ | Angle between $\boldsymbol{L} \times (\boldsymbol{S}_1 \times \boldsymbol{L})$ and $\boldsymbol{L} \times (\boldsymbol{S}_2 \times \boldsymbol{L})$ [angle] |
| Effective inspiral spin parameter | $\chi_\mathrm{eff}$ | See Equation (15) [dimensionless] |
| Effective precession spin parameter | $\chi_\mathrm{p}$ | See Equation (16) [dimensionless] |
| Orbital inclination angle | $\iota$ | Angle between $\boldsymbol{L}$ and the direction to the Earth $\boldsymbol{N}$ (see Figure 5) [angle] |
| Source inclination angle | $\theta_{JN}$ | Angle between $\boldsymbol{J}$ and the direction to the Earth $\boldsymbol{N}$ [angle] |
| Viewing angle | $\Theta$ | $\Theta = \min\{\theta_{JN}, \pi - \theta_{JN}\}$ [angle] |
| Orbital phase | $\phi$ | Phase of a binary's orbit, the angle on the orbital plane between the separation vector (the position vector of the primary minus the position vector of the secondary) and the line of nodes, $\boldsymbol{L} \times \boldsymbol{N}$ (see Figure 5) [angle] |
| Coalescence phase | $\phi_\mathrm{c}$ | Orbital phase, the angle on the orbital plane between the separation vector (the position vector of the primary minus the position vector of the secondary) and the line of nodes, $\boldsymbol{L} \times \boldsymbol{N}$, at a point in the evolution corresponding to the point in the waveform used to define $t_\mathrm{geo}$ (see Table 2) [angle] |
| Angular diameter distance | $D_\mathrm{A}$ | An object of transverse length $x$ is observed to subtend an angle in radians of $x/D_\mathrm{A}$ when both the object and observer are at rest relative to a homogeneous cosmology (Hogg 1999) [length] |
| Transverse comoving distance | $D_\mathrm{M}$ | Areal radius of a sphere centered on a point in an isotropic cosmology, defined so the sphere has area $4\pi D_\mathrm{M}^2$ (Hogg 1999) [length] |
| Luminosity distance | $D_\mathrm{L}$ | A source of isotropic radiation having luminosity $\ell_\mathrm{iso}$ is observed to have flux $\ell_\mathrm{iso}/(4\pi D_\mathrm{L}^2)$ when both the source and observer are at rest relative to a homogeneous cosmology (Hogg 1999) [length] |
| Redshift | $z$ | The fractional difference between the frequency of a wave at emission at its source $f_\mathrm{src}$, and its observed frequency at a detector $f_\mathrm{det}$, $z = (f_\mathrm{src} - f_\mathrm{det})/f_\mathrm{det}$; the reference cosmology for the relationship between distance and the *cosmological* redshift is given in the text [dimensionless] |
| Primary and secondary dimensionless tidal deformabilities | $\Lambda_1, \Lambda_2$ | See Equation (18); $\Lambda_{1,2} = 0$ for a BH primary/secondary [dimensionless] |
| Effective tidal deformability | $\tilde{\Lambda}$ | See Equation (19); $\tilde{\Lambda} = 0$ for a BBH [dimensionless] |
| Primary and secondary dimensionless spin-induced quadrupole moments | $\kappa_1, \kappa_2$ | See Equation (21); $\kappa_{1,2} = 1$ for a BH primary/secondary [dimensionless] |
| Primary and secondary radii | $R_1, R_2$ | Areal radii of primary and secondary and objects, defined so their surface areas are $4\pi R_{1,2}^2$; used in defining NS compactness [length] |

**Table 3** *continued*

**Table 3** *(continued)*

| Parameter name | Symbol | Notes [Dimensions] |
|---|---|---|
| Primary and secondary compactness | $C_1, C_2$ | Dimensionless mass-to-radius ratios $C_{1,2} = Gm_1/(c^2 R_{1,2})$ of primary/secondary; $1/C_{1,2} = 1 + \sqrt{1+\chi_{1,2}^2}$ for Kerr BH primary/secondary [dimensionless] |
| Merger rate density | $\mathcal{R}$ | Rate of binary mergers per unit volume in the local universe; may be expressed as a function of cosmological redshift, $\mathcal{R}(z)$; the rate in the local Universe $\mathcal{R}(z = 0)$ can be notated $\mathcal{R}_0$; subscripts can be used if considering different populations, e.g., $\mathcal{R}_{\rm BNS}$, $\mathcal{R}_{\rm NSBH}$, and $\mathcal{R}_{\rm BBH}$ [time$^{-1}$ volume$^{-1}$] |

## 6. SYNOPSIS

This paper serves as an introduction to the collection of papers accompanying the LVK's GWTC-4.0. We have provided an overview of the GW detectors and observing runs of the LVK network and of the observed GWs from CBCs. The primary sequels to this introduction are a description of the methods used to perform searches for GWs in LVK data and to characterize source properties of identified signals (Abac et al. 2025a), and a summary of the main observations of GWTC-4.0, highlighting new CBC candidates and their estimated estimated masses and spins (Abac et al. 2025b). Other companion papers presenting science results from the analysis of the GWTC-4.0 candidates were described in Section 1. GWTC provides a prodigious census of over 200 merging BHs and NSs spanning two orders of magnitude in mass from $\sim 1\,M_\odot$ NSs to remnant BHs exceeding $100\,M_\odot$. Study of these observations will provide new insight into the nature of these objects, their population distribution, and their formation channels. These GW observations allow for sensitive tests of GR and provide information about the cosmological expansion history.

*Data availability:* Event data used within this work is openly available in the GWTC-4.0 online catalog, which is hosted at https://gwosc.org/GWTC-4.0 and documented further in Abac et al. (2025i). Data behind Figures 1, 2, and 3 can be found in LIGO–Virgo–KAGRA Collaboration (2025).

## ACKNOWLEDGEMENTS

This material is based upon work supported by NSF's LIGO Laboratory, which is a major facility fully funded by the National Science Foundation. The authors also gratefully acknowledge the support of the Science and Technology Facilities Council (STFC) of the United Kingdom, the Max-Planck-Society (MPS), and the State of Niedersachsen/Germany for support of the construction of Advanced LIGO and construction and operation of the GEO 600 detector. Additional support for Advanced LIGO was provided by the Australian Research Council. The authors gratefully acknowledge the Italian Istituto Nazionale di Fisica Nucleare (INFN), the French Centre National de la Recherche Scientifique (CNRS) and the Netherlands Organization for Scientific Research (NWO) for the construction and operation of the Virgo detector and the creation and support of the EGO consortium. The authors also gratefully acknowledge research support from these agencies as well as by the Council of Scientific and Industrial Research of India, the Department of Science and Technology, India, the Science & Engineering Research Board (SERB), India, the Ministry of Human Resource Development, India, the Spanish Agencia Estatal de Investigación (AEI), the Spanish Ministerio de Ciencia, Innovación y Universidades, the European Union NextGenerationEU/PRTR (PRTR-C17.I1), the ICSC - CentroNazionale di Ricerca in High Performance Computing, Big Data and Quantum Computing, funded by the European Union NextGenerationEU, the Comunitat Autonòma de les Illes Balears through the Conselleria d'Educació i Universitats, the Conselleria d'Innovació, Universitats, Ciència i Societat Digital de la Generalitat Valenciana and the CERCA Programme Generalitat de Catalunya, Spain, the Polish National Agency for Academic Exchange, the National Science Centre of Poland and the European Union - European Regional Development Fund; the Foundation for Polish Science (FNP), the Polish Ministry of Science and Higher Education, the Swiss National Science Foundation (SNSF), the Russian Science Foundation, the European Commission, the European Social Funds (ESF), the European Regional Development Funds (ERDF), the Royal Society, the Scottish Funding Council, the Scottish Universities Physics Alliance, the Hungarian Scientific Research Fund (OTKA), the French Lyon Institute of Origins (LIO), the Belgian Fonds de la Recherche Scientifique (FRS-FNRS), Actions de Recherche Concertées (ARC) and Fonds Wetenschappelijk Onderzoek - Vlaanderen (FWO), Belgium, the Paris Île-de-France Region, the National Research, Development and Innovation Office of Hungary (NKFIH), the National Research Foundation of Korea, the Natural Sciences and Engineering Research Council of Canada (NSERC), the Canadian Foundation for Innovation (CFI), the Brazilian Ministry of Science, Technology, and Innovations, the International Center for Theoretical Physics South American Institute for Fundamental Research (ICTP-SAIFR), the Research Grants Council of Hong Kong, the National Natural Science Foundation of China (NSFC), the Israel Science Foundation (ISF), the US-Israel Binational Science Fund (BSF), the Leverhulme

Trust, the Research Corporation, the National Science and Technology Council (NSTC), Taiwan, the United States Department of Energy, and the Kavli Foundation. The authors gratefully acknowledge the support of the NSF, STFC, INFN and CNRS for provision of computational resources.

This work was supported by MEXT, the JSPS Leading-edge Research Infrastructure Program, JSPS Grant-in-Aid for Specially Promoted Research 26000005, JSPS Grant-in-Aid for Scientific Research on Innovative Areas 2402: 24103006, 24103005, and 2905: JP17H06358, JP17H06361 and JP17H06364, JSPS Core-to-Core Program A. Advanced Research Networks, JSPS Grants-in-Aid for Scientific Research (S) 17H06133 and 20H05639, JSPS Grant-in-Aid for Transformative Research Areas (A) 20A203: JP20H05854, the joint research program of the Institute for Cosmic Ray Research, University of Tokyo, the National Research Foundation (NRF), the Computing Infrastructure Project of the Global Science experimental Data hub Center (GSDC) at KISTI, the Korea Astronomy and Space Science Institute (KASI), the Ministry of Science and ICT (MSIT) in Korea, Academia Sinica (AS), the AS Grid Center (ASGC) and the National Science and Technology Council (NSTC) in Taiwan under grants including the Science Vanguard Research Program, the Advanced Technology Center (ATC) of NAOJ, and the Mechanical Engineering Center of KEK.

Additional acknowledgements for support of individual authors may be found in the following document: https://dcc.ligo.org/LIGO-M2300033/public.  We request that citations to this article use 'A. G. Abac *et al.* (LIGO-Virgo-KAGRA Collaboration), ...' or similar phrasing, depending on journal convention.

*Facilities:* EGO:Virgo, GEO600, Kamioka:KAGRA, LIGO

*Software:* Plots were prepared with Matplotlib (Hunter 2007) and seaborn (Waskom 2021). Astropy (Price-Whelan et al. 2022), GWpy (Macleod et al. 2021), LALSuite (LIGO–Virgo–KAGRA Collaboration 2018; Wette 2020), NumPy (Harris et al. 2020), SciPy (Virtanen et al. 2020) were used for data processing in generating the figures and quantities in the manuscript.

## A. ACRONYMS AND GLOSSARY

This is a reference of frequently-used terms and acronyms.

**A+:** Advanced+ LIGO refers to a configuration of LIGO following a series of upgrades, some in advance of O4 (such as the addition of a new 300 m filter cavity for frequency-dependent vacuum squeezing), and some planned in advance of O5, such as installation of new optics with lower noise and loss (Abbott et al. 2020a; Cooper et al. 2023).

**A$^\sharp$:** LIGO A$^\sharp$ (A-sharp) is a proposed upgrade of the Advanced+ LIGO interferometers anticipated on a post-O5 timeline. The baseline A$^\sharp$ design is a room-temperature 1 μm laser wavelength interferometer upgrade with larger test masses having coatings with lower thermal noise, higher laser power, and increased levels of vacuum squeezing (Fritschel et al. 2024).

**AdV:** Advanced Virgo refers to an upgraded Virgo detector (Acernese et al. 2015) with an advanced interferometer. Virgo operated with the AdV configuration during O2 and O3.

**AdV+:** Advanced Virgo+ is an upgrade to the AdV detector to take place in two phases: the first phase for operation during O4 and the second phase for operation during O5.

**aLIGO:** Advanced LIGO refers to an upgraded LIGO configuration with advanced interferometers installed at both LHO and LLO. LIGO operated with the aLIGO configuration during O1, O2, O3, and O4 (Aasi et al. 2015a).

**BBH:** Binary black hole. A binary system where both components are BHs.

**BH:** Black hole.

**BHNS:** Black hole–neutron star binary specifically refers to systems in which the BH formed before the NS. See also NSBH.

**bKAGRA:** Baseline-design KAGRA is a configuration of the KAGRA detector as a cryogenic dual-recycled Fabry–Perot Michelson interferometer. bKAGRA phase-1 operation without power- or signal-recycling took place from 2018 April 28 to 2018 May 06 2018 (Akutsu et al. 2019).

**BNS:** Binary neutron star. A binary system where both components are NSs.

**CBC:** Compact binary coalescence. The gravitational-radiation-driven orbital decay resulting in merger of a binary system made of two compact objects (NSs or BHs).

**CI:** Credible interval. See CL.

**CL:** Credible level. Given a $n = 1$ univariate or $n$-dimensional multivariate random variable $\boldsymbol{x}$ having probability density function (PDF) $p(\boldsymbol{x})$ and a $n$-dimensional region $R^n_\alpha$, then the CL $\alpha$ of the region $R^n_\alpha$ is the probability of $\boldsymbol{x}$ lying in $R^n_\alpha$, $\alpha = P(\boldsymbol{x} \in R^n_\alpha) = \int_{R^n_\alpha} p(\boldsymbol{x})\, d^n x$. The region $R^n_\alpha$ is then known as a $100\alpha\%$ CL *credible region*, with special cases: *credible interval* (CI) if $n = 1$, *credible area* if $n = 2$, or *credible volume* if $n = 3$. When $n > 1$ we normally take $R^n_\alpha$ to be the region having the smallest volume that has CL $\alpha$ (the *highest density region*). When $n = 1$ (CI), $R^1_\alpha$ is normally chosen to be an *equal-tailed interval* (also known as a symmetric interval), from the $\alpha/2$ quantile to the $1 - \alpha/2$ quantile, but sometimes the smallest *highest density interval* is used instead.

**EoS:** Equation of state of a neutron star. For cold neutron stars (having temperature below the Fermi temperature), the equation of state (EoS) is of a barotropic fluid, a relationship between the energy density of the fluid and its pressure.

**FAR:** False alarm rate. Often used as a detection threshold, the probability of any one or more of a sequence of statistical tests performed over a duration $T$ erroneously rejecting a null hypothesis is $1 - \exp(-T \times \mathrm{FAR})$. FAR therefore has dimensions of time$^{-1}$. When interpreted as a measure of a detection significance of a candidate detection, this is the rate at which noise alone would produce more significant candidates.

**GEO:** The GEO 600 GW detector is a British–German L-shaped interferometric GW detector with 600 m arms located near Hannover, Germany (Willke et al. 2002).

**GR:** General relativity. Einstein's theory of gravitation.

**GW:** Gravitational wave. See Sections 1 and 5.1.

**GWOSC:** The Gravitational Wave Open Science Center (formerly known as the LIGO open science center) was created to provide public access to GW data products (Abbott et al. 2021d). The GWOSC online data and resources can be found at https://gwosc.org.

**GWTC:** The Gravitational-Wave Transient Catalog is the electronic catalog of GW transients observed by LIGO, Virgo, and KAGRA detectors produced by the LVK.

**IFAR:** Inverse false alarm rate. The reciprocal of FAR, $\mathrm{IFAR} = (\mathrm{FAR})^{-1}$, having dimensions of time. A larger IFAR implies a more significant candidate, while a larger FAR implies a less significant candidate.

**IFO:** Interferometer, a type of detector that uses laser interferometry to measure changes in the lengths of optical paths induced by GWs.

**IGWN:** The International GW Observatory Network is a self-governing consortium using ground-based GW interferometers to explore the fundamental physics of gravity and to observe the Universe. The observatory network includes the KAGRA, LHO, LLO, and Virgo detectors. In addition, the GEO detector serves as a technology testbed and operates in an *astrowatch* mode outside of other detectors' observing periods.

**iKAGRA:** Initial-phase KAGRA is a configuration of the KAGRA detector as a simple Michelson interferometer that consists of two end test masses and a beam splitter. iKAGRA was operated from 2016 March 25 to the 2016 March 31 and from 2016 April 11 to 2016 April 25 (Akutsu et al. 2018).

**IMBH:** Intermediate-mass black hole. A BH in the mass range $\sim 10^2\, M_\odot$ to $\sim 10^5\, M_\odot$.

**KAGRA:** KAGRA is a Japanese L-shaped interferometric GW detector with 3 km arms located underground at the Kamioka Observatory in Japan (Akutsu et al. 2021).

**KAGRA Collaboration:** The KAGRA Collaboration manages the building, operation, and development of the KAGRA detector.

**LHO:** The LIGO Hanford Observatory, one of the two LIGO observatories, located in Hanford, Washington, is an L-shaped interferometric GW detector with 4 km arms.

**LIGO:** The Laser Interferometer Gravitational-Wave Observatory consists of two widely spaced installations within the United States: one in Hanford, Washington (LHO) and the other in Livingston, Louisiana (LLO). LIGO is operated by the LIGO Laboratory, a consortium of the California Institute of Technology and the Massachusetts Institute of Technology funded by the U.S. National Science Foundation.

**LLO:** The LIGO Livingston Observatory, one of the two LIGO observatories, located in Livingston, Louisiana, is an L-shaped interferometric GW detector with 4 km arms.

**LSC:** The LIGO Scientific Collaboration, founded in 1997, is a group of more that 1000 scientists that carries out science related to the LIGO detectors and their observations.

**LV:** The LIGO–Virgo Collaboration. Prior to O3b, all observational results were published by the LV.

**LVC:** The LIGO–Virgo Collaboration. The acronym LV is now preferred.

**LVK:** The LIGO–Virgo–KAGRA Collaboration.

**NS:** Neutron star.

**NSBH:** The general term for a neutron star–black hole binary: a binary system in which one component is a NS and the other is a BH. If used in distinction with BHNS, it refers to such systems in which the NS formed before the BH.

**NR:** Numerical relativity, the use of numerical methods to solve relativistic field equations.

**O1:** The first observing run began on 2015 September 12 and ended on 2016 January 19. The LHO and LLO detectors participated in this observing run.

**O2:** The second observing run began on 2016 November 30 and ended on 2016 August 25 during which the LHO and the LLO detectors were operating. On 2017 August 1, the AdV detector joined the observing run, forming a three-detector network.

**O3:** The third observing run began on 2019 April 1 and ended on 2020 March 27 during which the LHO, LLO, and Virgo detectors were operating. A commissioning break from 2019 October 1 to 2019 November 1 divided O3 into two parts, O3a and O3b. A subsequent short run, O3GK, from 2020 April 7 to 2020 April 21 with GEO and KAGRA observing followed O3b.

**O3a:** The first, pre-commissioning-break, part of O3, from 2019 April 1 until 2019 October 1, during which the LHO, LLO, and Virgo detectors were operating.

**O3b:** The second, post-commissioning-break, part of O3, from 2019 November 1 until 2020 March 27, during which the LHO, LLO, and Virgo detectors were operating. O3b was planned to continue until 2020 April 30, but ended early due to the COVID-19 pandemic.

**O3GK:** A short observing run after O3b from 2020 April 7 to 2020 April 21, during which the KAGRA and GEO detectors were observing. KAGRA had intended to join LIGO and Virgo at the end of O3 but the early end of O3b made this impossible.

**O4:** The fourth observing run began on 2023 May 24 and is planned to continue into late 2025. It is divided into parts, the first of which, O4a, covered the period from 2023 May 24 until a commissioning break from 2024 January 16 until 2024 April 10. During O4a, LHO and LLO were observing. Following the break, observing continued in O4b from 2024 April 10 until an original intended end date of 2025 January 23, with LHO, LLO, and Virgo observing. It was decided to continue O4 observing in a third period O4c, beginning 2025 January 23, lasting until late 2025.

**O4a:** The first part of the fourth observing run including data from 2023 May 24 until a commissioning break that began on 2024 January 16. During O4a, LHO and LLO were observing.

**O4b:** The second part of the fourth observing run starting at the end of a commissioning break on 2024 April 10 and ending on the originally-planned O4 end date of 2025 January 23. During O4b, LHO, LLO, and Virgo were observing. It was decided to continue O4 observations with a third part, O4c, immediately following the end of O4b on 2025 January 23.

**O4c:** The third part of the fourth observing run, extending the run beyond its intended end date of 2025 January 23 through late 2025. A commissioning break in O4c took place between 2025 April 01 and 2025 June 11.

**O5:** The fifth observing run is the planned future observing run to follow O4.

**PDF:** Probability density function. Given a $n = 1$ univariate or $n$-dimensional multivariate random variable $\boldsymbol{x}$, the probability of $\boldsymbol{x}$ lying in a $n$-dimensional region $R^n$ is $P(\boldsymbol{x} \in R^n) = \int_{R^n} p(\boldsymbol{x})\, d^n x$, where $p(\boldsymbol{x})$ is the PDF.

**PE:** Parameter estimation, the process of measuring the parameters that describe the source of a signal, e.g., the masses and spins of the binary components of a CBC, from the observational data.

**PN:** Post-Newtonian, a perturbative method of obtaining solutions to relativistic field equations based on slow-motion and weak-field expansion of the spacetime metric and the stress–energy source.

**PSD:** Power spectral density. See Appendix B.

**SNR:** Signal-to-noise ratio. See Appendix B.

**Virgo:** The Virgo detector is a European L-shaped interferometric GW detector with 3 km arms located near Cascina, Italy (near Pisa).

**Virgo_nEXT:** Virgo_nEXT is a planned, post-O5, major upgrade of Virgo to fill the gap between the current phase, AdV+, and next-generation detectors.

**Virgo Collaboration:** The Virgo Collaboration manages the building, operation, and development of the Virgo detector.

## B. CONVENTIONS FOR DATA ANALYSIS

This appendix serves to define the data analysis conventions that will be used throughout the GWTC-4.0 companion papers. For a general introduction to data analysis we refer the reader to Abbott et al. (2020c) and references therein.

Time series $a(t)$ and frequency series $\tilde{a}(t)$ are related to each other by our conventions for the *Fourier transform*

$$\tilde{a}(f) = \int_{-\infty}^{+\infty} a(t) \exp(-2\pi i f t)\, dt \tag{23}$$

and its inverse transform

$$a(t) = \int_{-\infty}^{+\infty} \tilde{a}(f) \exp(+2\pi i f t)\, df\,. \tag{24}$$

With these conventions, the dimension of $\tilde{a}$ are $[\tilde{a}] = [a] \times$ time.

Detector noise is often taken to be a stochastic Gaussian process. If $n(t)$ is a real-valued stochastic Gaussian process then the *one-sided power spectral density* (PSD) $S_n(f)$ is formally defined by

$$\langle \tilde{n}^*(f')\tilde{n}(f)\rangle = \frac{1}{2} S_n(f)\delta(f - f')\,, \tag{25}$$

where $\langle\cdot\rangle$ is a statistical ensemble average of realizations of $n(t)$ and $\tilde{n}^*$ is the complex conjugate of $\tilde{n}$. The one-sided PSD is defined only for $f \geq 0$. With these conventions, the dimensions of $S_n$ are $[S_n] = [n]^2 \times$ time. Real detector noise is neither entirely stationary nor Gaussian (Abbott et al. 2020c). However, it is often sufficient to assume $n(t)$ is ergodic such that

$$S_n(f) = \lim_{T\to\infty} \frac{2}{T} \left| \int_{-T/2}^{T/2} n(t) \exp(-2\pi i f t)\, dt \right|^2\,. \tag{26}$$

The factor of two in the one-sided PSD ensures that the integrated power is

$$\int_0^{\infty} S_n(f)\, df = \lim_{T\to\infty} \frac{1}{T} \int_{-T/2}^{T/2} n^2(t)\, dt\,. \tag{27}$$

The *amplitude spectral density* is defined to be the square-root of the power spectral density, $S_n^{1/2}(f)$.

We often use a detector-noise-weighted inner product between two real-valued time series, $a(t)$ and $b(t)$, which is defined as

$$\langle a|b\rangle = 4\,\mathrm{Re} \int_0^{+\infty} \frac{\tilde{a}^*(f)\tilde{b}(f)}{S_n(f)}\, df \tag{28a}$$

$$= \int_{-\infty}^{+\infty} \frac{\tilde{a}^*(f)\tilde{b}(f)}{(1/2)S_n(|f|)}\, df\,, \tag{28b}$$

where $S_n(f)$ is the detector's one-sided PSD for the readout noise from that detector. The second form, Equation (28b), is an appropriate generalization of the inner product for complex-valued time series.

Since GW detectors are insensitive at very low frequencies, the mean of the detector readout is arbitrary and so we take the detector noise to have zero-mean, $\langle n(t)\rangle = 0$. Gaussian noise is then entirely characterized by its PSD and its distribution is given by the probability density

$$p(n) = \frac{1}{W} \exp\left(-\frac{1}{2}\langle n|n\rangle\right)\,, \tag{29}$$

where $W$ is a usually-neglected normalizing constant, the path integral $W = \int \exp(-\langle n|n\rangle/2)\, \mathcal{D}n$.

Consider a template waveform $u(t)$ that is unit-normalized, $\langle u|u\rangle = 1$, which is expected to match a hypothetical signal in detector data $d(t)$. The *matched filter signal-to-noise ratio* (SNR) is

$$\rho_{\mathrm{mf}} = \langle u|d\rangle \,. \tag{30}$$

If data $d(t) = n(t) + h(t)$ contains Gaussian noise $n(t)$ plus a signal $h(t)$ that is perfectly matched by the template waveform, $h(t) \propto u(t)$, then $\rho_{\mathrm{mf}}$ is a random variable having a normal distribution with unit variance and mean equal to the *optimal* SNR

$$\rho_{\mathrm{opt}} = \sqrt{\langle h|h\rangle} \,. \tag{31}$$

The *likelihood* that detector data $d(t)$ contains a signal $h(t)$ is given by Equation (29) with $n(t) = d(t) - h(t)$,

$$p(d|h) = \frac{1}{W}\exp\left(-\frac{1}{2}\langle d-h|d-h\rangle\right) \tag{32a}$$

$$= \frac{\exp\left(-\langle d|d\rangle\rangle/2\right)}{W}\exp\left(\langle h|d\rangle - \frac{1}{2}\langle h|h\rangle\right) \tag{32b}$$

$$= p(d|\varnothing)\exp\left(\rho_{\mathrm{mf}}\rho_{\mathrm{opt}} - \frac{1}{2}\rho_{\mathrm{opt}}^2\right) , \tag{32c}$$

where $\rho_{\mathrm{mf}}$ is the matched-filter SNR with unit-normalized template $u(t) \propto h(t)$ and $p(d|\varnothing) = W^{-1}\exp\left(-\langle d|d\rangle/2\right)$ is the likelihood under the no-signal hypothesis, $d(t) = n(t)$. The likelihood is viewed as a functional of $h(t)$ for a given realization of detector data $d(t)$. The second factor in Equation (32c) is the signal-to-noise *likelihood ratio* $p(d|h)/p(d|\varnothing)$. Note that the likelihood ratio is a monotonically-increasing function of the matched-filter SNR and so $\rho_{\mathrm{mf}}$ is the uniformly most powerful test for a known signal in Gaussian detector noise (Neyman & Pearson 1933). If the amplitude of the signal is unknown, $h(t) = \rho_{\mathrm{opt}}u(t)$ with unknown $\rho_{\mathrm{opt}}$, then the likelihood is maximized for $\rho_{\mathrm{opt}} = \rho_{\mathrm{mf}}$ and

$$\max_{\rho_{\mathrm{opt}}} p(d|\rho_{\mathrm{opt}}u) = p(d|\varnothing)\exp\left(\frac{1}{2}\rho_{\mathrm{mf}}^2\right) . \tag{33}$$

For the Newtonian inspiral of Section 5.2.1, the signal observed in a detector can be obtained in the frequency domain under the stationary phase approximation as (Sathyaprakash & Dhurandhar 1991; Cutler et al. 1993)

$$\tilde{h}(f) = -\sqrt{\frac{5\pi}{24}}\frac{G\mathcal{M}}{c^3}\frac{G\mathcal{M}}{c^2 D_{\mathrm{eff}}}\left(\frac{\pi G\mathcal{M}f}{c^3}\right)^{-7/6}\exp\left(-i\Psi(f)\right) , \tag{34}$$

where $\Psi(f)$ is the stationary phase function and

$$D_{\mathrm{eff}} = r\left[F_+^2(\vartheta,\varphi,\psi)\left(\frac{1}{2}+\frac{1}{2}\cos^2\iota\right)^2 + F_\times^2(\vartheta,\varphi,\psi)\cos^2\iota\right]^{-1/2} . \tag{35}$$

is the *effective distance* (Allen et al. 2012), which is related to the distance to the binary $r$ by a factor that accounts for the orientation angles that describe the position of the source on the sky $(\vartheta,\varphi)$, its inclination $\iota$, and polarization angle $\psi$. Since $F_+^2 + F_\times^2 \le 1$ (with equality for a source on the zenith or nadir of an L-shaped interferometric detector), $D_{\mathrm{eff}} \ge r$ (with equality only if $\iota = 0$ or $\iota = \pi$). The optimal SNR for such a signal is

$$\rho_{\mathrm{opt}} = \sqrt{\frac{5}{6\pi}}\frac{G\mathcal{M}}{c^2 D_{\mathrm{eff}}}\left(\frac{\pi G\mathcal{M}}{c^3}\right)^{-1/6}\sqrt{\int_0^{+\infty}\frac{f^{-7/3}}{S_n(f)}\,df} \,. \tag{36}$$

The *horizon distance* $D_{\mathrm{hor}}$ (Allen et al. 2012) of a source is the effective distance of a signal from such a source that has SNR $\rho_{\mathrm{opt}}$ equal to some detection threshold $\rho_{\mathrm{th}}$. Such sources would not be expected to be detected beyond the horizon distance, but not all nearer sources will be detected either. The *sensitive volume* (Finn & Chernoff 1993; Chen et al. 2021) is a measure of the effective volume of space in which randomly isotropically-oriented and homogeneously-distributed identical sources will produce signals in the detector with SNR $\rho_{\mathrm{opt}}$ greater than the threshold $\rho_{\mathrm{th}}$,

$$V = \frac{\int_{\rho_{\mathrm{opt}}>\rho_{\mathrm{th}}} r^2\,\sin\vartheta\,\sin\iota\,dr\,d\vartheta\,d\varphi\,d\iota\,d\psi}{\int\sin\iota\,d\iota\,d\psi} \tag{37a}$$

$$= 0.086\,084 \times \frac{4}{3}\pi D_{\mathrm{hor}}^3 \,. \tag{37b}$$

If the merger rate density is $\mathcal{R}$ then the expected number of detections in time $T$ is $\mathcal{R}VT$. For a standard measure of detector sensitivity, a binary source of two $1.4\,M_\odot$ objects ($\mathcal{M} = 2^{-1/5} \times 1.4\,M_\odot \approx 1.22\,M_\odot$) is considered and a threshold SNR of $\rho_{\rm th} = 8$ is adopted (Chen et al. 2021). The sensitive volume is converted into an equivalent spherical radius as $V = (4\pi/3)R^3$ to obtain the BNS range $R = D_{\rm hor}/2.264\,78$, Equation (1).

## All Authors and Affiliations

A. G. Abac[1], I. Abouelfettouh[2], F. Acernese[3,4], K. Ackley[5], S. Adhicary[6], D. Adhikari[7,8], N. Adhikari[9], R. X. Adhikari[10], V. K. Adkins[11], S. Afroz[12], D. Agarwal[13,14], M. Agathos[15], M. Aghaei Abchouyeh[16], O. D. Aguiar[17], S. Ahmadzadeh[18], L. Aiello[19,20], A. Ain[21], P. Ajith[22], S. Akcay[23], T. Akutsu[24,25], S. Albanesi[26,27], R. A. Alfaidi[28], A. Al-Jodah[29], C. Allené[30], A. Allocca[31,4], S. Al-Shammari[32], P. A. Altin[33], S. Alvarez-Lopez[34], O. Amarasinghe[32], A. Amato[35,36], C. Amra[37], A. Ananyeva[10], S. B. Anderson[10], W. G. Anderson[10], M. Andia[38], M. Ando[39,40], T. Andrade[41], M. Andrés-Carcasona[42], T. Andrić[7,8,43], J. Anglin[44], S. Ansoldi[45,46], J. M. Antelis[47], S. Antier[48], M. Aoumi[49], E. Z. Appavuravther[50,51], S. Appert[10], S. K. Apple[52], K. Arai[10], A. Araya[53], M. C. Araya[10], M. Arca Sedda[43], J. S. Areeda[54], L. Argianas[55], N. Aritomi[2], F. Armato[56,57], S. Armstrong[58], N. Arnaud[38,59], M. Arogeti[60], S. M. Aronson[11], G. Ashton[61], Y. Aso[24,62], M. Assiduo[63,64], S. Assis de Souza Melo[59], S. M. Aston[65], P. Astone[66], F. Attadio[67,66], F. Aubin[68], K. AultONeal[69], G. Avallone[70], S. Babak[71], F. Badaracco[56], C. Badger[72], S. Bae[73], S. Bagnasco[27], E. Bagui[74], L. Baiotti[75], R. Bajpai[24], T. Baka[76], T. Baker[77], M. Ball[78], G. Ballardin[59], S. W. Ballmer[79], S. Banagiri[80], B. Banerjee[43], D. Bankar[14], T. M. Baptiste[11], P. Baral[9], J. C. Barayoga[10], B. C. Barish[10], D. Barker[2], N. Barman[14], P. Barneo[41,81], F. Barone[82,4], B. Barr[28], L. Barsotti[34], M. Barsuglia[71], D. Barta[83], A. M. Bartoletti[84], M. A. Barton[28], I. Bartos[44], S. Basak[22], A. Basalaev[85], R. Bassiri[86], A. Basti[87,88], D. E. Bates[32], M. Bawaj[89,50], P. Baxi[90], J. C. Bayley[28], A. C. Baylor[9], P. A. Baynard II[60], M. Bazzan[91,92], V. M. Bedakihale[93], F. Beirnaert[94], M. Bejger[95], D. Belardinelli[20], A. S. Bell[28], D. S. Bellie[80], L. Bellizzi[88,87], W. Benoit[96], I. Bentara[97], J. D. Bentley[85], M. Ben Yaala[58], S. Bera[98], F. Bergamin[7,8], B. K. Berger[86], S. Bernuzzi[26], M. Beroiz[10], C. P. L. Berry[28], D. Bersanetti[56], A. Bertolini[36], J. Betzwieser[65], D. Beveridge[29], G. Bevilacqua[99], N. Bevins[55], R. Bhandare[100], S. A. Bhat[14], R. Bhatt[10], D. Bhattacharjee[101,102], S. Bhaumik[44], S. Bhowmick[103], V. Biancalana[99], A. Bianchi[36,104], I. A. Bilenko[105], G. Billingsley[10], A. Binetti[106], S. Bini[107,108], C. Binu[109], O. Birnholtz[110], S. Biscoveanu[80], A. Bisht[8], M. Bitossi[59,88], M.-A. Bizouard[48], S. Blaber[111], J. K. Blackburn[10], L. A. Blagg[78], C. D. Blair[29,65], D. G. Blair[29], F. Bobba[70,112], N. Bode[7,8], G. Boileau[48], M. Boldrini[66,67], G. N. Bolingbroke[113], A. Bolliand[114,37], L. D. Bonavena[44,91], R. Bondarescu[41], F. Bondu[115], E. Bonilla[86], M. S. Bonilla[54], A. Bonino[116], R. Bonnand[30,114], P. Booker[7,8], A. Borchers[7,8], S. Borhanian[6], V. Boschi[88], S. Bose[117], V. Bossilkov[65], A. Boudon[97], A. Bozzi[59], C. Bradaschia[88], P. R. Brady[9], A. Branch[65], M. Branchesi[43,118], I. Braun[101], T. Briant[119], A. Brillet[48], M. Brinkmann[7,8], P. Brockill[9], E. Brockmueller[7,8], A. F. Brooks[10], B. C. Brown[44], D. D. Brown[113], M. L. Brozzetti[89,50], S. Brunett[10], G. Bruno[13], R. Bruntz[120], J. Bryant[116], Y. Bu[121], F. Bucci[64], J. Buchanan[120], O. Bulashenko[41,81], T. Bulik[122], H. J. Bulten[36], A. Buonanno[123,1], K. Burtnyk[2], R. Buscicchio[124,125], D. Buskulic[30], C. Buy[126], R. L. Byer[86], G. S. Cabourn Davies[77], G. Cabras[45,46], R. Cabrita[13], V. Cáceres-Barbosa[6], L. Cadonati[60], G. Cagnoli[127], C. Cahillane[79], A. Calafat[98], J. Calderón Bustillo[128], T. A. Callister[129], E. Calloni[31,4], M. Canepa[57,56], G. Caneva Santoro[42], K. C. Cannon[40], H. Cao[34], L. A. Capistran[130], E. Capocasa[71], E. Capote[2], G. Capurri[88,87], G. Carapella[70,112], F. Carbognani[59], M. Carlassara[7,8], J. B. Carlin[121], T. K. Carlson[131], M. F. Carney[101], M. Carpinelli[124,132,59], G. Carrillo[78], J. J. Carter[7,8], G. Carullo[133], J. Casanueva Diaz[59], C. Casentini[134,19,20], S. Y. Castro-Lucas[103], S. Caudill[131,36,76], M. Cavaglià[102], R. Cavalieri[59], G. Cella[88], P. Cerdá-Durán[135,136], E. Cesarini[20], W. Chaibi[48], P. Chakraborty[7,8], S. Chakraborty[100], S. Chalathadka Subrahmanya[85], J. C. L. Chan[137], M. Chan[111], R.-J. Chang[138], S. Chao[139,140], E. L. Charlton[120], P. Charlton[141], E. Chassande-Mottin[71], C. Chatterjee[142], Debarati Chatterjee[14], Deep Chatterjee[34], M. Chaturvedi[100], S. Chaty[71], K. Chatziioannou[10], C. Checchia[99], A. Chen[15], A. H.-Y. Chen[143], D. Chen[144], H. Chen[139], H. Y. Chen[145], S. Chen[142], Y. Chen[139], Yanbei Chen[146], Yitian Chen[147], H. P. Cheng[148], P. Chessa[89,50], H. T. Cheung[90], S. Y. Cheung[149], F. Chiadini[150,112], G. Chiarini[92], R. Chierici[97], A. Chincarini[56], M. L. Chiofalo[87,88], A. Chiummo[4,59], C. Chou[143], S. Choudhary[29], N. Christensen[48], S. S. Y. Chua[33], P. Chugh[149], G. Ciani[107,108], P. Ciecielag[95], M. Cieślar[122], M. Cifaldi[20], R. Ciolfi[151,92], F. Clara[2], J. A. Clark[10,60], J. Clarke[32], T. A. Clarke[149], P. Clearwater[152], S. Clesse[74], S. M. Clyne[153], E. Coccia[43,118,42], E. Codazzo[154], P.-F. Cohadon[119], S. Colace[57], E. Colangeli[77], M. Colleoni[98], C. G. Collette[155], J. Collins[65], S. Colloms[28], A. Colombo[156,125], C. M. Compton[2], G. Connolly[78], L. Conti[92], T. R. Corbitt[11], I. Cordero-Carrión[157], S. Corezzi[89,50], N. J. Cornish[158], A. Corsi[159], S. Cortese[59], R. Cottingham[65], M. W. Coughlin[96], A. Couineaux[66], J.-P. Coulon[48], J.-F. Coupechoux[97], P. Couvares[10,60], D. M. Coward[29], R. Coyne[153], K. Craig[58], J. D. E. Creighton[9], T. D. Creighton[160], P. Cremonese[98], A. W. Criswell[96], S. Crook[65], R. Crouch[2], J. Csizmazia[2], J. R. Cudell[161], T. J. Cullen[10], A. Cumming[28], E. Cuoco[162,163], M. Cusinato[135], P. Dabadie[127], L. V. Da Conceição[164], T. Dal Canton[38], S. Dall'Osso[66], S. Dal Pra[165], G. Dálya[126], B. D'Angelo[56], S. Danilishin[35,36], S. D'Antonio[20], K. Danzmann[8,7,8], K. E. Darroch[120], L. P. Dartez[65], A. Dasgupta[93], S. Datta[166], V. Dattilo[59], A. Daumas[71], N. Davari[167,132], I. Dave[100], A. Davenport[103], M. Davier[38], T. F. Davies[29], D. Davis[10], L. Davis[29], M. C. Davis[96], P. Davis[168,169], M. Dax[1], J. De Bolle[94], M. Deenadayalan[14], J. Degallaix[170], U. Deka[22], M. De Laurentis[31,4], S. Deléglise[119], F. De Lillo[21], D. Dell'Aquila[167,132], F. Della Valle[99], W. Del Pozzo[87,88], F. De Marco[67,66], G. Demasi[171,64], F. De Matteis[19,20], V. D'Emilio[10], N. Demos[34], A. Depasse[13], N. DePergola[55], R. De Pietri[172,173], R. De Rosa[31,4], C. De Rossi[59], M. Desai[34], R. DeSalvo[174],

A. DeSimone,[175] R. De Simone,[150] A. Dhani,[1] R. Diab,[44] M. C. Díaz,[160] M. Di Cesare,[31,4] G. Dideron,[176] N. A. Didio,[79] T. Dietrich,[1] L. Di Fiore,[4] C. Di Fronzo,[29] M. Di Giovanni,[67,66] T. Di Girolamo,[31,4] D. Diksha,[36,35] A. Di Michele,[89] J. Ding,[34,71,177] S. Di Pace,[67,66] I. Di Palma,[67,66] F. Di Renzo,[97] Divyajyoti,[178] A. Dmitriev,[116] Z. Doctor,[80] N. Doerksen,[164] E. Dohmen,[2] D. Dominguez,[179] L. D'Onofrio,[66] F. Donovan,[34] K. L. Dooley,[32] T. Dooney,[76] S. Doravari,[14] O. Dorosh,[180] M. Drago,[67,66] J. C. Driggers,[2] J.-G. Ducoin,[181,71] L. Dunn,[121] U. Dupletsa,[43] D. D'Urso,[167,154] H. Duval,[182] S. E. Dwyer,[2] C. Eassa,[2] M. Ebersold,[30] T. Eckhardt,[85] G. Eddolls,[79] B. Edelman,[78] T. B. Edo,[10] O. Edy,[77] A. Effler,[65] J. Eichholz,[33] H. Einsle,[48] M. Eisenmann,[24] R. A. Eisenstein,[34] A. Ejlli,[32] M. Emma,[61] K. Endo,[183] R. Enficiaud,[1] A. J. Engl,[86] L. Errico,[31,4] R. Espinosa,[160] M. Esposito,[4,31] R. C. Essick,[184] H. Estellés,[1] T. Etzel,[10] M. Evans,[34] T. Evstafyeva,[185] B. E. Ewing,[6] J. M. Ezquiaga,[137] F. Fabrizi,[63,64] F. Faedi,[64,63] V. Fafone,[19,20] S. Fairhurst,[32] A. M. Farah,[129] B. Farr,[78] W. M. Farr,[186,187] G. Favaro,[91] M. Favata,[188] M. Fays,[161] M. Fazio,[58] J. Feicht,[10] M. M. Fejer,[86] R. Felicetti,[189] E. Fenyvesi,[83,190] D. L. Ferguson,[145] T. Fernandes,[191,135] D. Fernando,[109] S. Ferraiuolo,[192,67,66] I. Ferrante,[87,88] T. A. Ferreira,[11] F. Fidecaro,[87,88] P. Figura,[95] A. Fiori,[88,87] I. Fiori,[59] M. Fishbach,[184] R. P. Fisher,[120] R. Fittipaldi,[193,112] V. Fiumara,[194,112] R. Flaminio,[30] S. M. Fleischer,[195] L. S. Fleming,[18] E. Floden,[96] H. Fong,[111] J. A. Font,[135,136] C. Foo,[1] B. Fornal,[196] P. W. F. Forsyth,[33] K. Franceschetti,[172] N. Franchini,[197] S. Frasca,[67,66] F. Frasconi,[88] A. Frattale Mascioli,[67,66] Z. Frei,[198] A. Freise,[36,104] O. Freitas,[191,135] R. Frey,[78] W. Frischhertz,[65] P. Fritschel,[34] V. V. Frolov,[65] G. G. Fronzé,[27] M. Fuentes-Garcia,[10] S. Fujii,[199] T. Fujimori,[200] P. Fulda,[44] M. Fyffe,[65] B. Gadre,[76] J. R. Gair,[1] S. Galaudage,[201] V. Galdi,[174] H. Gallagher,[109] B. Gallego,[202] R. Gamba,[6,26] A. Gamboa,[1] D. Ganapathy,[34] A. Ganguly,[14] B. Garaventa,[56,57] J. García-Bellido,[203] C. García Núñez,[18] C. García-Quirós,[204] J. W. Gardner,[33] K. A. Gardner,[111] J. Gargiulo,[59] A. Garron,[98] F. Garufi,[31,4] P. A. Garver,[86] C. Gasbarra,[19,20] B. Gateley,[2] F. Gautier,[205] V. Gayathri,[9] T. Gayer,[79] G. Gemme,[56] A. Gennai,[88] V. Gennari,[126] J. George,[100] R. George,[145] O. Gerberding,[85] L. Gergely,[206] Archisman Ghosh,[94] Sayantan Ghosh,[207] Shaon Ghosh,[188] Shrobana Ghosh,[7,8] Suprovo Ghosh,[14] Tathagata Ghosh,[14] J. A. Giaime,[11,65] K. D. Giardina,[65] D. R. Gibson,[18] D. T. Gibson,[185] C. Gier,[58] S. Gkaitatzis,[87,88] J. Glanzer,[10] F. Glotin,[38] J. Godfrey,[78] P. Godwin,[10] A. S. Goettel,[32] E. Goetz,[111] J. Golomb,[10] S. Gomez Lopez,[67,66] B. Goncharov,[43] Y. Gong,[208] G. González,[11] P. Goodarzi,[209] S. Goode,[149] A. W. Goodwin-Jones,[10,29] M. Gosselin,[59] R. Gouaty,[30] D. W. Gould,[33] K. Govorkova,[34] S. Goyal,[1] B. Grace,[33] A. Grado,[89,50] V. Graham,[28] A. E. Granados,[96] M. Granata,[170] V. Granata,[70] S. Gras,[34] P. Grassia,[10] A. Gray,[96] C. Gray,[2] R. Gray,[28] G. Greco,[50] A. C. Green,[36,104] S. M. Green,[77] S. R. Green,[210] A. M. Gretarsson,[69] E. M. Gretarsson,[69] D. Griffith,[10] W. L. Griffiths,[32] H. L. Griggs,[60] G. Grignani,[89,50] C. Grimaud,[30] H. Grote,[32] S. Grunewald,[1] D. Guerra,[135] D. Guetta,[211] G. M. Guidi,[63,64] A. R. Guimaraes,[11] H. K. Gulati,[93] F. Gulminelli,[168,169] A. M. Gunny,[34] H. Guo,[212] W. Guo,[29] Y. Guo,[36,35] Anchal Gupta,[10] Anuradha Gupta,[213] I. Gupta,[6] N. C. Gupta,[93] P. Gupta,[36,76] S. K. Gupta,[44] T. Gupta,[158] V. Gupta,[96] N. Gupte,[1] J. Gurs,[85] N. Gutierrez,[170] F. Guzman,[130] D. Haba,[179] M. Haberland,[1] S. Haino,[214] E. D. Hall,[34] R. Hamburg,[215] E. Z. Hamilton,[98] G. Hammond,[28] W.-B. Han,[216] M. Haney,[36,204] J. Hanks,[2] C. Hanna,[6] M. D. Hannam,[32] O. A. Hannuksela,[217] A. G. Hanselman,[129] H. Hansen,[2] J. Hanson,[65] R. Harada,[40] A. R. Hardison,[175] S. Harikumar,[180] K. Haris,[36,76] T. Harmark,[133] J. Harms,[43,118] G. M. Harry,[218] I. W. Harry,[77] J. Hart,[101] B. Haskell,[95] C.-J. Haster,[219] K. Haughian,[28] H. Hayakawa,[49] K. Hayama,[220] R. Hayes,[32] M. C. Heintze,[65] J. Heinze,[116] J. Heinzel,[34] H. Heitmann,[48] A. Heffernan,[98] F. Hellman,[221] A. F. Helmling-Cornell,[78] G. Hemming,[59] O. Henderson-Sapir,[113] M. Hendry,[28] I. S. Heng,[28] M. H. Hennig,[28] C. Henshaw,[60] M. Heurs,[7,8] A. L. Hewitt,[185,222] J. Heyns,[34] S. Higginbotham,[32] S. Hild,[35,36] S. Hill,[28] Y. Himemoto,[223] N. Hirata,[24] C. Hirose,[224] S. Hochheim,[7,8] D. Hofman,[170] N. A. Holland,[36,104] D. E. Holz,[129] L. Honet,[74] C. Hong,[86] S. Hoshino,[224] J. Hough,[28] S. Hourihane,[10] N. T. Howard,[142] E. J. Howell,[29] C. G. Hoy,[77] C. A. Hrishikesh,[19] H.-F. Hsieh,[139] H.-Y. Hsieh,[139] C. Hsiung,[225] W.-F. Hsu,[106] Q. Hu,[28] H. Y. Huang,[140] Y. Huang,[6] Y. T. Huang,[79] A. D. Huddart,[226] B. Hughey,[69] D. C. Y. Hui,[227] V. Hui,[30] S. Husa,[98] R. Huxford,[6] L. Iampieri,[67,66] G. A. Iandolo,[35] M. Ianni,[20,19] A. Ierardi,[43] A. Iess,[228,88] H. Imafuku,[40] K. Inayoshi,[229] Y. Inoue,[140] G. Iorio,[91] P. Iosif,[189,46] M. H. Iqbal,[33] J. Irwin,[28] R. Ishikawa,[230] M. Isi,[186,187] Y. Itoh,[231,200] H. Iwanaga,[231] M. Iwaya,[199] B. R. Iyer,[22] C. Jacquet,[126] P.-E. Jacquet,[119] S. J. Jadhav,[232] S. P. Jadhav,[152] T. Jain,[185] A. L. James,[10] P. A. James,[120] R. Jamshidi,[155] A. Jan,[145] K. Jani,[142] J. Janquart,[13] K. Janssens,[21,48] N. N. Janthalur,[232] S. Jaraba,[203] P. Jaranowski,[233] R. Jaume,[98] W. Javed,[32] A. Jennings,[2] W. Jia,[34] J. Jiang,[148] S. J. Jin,[29] C. Johanson,[131] G. R. Johns,[120] N. A. Johnson,[44] N. K. Johnson-McDaniel,[213] M. C. Johnston,[219] R. Johnston,[28] N. Johny,[7,8] D. H. Jones,[33] D. I. Jones,[234] E. J. Jones,[11] R. Jones,[28] S. Jose,[178] P. Joshi,[6] S. K. Joshi,[14] J. Ju,[235] L. Ju,[29] K. Jung,[236] J. Junker,[33] V. Juste,[74] H. B. Kabagoz,[65] T. Kajita,[237] I. Kaku,[231] V. Kalogera,[80] M. Kalomenopoulos,[219] M. Kamiizumi,[49] N. Kanda,[200,231] S. Kandhasamy,[14] G. Kang,[238] N. C. Kannachel,[149] J. B. Kanner,[10] S. J. Kapadia,[14] D. P. Kapasi,[33] S. Karat,[10] R. Kashyap,[6] M. Kasprzack,[10] W. Kastaun,[7,8] T. Kato,[199] E. Katsavounidis,[34] W. Katzman,[65] R. Kaushik,[100] K. Kawabe,[2] R. Kawamoto,[231] A. Kazemi,[96] D. Keitel,[98] J. Kennington,[6] R. Kesharwani,[14] J. S. Key,[239] R. Khadela,[7,8] S. Khadka,[86] F. Y. Khalili,[105] F. Khan,[7,8] I. Khan,[240,37] T. Khanam,[159] M. Khursheed,[100] N. M. Khusid,[186,187] W. Kiendrebeogo,[48,241] N. Kijbunchoo,[113] C. Kim,[242] J. C. Kim,[243] K. Kim,[244] M. H. Kim,[235] S. Kim,[227] Y.-M. Kim,[244] C. Kimball,[80] M. Kinley-Hanlon,[28] M. Kinnear,[32] J. S. Kissel,[2] S. Klimenko,[44] A. M. Knee,[111] N. Knust,[7,8] K. Kobayashi,[199] P. Koch,[7,8] S. M. Koehlenbeck,[86] G. Koekoek,[36,35] K. Kohri,[245,246] K. Kokeyama,[32] S. Koley,[43] P. Kolitsidou,[116] K. Komori,[40,39] A. K. H. Kong,[139] A. Kontos,[247] M. Korobko,[85] R. V. Kossak,[7,8] X. Kou,[96] A. Koushik,[21]

N. Kouvatsos,[72] M. Kovalam,[29] D. B. Kozak,[10] S. L. Kranzhoff,[35,36] V. Kringel,[7,8] N. V. Krishnendu,[116] A. Królak,[248,180] K. Kruska,[7,8] J. Kubisz,[249] G. Kuehn,[7,8] S. Kulkarni,[213] A. Kulur Ramamohan,[33] A. Kumar,[232] Praveen Kumar,[128] Prayush Kumar,[22] Rahul Kumar,[2] Rakesh Kumar,[93] J. Kume,[250,251,40] K. Kuns,[34] N. Kuntimaddi,[32] S. Kuroyanagi,[203,252] S. Kuwahara,[40] K. Kwak,[236] K. Kwan,[33] J. Kwok,[185] G. Lacaille,[28] P. Lagabbe,[30,107] D. Laghi,[126] S. Lai,[143] E. Lalande,[253] M. Lalleman,[21] P. C. Lalremruati,[254] M. Landry,[2] B. B. Lane,[34] R. N. Lang,[34] J. Lange,[145] R. Langgin,[219] B. Lantz,[86] A. La Rana,[66] I. La Rosa,[98] J. Larsen,[195] A. Lartaux-Vollard,[38] P. D. Lasky,[149] J. Lawrence,[160,255] M. N. Lawrence,[11] M. Laxen,[65] C. Lazarte,[135] A. Lazzarini,[10] C. Lazzaro,[256,154] P. Leaci,[67,66] L. Leali,[96] Y. K. Lecoeuche,[111] H. M. Lee,[243] H. W. Lee,[257] J. Lee,[79] K. Lee,[235] R.-K. Lee,[139] R. Lee,[34] Sungho Lee,[258] Sunjae Lee,[235] Y. Lee,[140] I. N. Legred,[10] J. Lehmann,[7,8] L. Lehner,[176] M. Le Jean,[170] A. Lemaître,[259] M. Lenti,[64,171] M. Leonardi,[107,108,24] M. Lequime,[37] N. Leroy,[38] M. Lesovsky,[10] N. Letendre,[30] M. Lethuillier,[97] Y. Levin,[149] K. Leyde,[71,77] A. K. Y. Li,[10] K. L. Li,[138] T. G. F. Li,[106] X. Li,[146] Y. Li,[80] Z. Li,[28] A. Lihos,[120] C-Y. Lin,[260] E. T. Lin,[139] L. C.-C. Lin,[138] Y.-C. Lin,[139] C. Lindsay,[18] S. D. Linker,[202] T. B. Littenberg,[261] A. Liu,[217] G. C. Liu,[225] Jian Liu,[29] F. Llamas Villarreal,[160] J. Llobera-Querol,[98] R. K. L. Lo,[137] J.-P. Locquet,[106] M. R. Loizou,[131] L. T. London,[72,34] A. Longo,[63,64] D. Lopez,[161,204] M. Lopez Portilla,[76] A. Lorenzo-Medina,[128] V. Loriette,[38] M. Lormand,[65] G. Losurdo,[228,88] E. Lotti,[131] T. P. Lott IV,[60] J. D. Lough,[7,8] H. A. Loughlin,[34] C. O. Lousto,[109] N. Low,[121] M. J. Lowry,[120] N. Lu,[33] L. Lucchesi,[88] H. Lück,[8,7,8] D. Lumaca,[20] A. P. Lundgren,[77] A. W. Lussier,[253] L.-T. Ma,[139] S. Ma,[176] R. Macas,[77] A. Macedo,[54] M. MacInnis,[34] R. R. Maciy,[7,8] D. M. Macleod,[32] I. A. O. MacMillan,[10] A. Macquet,[38] D. Macri,[34] K. Maeda,[183] S. Maenaut,[106] S. S. Magare,[14] R. M. Magee,[10] E. Maggio,[1] R. Maggiore,[36,104] M. Magnozzi,[56,57] M. Mahesh,[85] M. Maini,[153] S. Majhi,[14] E. Majorana,[67,66] C. N. Makarem,[10] D. Malakar,[102] J. A. Malaquias-Reis,[17] U. Mali,[184] S. Maliakal,[10] A. Malik,[100] L. Mallick,[164,184] A. Malz,[61] N. Man,[48] V. Mandic,[96] V. Mangano,[66,67] B. Mannix,[78] G. L. Mansell,[79] G. Mansingh,[218] M. Manske,[9] M. Mantovani,[59] M. Mapelli,[91,92,262] F. Marchesoni,[51,50,263] C. Marinelli,[99] D. Marín Pina,[41,81,264] F. Marion,[30] S. Márka,[265] Z. Márka,[265] A. S. Markosyan,[86] A. Markowitz,[10] E. Maros,[10] S. Marsat,[126] F. Martelli,[63,64] I. W. Martin,[28] R. M. Martin,[188] B. B. Martinez,[130] M. Martinez,[42,266] V. Martinez,[127] A. Martini,[107,108] J. C. Martins,[17] D. V. Martynov,[116] E. J. Marx,[34] L. Massaro,[35,36] A. Masserot,[30] M. Masso-Reid,[28] M. Mastrodicasa,[66,67] S. Mastrogiovanni,[66] T. Matcovich,[50] M. Matiushechkina,[7,8] M. Matsuyama,[231] N. Mavalvala,[34] N. Maxwell,[2] G. McCarrol,[65] R. McCarthy,[2] D. E. McClelland,[33] S. McCormick,[65] L. McCuller,[10] S. McEachin,[120] C. McElhenny,[120] G. I. McGhee,[28] J. McGinn,[28] K. B. M. McGowan,[142] J. McIver,[111] A. McLeod,[29] T. McRae,[33] D. Meacher,[9] Q. Meijer,[76] A. Melatos,[121] M. Melching,[7,8] S. Mellaerts,[106] C. S. Menoni,[103] F. Mera,[2] R. A. Mercer,[9] L. Mereni,[170] K. Merfeld,[159] E. L. Merilh,[65] J. R. Mérou,[98] J. D. Merritt,[78] M. Merzougui,[48] C. Messenger,[28] C. Messick,[9] B. Mestichelli,[43] M. Meyer-Conde,[267] F. Meylahn,[7,8] A. Mhaske,[14] A. Miani,[107,108] H. Miao,[268] I. Michaloliakos,[44] C. Michel,[170] Y. Michimura,[10,40] H. Middleton,[116] S. J. Miller,[10] M. Millhouse,[60] E. Milotti,[189,46] V. Milotti,[91] Y. Minenkov,[20] N. Mio,[269] Ll. M. Mir,[42] L. Mirasola,[154,256] M. Miravet-Tenés,[135] C.-A. Miritescu,[42] A. K. Mishra,[22] A. Mishra,[22] C. Mishra,[178] T. Mishra,[44] A. L. Mitchell,[36,104] J. G. Mitchell,[69] S. Mitra,[14] V. P. Mitrofanov,[105] R. Mittleman,[34] O. Miyakawa,[49] S. Miyamoto,[199] S. Miyoki,[49] G. Mo,[34] L. Mobilia,[63,64] S. R. P. Mohapatra,[10] S. R. Mohite,[6] M. Molina-Ruiz,[221] C. Mondal,[168] M. Mondin,[202] M. Montani,[63,64] C. J. Moore,[185] D. Moraru,[2] A. More,[14] S. More,[14] E. A. Moreno,[34] G. Moreno,[2] S. Morisaki,[40,199] Y. Moriwaki,[183] G. Morras,[203] A. Moscatello,[91] M. Mould,[34] P. Mourier,[98,270] B. Mours,[68] C. M. Mow-Lowry,[36,104] F. Muciaccia,[67,66] D. Mukherjee,[261] Samanwaya Mukherjee,[14] Soma Mukherjee,[160] Subroto Mukherjee,[93] Suvodip Mukherjee,[12,176,271] N. Mukund,[34] A. Mullavey,[65] H. Mullock,[111] J. Munch,[113] J. Mundi,[218] C. L. Mungioli,[29] Y. Murakami,[199] M. Murakoshi,[230] P. G. Murray,[28] S. Muusse,[33] D. Nabari,[107,108] S. L. Nadji,[7,8] A. Nagar,[27,272] N. Nagarajan,[28] K. Nakagaki,[49] K. Nakamura,[24] H. Nakano,[273] M. Nakano,[10] D. Nanadoumgar-Lacroze,[42] D. Nandi,[11] V. Napolano,[59] P. Narayan,[213] I. Nardecchia,[20] T. Narikawa,[199] H. Narola,[76] L. Naticchioni,[66] R. K. Nayak,[254] A. Nela,[28] A. Nelson,[130] T. J. N. Nelson,[65] M. Nery,[7,8] A. Neunzert,[2] S. Ng,[54] L. Nguyen Quynh,[274,275] S. A. Nichols,[11] A. B. Nielsen,[276] G. Nieradka,[95] Y. Nishino,[24,277] A. Nishizawa,[278] S. Nissanke,[271,36] E. Nitoglia,[97] W. Niu,[6] F. Nocera,[59] M. Norman,[32] C. North,[32] J. Novak,[114,279,280] J. F. Nuño Siles,[203] L. K. Nuttall,[77] K. Obayashi,[230] J. Oberling,[2] J. O'Dell,[226] M. Oertel,[279,114,281,280] A. Offermans,[106] G. Oganesyan,[43,118] J. J. Oh,[282] K. Oh,[227] T. O'Hanlon,[65] M. Ohashi,[49] M. Ohkawa,[224] F. Ohme,[7,8] R. Oliveri,[114,281,280] R. Omer,[96] B. O'Neal,[120] K. Oohara,[283,284] B. O'Reilly,[65] R. Oram,[10] N. D. Ormsby,[120] M. Orselli,[50,89] R. O'Shaughnessy,[109] S. O'Shea,[28] Y. Oshima,[39] S. Oshino,[49] C. Osthelder,[10] I. Ota,[11] D. J. Ottaway,[113] A. Ouzriat,[97] H. Overmier,[65] B. J. Owen,[285] A. E. Pace,[6] R. Pagano,[11] M. A. Page,[24] A. Pai,[207] L. Paiella,[43] A. Pal,[286] S. Pal,[254] M. A. Palaia,[88,87] M. Pálfi,[198] P. P. Palma,[67,19,20] C. Palomba,[66] P. Palud,[71] J. Pan,[29] K. C. Pan,[139] R. Panai,[154,91] P. K. Panda,[232] Shiksha Pandey,[6] Swadha Pandey,[34] P. T. H. Pang,[36,76] F. Pannarale,[67,66] K. A. Pannone,[54] B. C. Pant,[100] F. H. Panther,[29] F. Paoletti,[88] A. Paolone,[66,287] A. Papadopoulos,[28] E. E. Papalexakis,[209] L. Papalini,[88,87] G. Papigkiotis,[288] A. Paquis,[38] A. Parisi,[89,50] B.-J. Park,[258] J. Park,[289] W. Parker,[65] G. Pascale,[7,8] D. Pascucci,[94] A. Pasqualetti,[59] R. Passaquieti,[87,88] L. Passenger,[149] D. Passuello,[88] O. Patane,[2] D. Pathak,[14] L. Pathak,[14] A. Patra,[32] B. Patricelli,[87,88] A. S. Patron,[11] B. G. Patterson,[32] K. Paul,[178] S. Paul,[78] E. Payne,[10] T. Pearce,[32] M. Pedraza,[10] A. Pele,[10] F. E. Peña Arellano,[290] S. Penn,[291] M. D. Penuliar,[54] A. Perego,[107,108] Z. Pereira,[131] J. J. Perez,[44] C. Périgois,[151,92,91] G. Perna,[91] A. Perreca,[107,108] J. Perret,[71] S. Perriès,[97] J. W. Perry,[36,104] D. Pesios,[288] S. Petracca,[174] C. Petrillo,[89]

H. P. Pfeiffer ,[1] H. Pham,[65] K. A. Pham ,[96] K. S. Phukon ,[116] H. Phurailatpam,[217] M. Piarulli,[126] L. Piccari ,[67,66] O. J. Piccinni ,[33] M. Pichot ,[48] M. Piendibene ,[87,88] F. Piergiovanni ,[63,64] L. Pierini ,[66] G. Pierra ,[97] V. Pierro ,[292,112] M. Pietrzak,[95] M. Pillas ,[161] F. Pilo ,[88] L. Pinard,[170] I. M. Pinto ,[292,112,293,31] M. Pinto,[59] B. J. Piotrzkowski ,[9] M. Pirello,[2] M. D. Pitkin ,[185,222] A. Placidi ,[50] E. Placidi ,[67,66] M. L. Planas ,[98] W. Plastino ,[294,20] C. Plunkett ,[34] R. Poggiani ,[87,88] E. Polini ,[34] L. Pompili ,[1] J. Poon,[217] E. Porcelli,[36] E. K. Porter,[71] C. Posnansky ,[6] R. Poulton ,[59] J. Powell ,[152] M. Pracchia ,[161] B. K. Pradhan ,[14] T. Pradier ,[68] A. K. Prajapati,[93] K. Prasai,[86] R. Prasanna,[232] P. Prasia,[14] G. Pratten ,[116] G. Principe ,[189,46] M. Principe,[174] G. A. Prodi ,[107,108] L. Prokhorov ,[116] P. Prosperi,[88] P. Prosposito,[19,20] A. C. Providence,[69] A. Puecher,[36,76] J. Pullin ,[11] M. Punturo ,[50] P. Puppo,[66] M. Pürrer ,[153] H. Qi ,[15] J. Qin ,[33] G. Quéméner ,[169,114] V. Quetschke,[160] P. J. Quinonez,[69] F. J. Raab ,[2] I. Rainho,[135] S. Raja,[100] C. Rajan,[100] B. Rajbhandari ,[109] K. E. Ramirez ,[65] F. A. Ramis Vidal ,[98] A. Ramos-Buades,[36,1] D. Rana,[14] S. Ranjan ,[60] K. Ransom,[65] P. Rapagnani ,[67,66] B. Ratto,[69] A. Ray ,[9] V. Raymond ,[32] M. Razzano ,[87,88] J. Read,[54] M. Recaman Payo,[106] T. Regimbau,[30] L. Rei ,[56] S. Reid,[58] D. H. Reitze ,[10] P. Relton ,[32] A. I. Renzini ,[10,124] B. Revenu ,[295,38] R. Reyes,[202] A. S. Rezaei ,[66,67] F. Ricci,[67,66] M. Ricci ,[66,67] A. Ricciardone ,[87,88] J. W. Richardson ,[209] M. Richardson,[113] A. Rijal,[69] K. Riles ,[90] H. K. Riley,[32] S. Rinaldi ,[262,91] J. Rittmeyer,[85] C. Robertson,[226] F. Robinet,[38] M. Robinson,[2] A. Rocchi ,[20] L. Rolland ,[30] J. G. Rollins ,[10] A. E. Romano ,[296] R. Romano ,[3,4] A. Romero ,[30] I. M. Romero-Shaw,[185] J. H. Romie,[65] S. Ronchini ,[6,43,118] T. J. Roocke ,[113] L. Rosa,[4,31] T. J. Rosauer,[209] C. A. Rose,[60] D. Rosińska ,[122] M. P. Ross ,[52] M. Rossello-Sastre ,[98] S. Rowan ,[28] S. Roy ,[13] S. K. Roy ,[186,187] D. Rozza ,[124,125] P. Ruggi,[59] N. Ruhama,[236] E. Ruiz Morales ,[297,203] K. Ruiz-Rocha,[142] S. Sachdev ,[60] T. Sadecki,[2] J. Sadiq ,[128] P. Saffarieh ,[36,104] S. Safi-Harb,[164] M. R. Sah ,[12] S. Saha ,[139] T. Sainrat ,[68] S. Sajith Menon ,[211,67,66] K. Sakai,[298] M. Sakellariadou ,[72] S. Sakon ,[6] O. S. Salafia ,[156,125,124] F. Salces-Carcoba ,[10] L. Salconi,[59] M. Saleem ,[96] F. Salemi ,[67,66] M. Sallé ,[36] S. U. Salunkhe,[14] S. Salvador ,[169,168] A. Samajdar ,[76,36] A. Sanchez,[2] E. J. Sanchez,[10] J. H. Sanchez ,[80] L. E. Sanchez,[10] N. Sanchis-Gual ,[135] J. R. Sanders,[175] E. M. Sänger ,[1] F. Santoliquido,[43] F. Sarandrea,[27] T. R. Saravanan,[14] N. Sarin,[149] P. Sarkar,[7,8] S. Sasaoka ,[179] A. Sasli ,[288] P. Sassi ,[50,89] B. Sassolas ,[170] B. S. Sathyaprakash ,[6,32] R. Sato,[224] Y. Sato,[183] O. Sauter ,[44] R. L. Savage ,[2] T. Sawada ,[49] H. L. Sawant,[14] S. Sayah,[30] V. Scacco,[19,20] D. Schaetzl,[10] M. Scheel,[146] A. Schiebelbein,[184] M. G. Schiworski ,[79] P. Schmidt ,[116] S. Schmidt ,[76] R. Schnabel ,[85] M. Schneewind,[7,8] R. M. S. Schofield,[78] K. Schouteden,[106] B. W. Schulte,[7,8] B. F. Schutz,[32,7,8] E. Schwartz ,[86] M. Scialpi,[299] J. Scott ,[28] S. M. Scott ,[33] R. M. Sedas ,[65] T. C. Seetharamu,[28] M. Seglar-Arroyo ,[42] Y. Sekiguchi ,[300] D. Sellers,[65] A. S. Sengupta ,[301] D. Sentenac,[59] E. G. Seo ,[28] J. W. Seo ,[106] V. Sequino,[31,4] M. Serra ,[66] G. Servignat ,[71,281] A. Sevrin,[182] T. Shaffer,[2] U. S. Shah ,[60] M. S. Shahriar ,[80] M. A. Shaikh ,[243] L. Shao ,[229] A. Sharma ,[301] A. K. Sharma,[22] P. Sharma,[100] S. Sharma Chaudhary,[102] M. R. Shaw,[32] P. Shawhan ,[123] N. S. Shcheblanov ,[302,259] Y. Shikano ,[303,304] M. Shikauchi,[40] K. Shimode ,[49] H. Shinkai ,[305] J. Shiota,[230] S. Shirke,[14] D. H. Shoemaker ,[34] D. M. Shoemaker ,[145] R. W. Short,[2] S. ShyamSundar,[100] A. Sider,[155] H. Siegel ,[186,187] D. Sigg ,[2] L. Silenzi ,[50,51] M. Simmonds,[113] L. P. Singer ,[306] A. Singh,[213] D. Singh ,[6] M. K. Singh ,[22] N. Singh ,[98] S. Singh,[179,62] A. Singha ,[35,36] A. M. Sintes ,[98] V. Sipala,[167,154] V. Skliris ,[32] B. J. J. Slagmolen ,[33] D. A. Slater,[195] T. J. Slaven-Blair,[29] J. Smetana,[116] J. R. Smith ,[54] L. Smith ,[28,189] R. J. E. Smith ,[149] W. J. Smith ,[142] K. Somiya ,[179] I. Song ,[139] K. Soni ,[14] S. Soni ,[34] V. Sordini ,[97] F. Sorrentino,[56] H. Sotani ,[307] A. Southgate,[32] F. Spada ,[88] V. Spagnuolo ,[35,36] A. P. Spencer ,[28] M. Spera ,[46,308] P. Spinicelli ,[59] C. A. Sprague,[274] A. K. Srivastava,[93] F. Stachurski ,[28] D. A. Steer ,[309] N. Steinle ,[164] J. Steinlechner,[35,36] S. Steinlechner ,[35,36] N. Stergioulas ,[288] P. Stevens,[38] S. P. Stevenson,[152] F. Stolzi ,[99] M. StPierre,[153] G. Stratta ,[310,134,66,311] M. D. Strong,[11] A. Strunk,[2] R. Sturani,[312] A. L. Stuver,[55,*] M. Suchenek,[95] S. Sudhagar ,[95] N. Sueltmann,[85] L. Suleiman ,[54] J.M. Sullivan ,[60] K. D. Sullivan,[11] J. Sun,[238] L. Sun ,[33] S. Sunil,[93] J. Suresh ,[48] B. J. Sutton,[72] P. J. Sutton ,[32] T. Suzuki ,[224] Y. Suzuki,[230] B. L. Swinkels ,[36] A. Syx,[68] M. J. Szczepańczyk ,[313,44] P. Szewczyk ,[122] M. Tacca ,[36] H. Tagoshi ,[199] S. C. Tait ,[10] H. Takahashi ,[267] R. Takahashi ,[24] A. Takamori ,[53] T. Takase,[49] K. Takatani,[231] H. Takeda ,[314] K. Takeshita,[179] C. Talbot,[129] M. Tamaki,[199] N. Tamanini ,[126] D. Tanabe,[140] K. Tanaka,[49] S. J. Tanaka ,[230] T. Tanaka ,[314] D. Tang,[29] S. Tanioka ,[79] D. B. Tanner,[44] W. Tanner,[7,8] L. Tao ,[209] R. D. Tapia,[6] E. N. Tapia San Martín ,[36] R. Tarafder,[10] C. Taranto,[19,20] A. Taruya ,[315] J. D. Tasson ,[316] J. G. Tau ,[109] R. Tenorio ,[98] H. Themann,[202] A. Theodoropoulos ,[135] M. P. Thirugnanasambandam,[14] L. M. Thomas ,[10] M. Thomas,[65] P. Thomas,[2] J. E. Thompson ,[234] S. R. Thondapu,[100] K. A. Thorne,[65] E. Thrane,[149] S. Tibrewal ,[145] J. Tissino ,[43] A. Tiwari,[14] P. Tiwari,[43] S. Tiwari ,[204] V. Tiwari ,[116] M. R. Todd,[79] A. M. Toivonen ,[96] K. Toland ,[28] A. E. Tolley ,[77] T. Tomaru ,[24] K. Tomita,[231] V. Tommasini,[10] T. Tomura ,[49] H. Tong ,[149] C. Tong-Yu,[140] A. Toriyama,[230] N. Toropov ,[116] A. Torres-Forné ,[135,136] C. I. Torrie,[10] M. Toscani ,[126] I. Tosta e Melo ,[317] E. Tournefier ,[30] M. Trad Nery,[48] A. Trapananti ,[51,50] F. Travasso ,[51,50] G. Traylor,[65] C. Trejo,[10] M. Trevor,[123] M. C. Tringali ,[59] A. Tripathee ,[90] G. Troian ,[189,46] A. Trovato ,[189,46] L. Trozzo,[4] R. J. Trudeau,[10] T. T. L. Tsang ,[32] S. Tsuchida ,[318] L. Tsukada ,[219] K. Turbang ,[182,21] M. Turconi ,[48] C. Turski,[94] H. Ubach ,[41,81] N. Uchikata ,[199] T. Uchiyama ,[49] R. P. Udall ,[10] T. Uehara ,[319] M. Uematsu,[231] S. Ueno,[230] V. Undheim ,[276] T. Ushiba ,[49] M. Vacatello ,[88,87] H. Vahlbruch ,[7,8] G. Vajente ,[10] A. Vajpeyi,[149] G. Valdes ,[320] J. Valencia ,[98] A. F. Valentini,[11] M. Valentini ,[104,36] S. A. Vallejo-Peña ,[296] S. Vallero,[27] V. Valsan ,[9] N. van Bakel,[36] M. van Beuzekom ,[36] M. van Dael ,[36,321] J. F. J. van den Brand ,[35,104,36] C. Van Den Broeck,[76,36] D. C. Vander-Hyde,[79] M. van der Sluys ,[36,76] A. Van de Walle,[38] J. van Dongen ,[36,104] K. Vandra,[55] H. van Haevermaet ,[21] J. V. van Heijningen ,[36,104] P. Van Hove ,[68] J. Vanier,[253] M. VanKeuren,[101] J. Vanosky,[2] M. H. P. M. van Putten ,[16] Z. Van Ranst ,[35,36] N. van Remortel ,[21]

M. Vardaro,[35,36] A. F. Vargas,[121] J. J. Varghese,[69] V. Varma,[131] A. N. Vazquez,[86] A. Vecchio,[116] G. Vedovato,[92] J. Veitch,[28] P. J. Veitch,[113] S. Venikoudis,[13] J. Venneberg,[7,8] P. Verdier,[97] M. Vereecken,[13] D. Verkindt,[30] B. Verma,[131] P. Verma,[180] Y. Verma,[100] S. M. Vermeulen,[10] F. Vetrano,[63] A. Veutro,[66,67] A. M. Vibhute,[2] A. Viceré,[63,64] S. Vidyant,[79] A. D. Viets,[84] A. Vijaykumar,[184] A. Vilkha,[109] V. Villa-Ortega,[128] E. T. Vincent,[60] J.-Y. Vinet,[48] S. Viret,[97] A. Virtuoso,[46] S. Vitale,[34] A. Vives,[78] H. Vocca,[89,50] D. Voigt,[85] E. R. G. von Reis,[2] J. S. A. von Wrangel,[7,8] L. Vujeva,[137] S. P. Vyatchanin,[105] J. Wack,[10] L. E. Wade,[101] M. Wade,[101] K. J. Wagner,[109] A. Wajid,[56,57] M. Walker,[120] G. S. Wallace,[58] L. Wallace,[10] E. J. Wang,[86] H. Wang,[39] J. Z. Wang,[90] W. H. Wang,[160] Y. F. Wang,[1] Z. Wang,[140] G. Waratkar,[207] J. Warner,[2] M. Was,[30] T. Washimi,[24] N. Y. Washington,[10] D. Watarai,[40] K. E. Wayt,[101] B. R. Weaver,[32] B. Weaver,[2] C. R. Weaving,[77] S. A. Webster,[28] N. L. Weickhardt,[85] M. Weinert,[7,8] A. J. Weinstein,[10] R. Weiss,[34] F. Wellmann,[7,8] L. Wen,[29] P. Wessels,[7,8] K. Wette,[33] J. T. Whelan,[109] B. F. Whiting,[44] C. Whittle,[10] E. G. Wickens,[77] J. B. Wildberger,[1] D. Wilken,[7,8,8] D. J. Willadsen,[84] K. Willetts,[32] D. Williams,[28] M. J. Williams,[77] N. S. Williams,[116] J. L. Willis,[10] B. Willke,[8,7,8] M. Wils,[106] C. W. Winborn,[102] J. Winterflood,[29] C. C. Wipf,[10] G. Woan,[28] J. Woehler,[35,36] N. E. Wolfe,[34] H. T. Wong,[140] I. C. F. Wong,[217,106] J. L. Wright,[33] M. Wright,[28] C. Wu,[139] D. S. Wu,[7,8] H. Wu,[139] E. Wuchner,[54] D. M. Wysocki,[9] V. A. Xu,[34] Y. Xu,[204] N. Yadav,[95] H. Yamamoto,[10] K. Yamamoto,[183] T. S. Yamamoto,[40] T. Yamamoto,[49] S. Yamamura,[199] R. Yamazaki,[230] T. Yan,[116] F. W. Yang,[322] F. Yang,[265] K. Z. Yang,[96] Y. Yang,[143] Z. Yarbrough,[11] H. Yasui,[49] S.-W. Yeh,[139] A. B. Yelikar,[109] X. Yin,[34] J. Yokoyama,[323,40,39] T. Yokozawa,[49] J. Yoo,[147] H. Yu,[146] S. Yuan,[29] H. Yuzurihara,[49] A. Zadrożny,[180] M. Zanolin,[69] M. Zeeshan,[109] T. Zelenova,[59] J.-P. Zendri,[92] M. Zeoli,[13] M. Zerrad,[37] M. Zevin,[80] A. C. Zhang,[265] L. Zhang,[10] R. Zhang,[148] T. Zhang,[116] Y. Zhang,[33] C. Zhao,[29] Yue Zhao,[322] Yuhang Zhao,[71] Y. Zheng,[102] H. Zhong,[96] R. Zhou,[221] X.-J. Zhu,[324] Z.-H. Zhu,[324,208] A. B. Zimmerman,[145] M. E. Zucker[34,10] And J. Zweizig[10]

The LIGO Scientific Collaboration, the Virgo Collaboration, and the KAGRA Collaboration

[1] *Max Planck Institute for Gravitational Physics (Albert Einstein Institute), D-14476 Potsdam, Germany*
[2] *LIGO Hanford Observatory, Richland, WA 99352, USA*
[3] *Dipartimento di Farmacia, Università di Salerno, I-84084 Fisciano, Salerno, Italy*
[4] *INFN, Sezione di Napoli, I-80126 Napoli, Italy*
[5] *University of Warwick, Coventry CV4 7AL, United Kingdom*
[6] *The Pennsylvania State University, University Park, PA 16802, USA*
[7] *Max Planck Institute for Gravitational Physics (Albert Einstein Institute), D-30167 Hannover, Germany*
[8] *Leibniz Universität Hannover, D-30167 Hannover, Germany*
[9] *University of Wisconsin-Milwaukee, Milwaukee, WI 53201, USA*
[10] *LIGO Laboratory, California Institute of Technology, Pasadena, CA 91125, USA*
[11] *Louisiana State University, Baton Rouge, LA 70803, USA*
[12] *Tata Institute of Fundamental Research, Mumbai 400005, India*
[13] *Université catholique de Louvain, B-1348 Louvain-la-Neuve, Belgium*
[14] *Inter-University Centre for Astronomy and Astrophysics, Pune 411007, India*
[15] *Queen Mary University of London, London E1 4NS, United Kingdom*
[16] *Department of Physics and Astronomy, Sejong University, 209 Neungdong-ro, Gwangjin-gu, Seoul 143-747, Republic of Korea*
[17] *Instituto Nacional de Pesquisas Espaciais, 12227-010 São José dos Campos, São Paulo, Brazil*
[18] *SUPA, University of the West of Scotland, Paisley PA1 2BE, United Kingdom*
[19] *Università di Roma Tor Vergata, I-00133 Roma, Italy*
[20] *INFN, Sezione di Roma Tor Vergata, I-00133 Roma, Italy*
[21] *Universiteit Antwerpen, 2000 Antwerpen, Belgium*
[22] *International Centre for Theoretical Sciences, Tata Institute of Fundamental Research, Bengaluru 560089, India*
[23] *University College Dublin, Belfield, Dublin 4, Ireland*
[24] *Gravitational Wave Science Project, National Astronomical Observatory of Japan, 2-21-1 Osawa, Mitaka City, Tokyo 181-8588, Japan*
[25] *Advanced Technology Center, National Astronomical Observatory of Japan, 2-21-1 Osawa, Mitaka City, Tokyo 181-8588, Japan*
[26] *Theoretisch-Physikalisches Institut, Friedrich-Schiller-Universität Jena, D-07743 Jena, Germany*
[27] *INFN Sezione di Torino, I-10125 Torino, Italy*
[28] *SUPA, University of Glasgow, Glasgow G12 8QQ, United Kingdom*
[29] *OzGrav, University of Western Australia, Crawley, Western Australia 6009, Australia*
[30] *Univ. Savoie Mont Blanc, CNRS, Laboratoire d'Annecy de Physique des Particules - IN2P3, F-74000 Annecy, France*
[31] *Università di Napoli "Federico II", I-80126 Napoli, Italy*
[32] *Cardiff University, Cardiff CF24 3AA, United Kingdom*
[33] *OzGrav, Australian National University, Canberra, Australian Capital Territory 0200, Australia*

[34]LIGO Laboratory, Massachusetts Institute of Technology, Cambridge, MA 02139, USA
[35]Maastricht University, 6200 MD Maastricht, Netherlands
[36]Nikhef, 1098 XG Amsterdam, Netherlands
[37]Aix Marseille Univ, CNRS, Centrale Med, Institut Fresnel, F-13013 Marseille, France
[38]Université Paris-Saclay, CNRS/IN2P3, IJCLab, 91405 Orsay, France
[39]Department of Physics, The University of Tokyo, 7-3-1 Hongo, Bunkyo-ku, Tokyo 113-0033, Japan
[40]University of Tokyo, Tokyo, 113-0033, Japan.
[41]Institut de Ciències del Cosmos (ICCUB), Universitat de Barcelona (UB), c. Martí i Franquès, 1, 08028 Barcelona, Spain
[42]Institut de Física d'Altes Energies (IFAE), The Barcelona Institute of Science and Technology, Campus UAB, E-08193 Bellaterra (Barcelona), Spain
[43]Gran Sasso Science Institute (GSSI), I-67100 L'Aquila, Italy
[44]University of Florida, Gainesville, FL 32611, USA
[45]Dipartimento di Scienze Matematiche, Informatiche e Fisiche, Università di Udine, I-33100 Udine, Italy
[46]INFN, Sezione di Trieste, I-34127 Trieste, Italy
[47]Tecnologico de Monterrey, Escuela de Ingeniería y Ciencias, Monterrey 64849, Mexico
[48]Université Côte d'Azur, Observatoire de la Côte d'Azur, CNRS, Artemis, F-06304 Nice, France
[49]Institute for Cosmic Ray Research, KAGRA Observatory, The University of Tokyo, 238 Higashi-Mozumi, Kamioka-cho, Hida City, Gifu 506-1205, Japan
[50]INFN, Sezione di Perugia, I-06123 Perugia, Italy
[51]Università di Camerino, I-62032 Camerino, Italy
[52]University of Washington, Seattle, WA 98195, USA
[53]Earthquake Research Institute, The University of Tokyo, 1-1-1 Yayoi, Bunkyo-ku, Tokyo 113-0032, Japan
[54]California State University Fullerton, Fullerton, CA 92831, USA
[55]Villanova University, Villanova, PA 19085, USA
[56]INFN, Sezione di Genova, I-16146 Genova, Italy
[57]Dipartimento di Fisica, Università degli Studi di Genova, I-16146 Genova, Italy
[58]SUPA, University of Strathclyde, Glasgow G1 1XQ, United Kingdom
[59]European Gravitational Observatory (EGO), I-56021 Cascina, Pisa, Italy
[60]Georgia Institute of Technology, Atlanta, GA 30332, USA
[61]Royal Holloway, University of London, London TW20 0EX, United Kingdom
[62]Astronomical course, The Graduate University for Advanced Studies (SOKENDAI), 2-21-1 Osawa, Mitaka City, Tokyo 181-8588, Japan
[63]Università degli Studi di Urbino "Carlo Bo", I-61029 Urbino, Italy
[64]INFN, Sezione di Firenze, I-50019 Sesto Fiorentino, Firenze, Italy
[65]LIGO Livingston Observatory, Livingston, LA 70754, USA
[66]INFN, Sezione di Roma, I-00185 Roma, Italy
[67]Università di Roma "La Sapienza", I-00185 Roma, Italy
[68]Université de Strasbourg, CNRS, IPHC UMR 7178, F-67000 Strasbourg, France
[69]Embry-Riddle Aeronautical University, Prescott, AZ 86301, USA
[70]Dipartimento di Fisica "E.R. Caianiello", Università di Salerno, I-84084 Fisciano, Salerno, Italy
[71]Université Paris Cité, CNRS, Astroparticule et Cosmologie, F-75013 Paris, France
[72]King's College London, University of London, London WC2R 2LS, United Kingdom
[73]Korea Institute of Science and Technology Information, Daejeon 34141, Republic of Korea
[74]Université libre de Bruxelles, 1050 Bruxelles, Belgium
[75]International College, Osaka University, 1-1 Machikaneyama-cho, Toyonaka City, Osaka 560-0043, Japan
[76]Institute for Gravitational and Subatomic Physics (GRASP), Utrecht University, 3584 CC Utrecht, Netherlands
[77]University of Portsmouth, Portsmouth, PO1 3FX, United Kingdom
[78]University of Oregon, Eugene, OR 97403, USA
[79]Syracuse University, Syracuse, NY 13244, USA
[80]Northwestern University, Evanston, IL 60208, USA
[81]Departament de Física Quàntica i Astrofísica (FQA), Universitat de Barcelona (UB), c. Martí i Franqués, 1, 08028 Barcelona, Spain
[82]Dipartimento di Medicina, Chirurgia e Odontoiatria "Scuola Medica Salernitana", Università di Salerno, I-84081 Baronissi, Salerno, Italy
[83]HUN-REN Wigner Research Centre for Physics, H-1121 Budapest, Hungary
[84]Concordia University Wisconsin, Mequon, WI 53097, USA
[85]Universität Hamburg, D-22761 Hamburg, Germany
[86]Stanford University, Stanford, CA 94305, USA
[87]Università di Pisa, I-56127 Pisa, Italy
[88]INFN, Sezione di Pisa, I-56127 Pisa, Italy
[89]Università di Perugia, I-06123 Perugia, Italy
[90]University of Michigan, Ann Arbor, MI 48109, USA

[91]Università di Padova, Dipartimento di Fisica e Astronomia, I-35131 Padova, Italy
[92]INFN, Sezione di Padova, I-35131 Padova, Italy
[93]Institute for Plasma Research, Bhat, Gandhinagar 382428, India
[94]Universiteit Gent, B-9000 Gent, Belgium
[95]Nicolaus Copernicus Astronomical Center, Polish Academy of Sciences, 00-716, Warsaw, Poland
[96]University of Minnesota, Minneapolis, MN 55455, USA
[97]Université Claude Bernard Lyon 1, CNRS, IP2I Lyon / IN2P3, UMR 5822, F-69622 Villeurbanne, France
[98]IAC3–IEEC, Universitat de les Illes Balears, E-07122 Palma de Mallorca, Spain
[99]Università di Siena, I-53100 Siena, Italy
[100]RRCAT, Indore, Madhya Pradesh 452013, India
[101]Kenyon College, Gambier, OH 43022, USA
[102]Missouri University of Science and Technology, Rolla, MO 65409, USA
[103]Colorado State University, Fort Collins, CO 80523, USA
[104]Department of Physics and Astronomy, Vrije Universiteit Amsterdam, 1081 HV Amsterdam, Netherlands
[105]Lomonosov Moscow State University, Moscow 119991, Russia
[106]Katholieke Universiteit Leuven, Oude Markt 13, 3000 Leuven, Belgium
[107]Università di Trento, Dipartimento di Fisica, I-38123 Povo, Trento, Italy
[108]INFN, Trento Institute for Fundamental Physics and Applications, I-38123 Povo, Trento, Italy
[109]Rochester Institute of Technology, Rochester, NY 14623, USA
[110]Bar-Ilan University, Ramat Gan, 5290002, Israel
[111]University of British Columbia, Vancouver, BC V6T 1Z4, Canada
[112]INFN, Sezione di Napoli, Gruppo Collegato di Salerno, I-80126 Napoli, Italy
[113]OzGrav, University of Adelaide, Adelaide, South Australia 5005, Australia
[114]Centre national de la recherche scientifique, 75016 Paris, France
[115]Univ Rennes, CNRS, Institut FOTON - UMR 6082, F-35000 Rennes, France
[116]University of Birmingham, Birmingham B15 2TT, United Kingdom
[117]Washington State University, Pullman, WA 99164, USA
[118]INFN, Laboratori Nazionali del Gran Sasso, I-67100 Assergi, Italy
[119]Laboratoire Kastler Brossel, Sorbonne Université, CNRS, ENS-Université PSL, Collège de France, F-75005 Paris, France
[120]Christopher Newport University, Newport News, VA 23606, USA
[121]OzGrav, University of Melbourne, Parkville, Victoria 3010, Australia
[122]Astronomical Observatory Warsaw University, 00-478 Warsaw, Poland
[123]University of Maryland, College Park, MD 20742, USA
[124]Università degli Studi di Milano-Bicocca, I-20126 Milano, Italy
[125]INFN, Sezione di Milano-Bicocca, I-20126 Milano, Italy
[126]L2IT, Laboratoire des 2 Infinis - Toulouse, Université de Toulouse, CNRS/IN2P3, UPS, F-31062 Toulouse Cedex 9, France
[127]Université de Lyon, Université Claude Bernard Lyon 1, CNRS, Institut Lumière Matière, F-69622 Villeurbanne, France
[128]IGFAE, Universidade de Santiago de Compostela, 15782 Spain
[129]University of Chicago, Chicago, IL 60637, USA
[130]University of Arizona, Tucson, AZ 85721, USA
[131]University of Massachusetts Dartmouth, North Dartmouth, MA 02747, USA
[132]INFN, Laboratori Nazionali del Sud, I-95125 Catania, Italy
[133]Niels Bohr Institute, Copenhagen University, 2100 København, Denmark
[134]Istituto di Astrofisica e Planetologia Spaziali di Roma, 00133 Roma, Italy
[135]Departamento de Astronomía y Astrofísica, Universitat de València, E-46100 Burjassot, València, Spain
[136]Observatori Astronòmic, Universitat de València, E-46980 Paterna, València, Spain
[137]Niels Bohr Institute, University of Copenhagen, 2100 Kóbenhavn, Denmark
[138]Department of Physics, National Cheng Kung University, No.1, University Road, Tainan City 701, Taiwan
[139]National Tsing Hua University, Hsinchu City 30013, Taiwan
[140]National Central University, Taoyuan City 320317, Taiwan
[141]OzGrav, Charles Sturt University, Wagga Wagga, New South Wales 2678, Australia
[142]Vanderbilt University, Nashville, TN 37235, USA
[143]Department of Electrophysics, National Yang Ming Chiao Tung University, 101 Univ. Street, Hsinchu, Taiwan
[144]Kamioka Branch, National Astronomical Observatory of Japan, 238 Higashi-Mozumi, Kamioka-cho, Hida City, Gifu 506-1205, Japan
[145]University of Texas, Austin, TX 78712, USA
[146]CaRT, California Institute of Technology, Pasadena, CA 91125, USA
[147]Cornell University, Ithaca, NY 14850, USA

[148]Northeastern University, Boston, MA 02115, USA
[149]OzGrav, School of Physics & Astronomy, Monash University, Clayton 3800, Victoria, Australia
[150]Dipartimento di Ingegneria Industriale (DIIN), Università di Salerno, I-84084 Fisciano, Salerno, Italy
[151]INAF, Osservatorio Astronomico di Padova, I-35122 Padova, Italy
[152]OzGrav, Swinburne University of Technology, Hawthorn VIC 3122, Australia
[153]University of Rhode Island, Kingston, RI 02881, USA
[154]INFN Cagliari, Physics Department, Università degli Studi di Cagliari, Cagliari 09042, Italy
[155]Université Libre de Bruxelles, Brussels 1050, Belgium
[156]INAF, Osservatorio Astronomico di Brera sede di Merate, I-23807 Merate, Lecco, Italy
[157]Departamento de Matemáticas, Universitat de València, E-46100 Burjassot, València, Spain
[158]Montana State University, Bozeman, MT 59717, USA
[159]Johns Hopkins University, Baltimore, MD 21218, USA
[160]The University of Texas Rio Grande Valley, Brownsville, TX 78520, USA
[161]Université de Liège, B-4000 Liège, Belgium
[162]DIFA- Alma Mater Studiorum Università di Bologna, Via Zamboni, 33 - 40126 Bologna, Italy
[163]Istituto Nazionale Di Fisica Nucleare - Sezione di Bologna, viale Carlo Berti Pichat 6/2, Bologna, Italy
[164]University of Manitoba, Winnipeg, MB R3T 2N2, Canada
[165]INFN-CNAF - Bologna, Viale Carlo Berti Pichat, 6/2, 40127 Bologna BO, Italy
[166]Chennai Mathematical Institute, Chennai 603103, India
[167]Università degli Studi di Sassari, I-07100 Sassari, Italy
[168]Université de Normandie, ENSICAEN, UNICAEN, CNRS/IN2P3, LPC Caen, F-14000 Caen, France
[169]Laboratoire de Physique Corpusculaire Caen, 6 boulevard du maréchal Juin, F-14050 Caen, France
[170]Université Claude Bernard Lyon 1, CNRS, Laboratoire des Matériaux Avancés (LMA), IP2I Lyon / IN2P3, UMR 5822, F-69622 Villeurbanne, France
[171]Università di Firenze, Sesto Fiorentino I-50019, Italy
[172]Dipartimento di Scienze Matematiche, Fisiche e Informatiche, Università di Parma, I-43124 Parma, Italy
[173]INFN, Sezione di Milano Bicocca, Gruppo Collegato di Parma, I-43124 Parma, Italy
[174]University of Sannio at Benevento, I-82100 Benevento, Italy and INFN, Sezione di Napoli, I-80100 Napoli, Italy
[175]Marquette University, Milwaukee, WI 53233, USA
[176]Perimeter Institute, Waterloo, ON N2L 2Y5, Canada
[177]Corps des Mines, Mines Paris, Université PSL, 60 Bd Saint-Michel, 75272 Paris, France
[178]Indian Institute of Technology Madras, Chennai 600036, India
[179]Graduate School of Science, Tokyo Institute of Technology, 2-12-1 Ookayama, Meguro-ku, Tokyo 152-8551, Japan
[180]National Center for Nuclear Research, 05-400 Świerk-Otwock, Poland
[181]Institut d'Astrophysique de Paris, Sorbonne Université, CNRS, UMR 7095, 75014 Paris, France
[182]Vrije Universiteit Brussel, 1050 Brussel, Belgium
[183]Faculty of Science, University of Toyama, 3190 Gofuku, Toyama City, Toyama 930-8555, Japan
[184]Canadian Institute for Theoretical Astrophysics, University of Toronto, Toronto, ON M5S 3H8, Canada
[185]University of Cambridge, Cambridge CB2 1TN, United Kingdom
[186]Stony Brook University, Stony Brook, NY 11794, USA
[187]Center for Computational Astrophysics, Flatiron Institute, New York, NY 10010, USA
[188]Montclair State University, Montclair, NJ 07043, USA
[189]Dipartimento di Fisica, Università di Trieste, I-34127 Trieste, Italy
[190]HUN-REN Institute for Nuclear Research, H-4026 Debrecen, Hungary
[191]Centro de Física das Universidades do Minho e do Porto, Universidade do Minho, PT-4710-057 Braga, Portugal
[192]Centre de Physique des Particules de Marseille, 163, avenue de Luminy, 13288 Marseille cedex 09, France
[193]CNR-SPIN, I-84084 Fisciano, Salerno, Italy
[194]Scuola di Ingegneria, Università della Basilicata, I-85100 Potenza, Italy
[195]Western Washington University, Bellingham, WA 98225, USA
[196]Barry University, Miami Shores, FL 33168, USA
[197]Centro de Astrofísica e Gravitação, Departamento de Física, Instituto Superior Técnico - IST, Universidade de Lisboa - UL, Av. Rovisco Pais 1, 1049-001 Lisboa, Portugal
[198]Eötvös University, Budapest 1117, Hungary
[199]Institute for Cosmic Ray Research, KAGRA Observatory, The University of Tokyo, 5-1-5 Kashiwa-no-Ha, Kashiwa City, Chiba 277-8582, Japan
[200]Nambu Yoichiro Institute of Theoretical and Experimental Physics (NITEP), Osaka Metropolitan University, 3-3-138 Sugimoto-cho, Sumiyoshi-ku, Osaka City, Osaka 558-8585, Japan
[201]Université Côte d'Azur, Observatoire de la Côte d'Azur, CNRS, Lagrange, F-06304 Nice, France
[202]California State University, Los Angeles, Los Angeles, CA 90032, USA

[203]Instituto de Fisica Teorica UAM-CSIC, Universidad Autonoma de Madrid, 28049 Madrid, Spain
[204]University of Zurich, Winterthurerstrasse 190, 8057 Zurich, Switzerland
[205]Laboratoire d'Acoustique de l'Université du Mans, UMR CNRS 6613, F-72085 Le Mans, France
[206]University of Szeged, Dóm tér 9, Szeged 6720, Hungary
[207]Indian Institute of Technology Bombay, Powai, Mumbai 400 076, India
[208]School of Physics and Technology, Wuhan University, Bayi Road 299, Wuchang District, Wuhan, Hubei, 430072, China
[209]University of California, Riverside, Riverside, CA 92521, USA
[210]University of Nottingham NG7 2RD, UK
[211]Ariel University, Ramat HaGolan St 65, Ari'el, Israel
[212]University of the Chinese Academy of Sciences / International Centre for Theoretical Physics Asia-Pacific, Bejing 100049, China
[213]The University of Mississippi, University, MS 38677, USA
[214]Institute of Physics, Academia Sinica, 128 Sec. 2, Academia Rd., Nankang, Taipei 11529, Taiwan
[215]Science and Technology Institute, Universities Space Research Association, Huntsville, AL 35805, USA
[216]Shanghai Astronomical Observatory, Chinese Academy of Sciences, 80 Nandan Road, Shanghai 200030, China
[217]The Chinese University of Hong Kong, Shatin, NT, Hong Kong
[218]American University, Washington, DC 20016, USA
[219]University of Nevada, Las Vegas, Las Vegas, NV 89154, USA
[220]Department of Applied Physics, Fukuoka University, 8-19-1 Nanakuma, Jonan, Fukuoka City, Fukuoka 814-0180, Japan
[221]University of California, Berkeley, CA 94720, USA
[222]University of Lancaster, Lancaster LA1 4YW, United Kingdom
[223]College of Industrial Technology, Nihon University, 1-2-1 Izumi, Narashino City, Chiba 275-8575, Japan
[224]Faculty of Engineering, Niigata University, 8050 Ikarashi-2-no-cho, Nishi-ku, Niigata City, Niigata 950-2181, Japan
[225]Department of Physics, Tamkang University, No. 151, Yingzhuan Rd., Danshui Dist., New Taipei City 25137, Taiwan
[226]Rutherford Appleton Laboratory, Didcot OX11 0DE, United Kingdom
[227]Department of Astronomy and Space Science, Chungnam National University, 9 Daehak-ro, Yuseong-gu, Daejeon 34134, Republic of Korea
[228]Scuola Normale Superiore, I-56126 Pisa, Italy
[229]Kavli Institute for Astronomy and Astrophysics, Peking University, Yiheyuan Road 5, Haidian District, Beijing 100871, China
[230]Department of Physical Sciences, Aoyama Gakuin University, 5-10-1 Fuchinobe, Sagamihara City, Kanagawa 252-5258, Japan
[231]Department of Physics, Graduate School of Science, Osaka Metropolitan University, 3-3-138 Sugimoto-cho, Sumiyoshi-ku, Osaka City, Osaka 558-8585, Japan
[232]Directorate of Construction, Services & Estate Management, Mumbai 400094, India
[233]Faculty of Physics, University of Białystok, 15-245 Białystok, Poland
[234]University of Southampton, Southampton SO17 1BJ, United Kingdom
[235]Sungkyunkwan University, Seoul 03063, Republic of Korea
[236]Department of Physics, Ulsan National Institute of Science and Technology (UNIST), 50 UNIST-gil, Ulju-gun, Ulsan 44919, Republic of Korea
[237]Institute for Cosmic Ray Research, The University of Tokyo, 5-1-5 Kashiwa-no-Ha, Kashiwa City, Chiba 277-8582, Japan
[238]Chung-Ang University, Seoul 06974, Republic of Korea
[239]University of Washington Bothell, Bothell, WA 98011, USA
[240]Aix Marseille Université, Jardin du Pharo, 58 Boulevard Charles Livon, 13007 Marseille, France
[241]Laboratoire de Physique et de Chimie de l'Environnement, Université Joseph KI-ZERBO, 9GH2+3V5, Ouagadougou, Burkina Faso
[242]Ewha Womans University, Seoul 03760, Republic of Korea
[243]Seoul National University, Seoul 08826, Republic of Korea
[244]Korea Astronomy and Space Science Institute, Daejeon 34055, Republic of Korea
[245]Institute of Particle and Nuclear Studies (IPNS), High Energy Accelerator Research Organization (KEK), 1-1 Oho, Tsukuba City, Ibaraki 305-0801, Japan
[246]Division of Science, National Astronomical Observatory of Japan, 2-21-1 Osawa, Mitaka City, Tokyo 181-8588, Japan
[247]Bard College, Annandale-On-Hudson, NY 12504, USA
[248]Institute of Mathematics, Polish Academy of Sciences, 00656 Warsaw, Poland
[249]Astronomical Observatory, Jagiellonian University, 31-007 Cracow, Poland
[250]Department of Physics and Astronomy, University of Padova, Via Marzolo, 8-35151 Padova, Italy
[251]Sezione di Padova, Istituto Nazionale di Fisica Nucleare (INFN), Via Marzolo, 8-35131 Padova, Italy
[252]Department of Physics, Nagoya University, ES building, Furocho, Chikusa-ku, Nagoya, Aichi 464-8602, Japan
[253]Université de Montréal/Polytechnique, Montreal, Quebec H3T 1J4, Canada
[254]Indian Institute of Science Education and Research, Kolkata, Mohanpur, West Bengal 741252, India
[255]Texas Tech University, Lubbock, TX 79409, USA
[256]Università degli Studi di Cagliari, Via Università 40, 09124 Cagliari, Italy
[257]Inje University Gimhae, South Gyeongsang 50834, Republic of Korea

[258]Technology Center for Astronomy and Space Science, Korea Astronomy and Space Science Institute (KASI), 776 Daedeokdae-ro, Yuseong-gu, Daejeon 34055, Republic of Korea
[259]NAVIER, École des Ponts, Univ Gustave Eiffel, CNRS, Marne-la-Vallée, France
[260]National Center for High-performance Computing, National Applied Research Laboratories, No. 7, R&D 6th Rd., Hsinchu Science Park, Hsinchu City 30076, Taiwan
[261]NASA Marshall Space Flight Center, Huntsville, AL 35811, USA
[262]Institut fuer Theoretische Astrophysik, Zentrum fuer Astronomie Heidelberg, Universitaet Heidelberg, Albert Ueberle Str. 2, 69120 Heidelberg, Germany
[263]School of Physics Science and Engineering, Tongji University, Shanghai 200092, China
[264]Institut d'Estudis Espacials de Catalunya, c. Gran Capità, 2-4, 08034 Barcelona, Spain
[265]Columbia University, New York, NY 10027, USA
[266]Institucio Catalana de Recerca i Estudis Avançats (ICREA), Passeig de Lluís Companys, 23, 08010 Barcelona, Spain
[267]Research Center for Space Science, Advanced Research Laboratories, Tokyo City University, 3-3-1 Ushikubo-Nishi, Tsuzuki-Ku, Yokohama, Kanagawa 224-8551, Japan
[268]Tsinghua University, Beijing 100084, China
[269]Institute for Photon Science and Technology, The University of Tokyo, 2-11-16 Yayoi, Bunkyo-ku, Tokyo 113-8656, Japan
[270]School of Physical & Chemical Sciences, University of Canterbury, Private Bag 4800, Christchurch 8041, New Zealand
[271]GRAPPA, Anton Pannekoek Institute for Astronomy and Institute for High-Energy Physics, University of Amsterdam, 1098 XH Amsterdam, Netherlands
[272]Institut des Hautes Etudes Scientifiques, F-91440 Bures-sur-Yvette, France
[273]Faculty of Law, Ryukoku University, 67 Fukakusa Tsukamoto-cho, Fushimi-ku, Kyoto City, Kyoto 612-8577, Japan
[274]Department of Physics and Astronomy, University of Notre Dame, 225 Nieuwland Science Hall, Notre Dame, IN 46556, USA
[275]Phenikaa Institute for Advanced Study (PIAS), Phenikaa University, To Huu street Yen Nghia Ward, Ha Dong District, Hanoi, Vietnam
[276]University of Stavanger, 4021 Stavanger, Norway
[277]Department of Astronomy, The University of Tokyo, 7-3-1 Hongo, Bunkyo-ku, Tokyo 113-0033, Japan
[278]Physics Program, Graduate School of Advanced Science and Engineering, Hiroshima University, 1-3-1 Kagamiyama, Higashihiroshima City, Hiroshima 903-0213, Japan
[279]Observatoire Astronomique de Strasbourg, 11 Rue de l'Université, 67000 Strasbourg, France
[280]Observatoire de Paris, 75014 Paris, France
[281]Laboratoire Univers et Théories, Observatoire de Paris, 92190 Meudon, France
[282]National Institute for Mathematical Sciences, Daejeon 34047, Republic of Korea
[283]Graduate School of Science and Technology, Niigata University, 8050 Ikarashi-2-no-cho, Nishi-ku, Niigata City, Niigata 950-2181, Japan
[284]Niigata Study Center, The Open University of Japan, 754 Ichibancho, Asahimachi-dori, Chuo-ku, Niigata City, Niigata 951-8122, Japan
[285]University of Maryland, Baltimore County, Baltimore, MD 21250, USA
[286]CSIR-Central Glass and Ceramic Research Institute, Kolkata, West Bengal 700032, India
[287]Consiglio Nazionale delle Ricerche - Istituto dei Sistemi Complessi, I-00185 Roma, Italy
[288]Department of Physics, Aristotle University of Thessaloniki, 54124 Thessaloniki, Greece
[289]Department of Astronomy, Yonsei University, 50 Yonsei-Ro, Seodaemun-Gu, Seoul 03722, Republic of Korea
[290]Department of Physics, University of Guadalajara, Av. Revolucion 1500, Colonia Olimpica C.P. 44430, Guadalajara, Jalisco, Mexico
[291]Hobart and William Smith Colleges, Geneva, NY 14456, USA
[292]Dipartimento di Ingegneria, Università del Sannio, I-82100 Benevento, Italy
[293]Museo Storico della Fisica e Centro Studi e Ricerche "Enrico Fermi", I-00184 Roma, Italy
[294]Dipartimento di Ingegneria Industriale, Elettronica e Meccanica, Università degli Studi Roma Tre, I-00146 Roma, Italy
[295]Subatech, CNRS/IN2P3 - IMT Atlantique - Nantes Université, 4 rue Alfred Kastler BP 20722 44307 Nantes CÉDEX 03, France
[296]Universidad de Antioquia, Medellín, Colombia
[297]Departamento de Física - ETSIDI, Universidad Politécnica de Madrid, 28012 Madrid, Spain
[298]Department of Electronic Control Engineering, National Institute of Technology, Nagaoka College, 888 Nishikatakai, Nagaoka City, Niigata 940-8532, Japan
[299]Dipartimento di Fisica e Scienze della Terra, Università Degli Studi di Ferrara, Via Saragat, 1, 44121 Ferrara FE, Italy
[300]Faculty of Science, Toho University, 2-2-1 Miyama, Funabashi City, Chiba 274-8510, Japan
[301]Indian Institute of Technology, Palaj, Gandhinagar, Gujarat 382355, India
[302]Laboratoire MSME, Cité Descartes, 5 Boulevard Descartes, Champs-sur-Marne, 77454 Marne-la-Vallée Cedex 2, France
[303]University of Tsukuba, 1-1-1, Tennodai, Tsukuba, Ibaraki 305-8573, Japan
[304]Institute for Quantum Studies, Chapman University, 1 University Dr., Orange, CA 92866, USA
[305]Faculty of Information Science and Technology, Osaka Institute of Technology, 1-79-1 Kitayama, Hirakata City, Osaka 573-0196, Japan
[306]NASA Goddard Space Flight Center, Greenbelt, MD 20771, USA
[307]iTHEMS (Interdisciplinary Theoretical and Mathematical Sciences Program), RIKEN, 2-1 Hirosawa, Wako, Saitama 351-0198, Japan
[308]Scuola Internazionale Superiore di Studi Avanzati, Via Bonomea, 265, I-34136, Trieste TS, Italy
[309]Laboratoire de Physique de l'École Normale Supérieure, ENS, (CNRS, Université PSL, Sorbonne Université, Université Paris Cité), F-75005 Paris, France
[310]Institut für Theoretische Physik, Johann Wolfgang Goethe-Universität, Max-von-Laue-Str. 1, 60438 Frankfurt am Main, Germany

[311]INAF, Osservatorio di Astrofisica e Scienza dello Spazio, I-40129 Bologna, Italy
[312]Universidade Estadual Paulista, 01140-070 São Paulo, Brazil
[313]Faculty of Physics, University of Warsaw, Ludwika Pasteura 5, 02-093 Warszawa, Poland
[314]Department of Physics, Kyoto University, Kita-Shirakawa Oiwake-cho, Sakyou-ku, Kyoto City, Kyoto 606-8502, Japan
[315]Yukawa Institute for Theoretical Physics (YITP), Kyoto University, Kita-Shirakawa Oiwake-cho, Sakyou-ku, Kyoto City, Kyoto 606-8502, Japan
[316]Carleton College, Northfield, MN 55057, USA
[317]University of Catania, Department of Physics and Astronomy, Via S. Sofia, 64, 95123 Catania CT, Italy
[318]National Institute of Technology, Fukui College, Geshi-cho, Sabae-shi, Fukui 916-8507, Japan
[319]Department of Communications Engineering, National Defense Academy of Japan, 1-10-20 Hashirimizu, Yokosuka City, Kanagawa 239-8686, Japan
[320]Texas A&M University, College Station, TX 77843, USA
[321]Eindhoven University of Technology, 5600 MB Eindhoven, Netherlands
[322]The University of Utah, Salt Lake City, UT 84112, USA
[323]Kavli Institute for the Physics and Mathematics of the Universe, WPI, The University of Tokyo, 5-1-5 Kashiwa-no-Ha, Kashiwa City, Chiba 277-8583, Japan
[324]Department of Astronomy, Beijing Normal University, Xinjiekouwai Street 19, Haidian District, Beijing 100875, China